\documentclass[numbered,authoryear,preprint,3p,times,10.5pt]{elsarticle}
\usepackage{tikz}
\usepackage[]{lineno}
\usetikzlibrary{arrows,intersections}
\usepackage{txfonts}
\usepackage{natbib}
\usepackage{epsfig}
\usepackage{float}
\usepackage{graphicx}
\usepackage{subfigure}
\usepackage{epstopdf}
\usepackage{amssymb}
\usepackage[colorlinks,linkcolor=blue,hyperindex,CJKbookmarks]{hyperref}
\usepackage{array}
\usepackage{latexsym,bm,amsmath}
\usepackage{amsfonts}
\usepackage{amsthm}
\usepackage{mathrsfs}
\usepackage{color}
\usepackage[toc,page,title,titletoc,header]{appendix}
\usepackage{threeparttable}
\usepackage{lscape}
\usepackage{multirow}
\usepackage{url}
\usepackage{amsmath}
\usepackage{extarrows}
\usepackage{lipsum}
\usepackage[utf8]{inputenc}
\usepackage{cuted}
\usepackage[margins]{trackchanges}
\graphicspath{{fig/}}

\newcommand{\EQ}{\begin{eqnarray}}
\newcommand{\EN}{\end{eqnarray}}
\newcommand{\EQQ}{\begin{eqnarray*}}
\newcommand{\ENN}{\end{eqnarray*}}

\newcommand{\bremark}{\begin{remark} \begin{rm} }
\newcommand{\eremark}{ \end{rm} \rule{1mm}{2mm}\end{remark} }

\newcommand{\basm}{\begin{assumption} \begin{rm}}
\newcommand{\easm}{\end{rm}\end{assumption}}

\newcommand{\btheorem}{\begin{theorem} \begin{rm} }
\newcommand{\etheorem}{ \end{rm} \rule{1mm}{2mm}\end{theorem} }

\newcommand{\blemma}{\begin{lemma} \begin{rm} }
\newcommand{\elemma}{ \end{rm} \rule{1mm}{2mm}\end{lemma} }

\newcommand{\bcorollary}{\medskip\begin{corollary} \begin{rm} }
\newcommand{\ecorollary}{ \end{rm} \rule{1mm}{2mm}\end{corollary} }

\newcommand{\bdefinition}{\medskip\begin{definition}\begin{rm} }
\newcommand{\edefinition}{ \end{rm} \rule{1mm}{2mm}\end{definition} }

\newcommand{\bproposition}{\medskip\begin{proposition} \begin{rm} }
\newcommand{\eproposition}{ \end{rm} \rule{1mm}{2mm}\end{proposition} }

\newcommand{\bexample}{\begin{example} \begin{rm} }
\newcommand{\eexample}{ \end{rm} \rule{1mm}{2mm}\end{example} }

\newcommand{\bproof}{\begin{proof} \begin{rm} }
\newcommand{\eproof}{ \end{rm} \rule{1mm}{2mm}\end{proof} }

\newcommand{\baplemma}{\begin{lemma} \begin{rm} }
\newcommand{\eaplemma}{ \end{rm} \rule{1mm}{2mm}\end{lemma} }

\newtheorem{theorem}{\bf Theorem}[section]
\newtheorem{lemma}{\bf Lemma}[section]
\newtheorem{definition}{\bf Definition}[section]
\newtheorem{remark}{\bf Remark}[section]
\newtheorem{corollary}{\bf Corollary}[section]
\newtheorem{proposition}{\bf Proposition}[section]
\newtheorem{example}{\bf Example}[section]
\newtheorem{assumption}{\bf Assumption}[section]

\usepackage{algorithm}
\usepackage{algpseudocode}
\usepackage{amssymb} 
\journal{Transportation Research Part C: Emerging Technologies}
\begin{document}
\begin{frontmatter}

\title{Data efficient reinforcement learning and adaptive optimal perimeter control of network traffic dynamics}
\author[polyu]{Can Chen}
\ead{can-caesar.chen@connect.polyu.hk}
\author[polyu]{Yunping Huang}
\ead{yunping.huang@connect.polyu.hk}
\author[polyu]{William H.K. Lam}
\ead{william.lam@polyu.edu.hk}
\author[szu]{Tianlu Pan}
\ead{luciatlpan@sina.cn}
\author[polyu]{Shu-Chien Hsu}
\ead{mark.hsu@polyu.edu.hk}
\author[chulau]{Agachai Sumalee}
\ead{agachai.s@chula.ac.th}
\author[sysu]{Renxin Zhong\corref{cor}}
\ead{zhrenxin@mail.sysu.edu.cn}
\cortext[cor]{Corresponding author}

\address[polyu]{Department of Civil and Environmental Engineering, The Hong Kong Polytechnic University, Hong Kong.}
\address[szu]{Research Institute for Smart Cities, School of Architecture and Urban Planning, Shenzhen University, China}
\address[chulau]{School of Integrated Innovation, Chulalongkorn University, Bangkok, Thailand}
\address[sysu]{School of Intelligent Systems Engineering, Sun Yat-Sen University, China.}

\begin{abstract}
Existing data-driven and feedback traffic control strategies do not consider the heterogeneity of real-time data measurements. Besides, traditional reinforcement learning (RL) methods for traffic control usually converge slowly for lacking data efficiency. Moreover, conventional optimal perimeter control schemes require exact knowledge of the system dynamics and thus would be fragile to endogenous uncertainties. To handle these challenges, this work proposes an integral reinforcement learning (IRL) based approach to learning the macroscopic traffic dynamics for adaptive optimal perimeter control. This work makes the following primary contributions to the transportation literature: (a) A continuous-time control is developed with discrete gain updates to adapt to the discrete-time sensor data. Different from the conventional RL approaches, the reinforcement interval of the proposed IRL method can be varying with respect to the real-time resolution of data measurements. Approximate optimization methods are carried out to address the curse of dimensionality of the optimal control problem with consideration of the resolution of data measurement. (b) To reduce the sampling complexity and use the available data more efficiently, the experience replay (ER) technique is introduced to the IRL algorithm. (c) The proposed method relaxes the requirement on model calibration in a ``model-free'' manner that enables robustness against modeling uncertainty and enhances the real-time performance via a data-driven RL algorithm. (d) The convergence of the IRL-based algorithms and the stability of the controlled traffic dynamics are proven via the Lyapunov theory. The optimal control law is parameterized and then approximated by neural networks (NN), which moderates the computational complexity. Both state and input constraints are considered while no model linearization is required. Numerical examples and simulation experiments are presented to verify the effectiveness and efficiency of the proposed method.
\end{abstract}


\end{frontmatter}

\vspace{-1cm}

\section{Introduction} \label{Intro}

Urbanization has induced dramatic growth in car usage in metropolises around the world, which results in growing traffic congestion, accidents and pollution.
Efficient utilization of existing infrastructures via appropriate traffic control schemes is crucial to handling the fast-growing travel demand.
Over the past decades, several traffic control strategies have been proposed and successfully implemented in practice \citep[see ][for an overview]{papageorgiou2003review}. Conventional traffic control methods such as SCOOT \citep{hunt1982scoot}, SCATS \citep{lowrie1982sydney} and Traffic-responsive Urban Control (TUC) such as ALINEA \citep[see Figure 12 in][]{papageorgiou2003review}, concentrate on link-level strategies. In the case of heterogeneous networks with multiple bottlenecks and heavily directional demand flows, local traffic-responsive metering controls such as TUC may not be optimal or might not achieve the stabilization of the system in a reasonable time period \citep{kouvelas2017enhancing}. Oversaturated traffic conditions with queues spilling back to upstream links and the huge spatial dimension would introduce significant challenges to the local adaptive real-time traffic signal control strategies at the link level, i.e., SCOOT and SCATS \citep{gayah2014impacts,zhong2018robust,ZHONG2018327}. Hence, under heavily saturated traffic conditions, traffic control strategies capturing network-level congestion should be devised to alleviate network congestion.

The network-level congestion can be significantly alleviated by identifying some critical intersections and regulating them effectively \citep{kouvelas2017enhancing}. This finding gives rise to the concept of perimeter control by leveraging the recent advances in the macroscopic fundamental diagrams (MFDs).
Pioneered by \cite{godfrey1969mechanism}, with its existence proven by \cite{daganzo2007urban} theoretically, the MFDs have been widely investigated \citep{haddad2012stability,haddad2013cooperative,keyvan2013urban,leclercq2014macroscopic, yildirimoglu2014approximating,saeedmanesh2017dynamic}.
The MFD intuitively describes a low-scatter relationship between the network vehicle accumulation and production, providing an analytically simple and computationally efficient framework for aggregate modeling of urban traffic network dynamics. Under the MFD framework, a heterogeneous urban traffic network is divided into several homogeneous regions with each admits a well-defined MFD \citep{ji2012spatial}.  Under certain regularity conditions, such as stationary (or slow-varying) and evenly distributed demand, well-defined MFDs were evidenced by both simulation-based experiments \citep{gartner2004analysis} and empirical investigations \citep{geroliminis2008existence}. In particular, \cite{loder2019understanding} empirically observed the existence of the MFDs and their critical point variations using billions of vehicle observations from more than 40 cities. Further analytical consideration and empirical evidence have been provided by \cite{daganzo2008analytical,helbing2009derivation,ji2010investigating, gayah2011clockwise,daganzo2011macroscopic}.   However, heterogeneous networks, in essence, do not exhibit a well-defined MFD. Such a network can be modeled by a set of differential equations governing the traffic flow conservation in conjunction with MFDs as long as it can be partitioned into homogeneous subregions with each admits a well-defined MFD \citep{ji2012spatial}.

Considerable research efforts have been dedicated to devising optimal network traffic control strategies based on MFDs. The perimeter control is believed to be a promising solution to address spatial dimension challenge while considering the network-scale traffic congestion. Recent studies showed that feedback-based gating/perimeter control is efficient in mitigating congestion in the protected urban networks by exploiting MFDs. One goal is to manipulate the regional accumulation to the desired equilibrium (e.g., to operate the protected regions around the critical accumulation that maximizes flow), i.e., set-point control. \cite{aboudolas2013perimeter} used linear-quadratic-integral (LQI) and linear-quadratic-regulator (LQR) to operate the MFD system to approach the equilibrium points while \cite{keyvan2012exploiting} utilized a proportional-integral (PI) controller considering system uncertainty. \cite{keyvan2013urban,keyvan2015controller,keyvan2015multiple} solved the set-point control problem using the PI controller with consideration of boundary queue in MFDs, and different kinds of uncertainty and disturbance were included in the simulations. \cite{haddad2016adaptive} proposed a transfer function embedded with time delay to deal with the set-point control problem. Another goal is to achieve the maximum trip completion flow or to minimize the total travel time of the road network by properly restricting the inflow traffic to the network. \cite{daganzo2007urban} applied the MFD framework to devise a control rule that maximizes the network trip completion rate. \cite{geroliminis2013optimal,ramezani2015dynamics} solved the optimal perimeter control problem within standard two-region MFD system by model predictive control (MPC) while \cite{haddad2013cooperative} implemented MPC on a mixed network. Other optimal perimeter controls of the MFD system using MPC were in a hierarchical scheme \citep{zhou2016two,fu2017hierarchical}. \cite{aalipour2018analytical} derived an analytical optimal control policy by solving the Hamilton-Jacobi-Bellman (HJB) equation for maximizing the trip completion rates. Apart from optimal control using the MFD framework, the robust perimeter control problem of the MFD-based network traffic was also addressed in previous studies using linear matrix inequalities, e.g., \cite{haddad2014robust,haddad2015robust}. All the above methods require linearization of the MFD function except for \cite{zhong2018robust,sirmatel2021stabilization}. \cite{sirmatel2019nonlinear} developed a nonlinear moving horizon estimation scheme for large-scale urban networks subject to measurement noises in state and inflow demand. \cite{li2021robust} proposed a sliding mode controller for two-region MFD based networks considering cordon queues and heterogeneous transfer flows. Other recent efforts were put to devising resilient perimeter control under cyberattacks \citep{mercader2021resilient}, real-time state estimation in multi-region MFD urban networks \citep{saeedmanesh2021extended}, multi-region extension for the M-model that captures the effects of remaining travel distance dynamics \citep{sirmatel2021modeling}, perimeter control for congested areas against state degradation risk \citep{ding2020perimeter}, optimal perimeter control considering coupled/decoupled controllers \citep{haddad2017optimalc} and aggregate boundary queue dynamics \citep{haddad2017optimalp}.

The aforementioned studies on perimeter control can be regarded as model-based traffic responsive control. Specifically, previous studies on the feedback-based perimeter control were derived under one common assumption that the model parameters can be accurately calibrated. For optimal perimeter control, in particular, it is generally assumed that perfect knowledge of the network is available and the parameters will not change during the planning horizon. Moreover, local linearization around the desired equilibrium is widely performed to simplify the control design.  Apart from model-based traffic responsive control, considerable research efforts have focused on adaptive perimeter control which must adapt to a controlled system with time-varying and/or uncertain parameters or external disturbances such as travel demand noise. By considering the boundary queue that can have a negative impact on upstream queue modeling, \cite{kouvelas2017enhancing} introduced an online adaptive parameters optimization algorithm for perimeter control.  \cite{HADDAD2020133} designed the distributed adaptive perimeter control laws with control gains varying with time considering state delays and interconnection delays. Since traffic networks are subject to various uncertainties, parameters of MFDs are uncertain and time-varying. Also, the travel demand and traffic control strategies can significantly affect the shape of the MFD \citep{geroliminis2012effect}. The performance of these control strategies increasingly deteriorates with increasing disturbance prediction and model errors \citep{zhong2014optimal,baldi2019simulation}. Nevertheless, as specified in \cite{kouvelas2017enhancing}, in many cases the adopted models are calibrated once and would not be re-calibrated regularly. This causes a defect in their field experiments. Despite the vast literature related to modeling and control with MFDs, the design of dynamic control policies to various exogenous disturbances that can affect the dynamics is seldom considered. To adapt the real-time observation and then the control to operate the traffic network optimally, it is necessary to keep adjusting the model parameters \citep{kouvelas2017enhancing}. However, this process can be a heavy computational burden and difficult to be implemented in real-time \citep{modares2014flow}. For an ever-changing traffic environment subject to various exogenous disturbances, a predefined model-based traffic responsive policy may become suboptimal or even impractical. Yet in the literature, to the best knowledge of the authors, few existing studies dealt with the problem of devising the adaptive control strategies for MFD systems with (partially) unknown system dynamics. \cite{lei2019data} and \cite{REN2020102618} devised a ``model-free'' perimeter controller for a multi-region MFD-based network via the iterative learning control by assuming recurrent traffic conditions that the traffic dynamics would not admit a significant change in a day-to-day time-scale during the learning period.

The emerging big data technology gives rise to data-driven approaches to solving the aforementioned difficulties.  Rooted in computer science, the reinforcement learning (RL) has attracted increasing attention recently for its success in video games \citep{mnih2015human} and Go \citep{silver2016mastering}. Under the RL setting, an agent optimizes a goal-oriented long-term reward via policy learning. At each step, the RL agent interacts with the environment and evaluates the performance of its action based on the feedback from the environment. The agent then tries to improve the performance of subsequent actions \citep{sutton2018reinforcement}. A reformulation of RL is called adaptive dynamic programming (ADP) in economics and management communities. The RL and ADP bridge the gap between optimal control and adaptive control. In an off-line manner, the RL and ADP provide an approximate solution to the optimal control problem obtained from the Pontryagin's minimum principle and the dynamic programming principle (i.e., the HJB equation). Solving the HJB equation takes the center stage in deriving optimal control strategies. However, the HJB equation is generally intractable to be solved by analytical approaches for strong nonlinearity, possible discontinuities in the solution and the curse of dimensionality. To handle the curse of dimensionality in optimal traffic signal control design for large-scale networks, \cite{baldi2019simulation} parametrized the solution of the HJB equation using an appropriate Lyapunov function. The simulation results showed that the approximately optimal traffic signal control design via low-complexity parametrization of the HJB equation can provide a satisfactory trade-off between computational complexity and network performance. However, there is a lack of analytical proof of the convergence as well as the explicit consideration of saturated constraints on the system state and input in \cite{baldi2019simulation}.  A conventional ADP based RL algorithm was proposed by \cite{SU2020102628} to provide an analytical optimal perimeter control law for the MFD dynamics. Both convergence and stability of the closed-loop system were achieved. However, conventional approximation techniques for solving the HJB equations require complete or partial knowledge of the system dynamics and are normally off-line. Thus, they cannot handle modeling uncertainties and be deployed for real-time applications.

The RL and ADP help the optimal control circumvent the requirement on complete knowledge of the system dynamics so that uncertainties and changes in dynamics can be incorporated into the optimal control framework. Compared with the off-line nature of the conventional optimal control framework, the RL and ADP can find the optimal solution online in real-time using data-driven mechanism meanwhile robustness and adaptiveness can be well achieved. Data-driven deep reinforcement learning has been incorporated in solving traffic optimization problems \citep{kheterpal2018flow}. RL in the data-driven control community gives rise to a promising solution for optimal perimeter control problems in a ``model-free'' manner. \cite{zhou2021model} proposed a deep RL based scheme for the two-region perimeter control problems, which can achieve comparable performances to the MPC approach.

Considering that traffic data are collected from sensors in a discrete-time manner, we would like to establish a continuous-time control (MFD dynamics) with discrete gain updates (adapting to the sensor data). Generally, the sample time interval of the data collected by a type of sensor is fixed. Thus, sensors of various types deployed in a traffic network would be heterogeneous with different resolutions of data measurements. It would be much better if the reinforcement intervals can be varying with respect to the real-time resolutions of data measurements, i.e., the reinforcement intervals can be selected online to ensure the data-driven RL algorithms do have rich data. Existing works utilizing RL such as \cite{SU2020102628,zhou2021model} do not consider such issues. Different from the traditional online RL approaches, the reinforcement intervals of the integral reinforcement learning (IRL)  need not be identical and can be adjusted online, which consequently is more suitable for real-world traffic data measurement and  allows adaptive online learning to guarantee real-time performance.  Based on the idea of IRL, an equivalent Bellman equation, namely the IRL Bellman equation was developed. An online policy iteration algorithm was developed for the optimal control problem of continuous-time systems via solving the IRL Bellman equation in \cite{VRABIE2009477}.
This adaptive optimal control does not explicitly employ the knowledge on system dynamics, i.e., ``model-free".

The actor-critic (AC) structure contributes significantly to the success of RL algorithms. In the AC structure, the actor deploys a control policy to the system or environment, while the critic evaluates the cost induced by the implemented control policy and provides reward signals to the actor. The actor-critic dual neural networks (AC-NN) can be used to circumvent the ``curse of dimensionality''.  Despite the adaptive learning capability, traditional RL approaches usually converge slowly for lacking  data efficiency, which is a major obstacle to real-time applications. Experience replay (ER) technique, also known as concurrent learning,  provides a promising approach to improve the efficiency of RL algorithms\footnote{Another benefit of ER is conquering the difficulty arising in the persistently exciting condition for nonlinear systems.}. The ER technique uses historical and current data simultaneously in a ``smart'' manner.  It has been found that the AC structure can be integrated with the ER technique to improve the data efficiency and convergent speed of RL algorithms \citep{modares2014flow}.

To handle the aforementioned challenges, this paper makes the following primary contributions to the transportation literature.
\begin{itemize}
    \item Robustness to heterogeneous data resolutions. Unlike the conventional RL algorithms, the reinforcement intervals of the proposed IRL approach can be selected online to adapt to heterogeneous real-time data resolutions. The introduction of the ER technique to RL algorithms can speed up their convergence when limited real-time data are available due to unexpected longer sample time intervals.
    \item Data efficiency. In the ER technique, a number of recent samples are stored in a database and are presented repeatedly to the underlying RL algorithm, which enhances their data efficiency. An easy-check rank condition is introduced to verify the data richness requirement and reduce sampling complexity.
    \item Model-free against modeling uncertainties. It is desirable for the controller to handle the modeling uncertainties. Unlike the previous studies which rely on exact knowledge of the underlying system dynamics, a key advantage of the proposed method is that the exact knowledge of the traffic model is no longer needed. Also, the proposed approach does not rely on the widely used model linearization.
    \item Incorporating real-time data-driven components for adaptiveness. It is necessary to enable the controller to adapt to the real-time traffic conditions, e.g., traffic incidents.  To this end, an online adaptive data-driven perimeter controller is devised.
    \item Convergence and stability guaranteed. Unlike many existing studies in the transportation literature that use RL algorithms without proof of convergence nor stability, this work guarantees the closed-loop stability of the overall system by leveraging the RL with Lyapunov theory. The input and state constraints are explicitly considered in the proposed IRL algorithms.
  \end{itemize}

This paper is organized as follows: \autoref{DPRL} provides a brief introduction to dynamic programming and IRL. \autoref{Prob} discusses the optimal perimeter control problem formulation.
\autoref{metdlg} develops a model-free data-driven IRL method for optimal adaptive perimeter control.
Then \autoref{ImpmNN} performs the implementations of the proposed online iterative learning algorithm with NNs. Numerical results are presented in \autoref{Simul} and a microscopic simulation experiment is provided in \autoref{SimEx}. Finally, \autoref{Conclu} provides concluding remarks. For convenience, we summarize the key notation used in this paper in \autoref{keynotation}.

\begin{table}[!htb]
    \caption{List of key notations}\label{keynotation}
    \centering
    \begin{threeparttable}
\begin{tabular}{ll}
\hline
Symbol         & Meaning \\
\hline%
$\mathbb{R}$   & The set of all real numbers \\
$\mathbb{R}^m$ & The Euclidean space of all real $m$-vectors \\
$\mathbb{R}^{n\times m}$ & The space of all $n\times m$ real matrices \\
$T$            & The transposition symbol \\
$\Omega$       & A compact set of $\mathbb{R}^n$ \\
$C^m(\Omega)$  & The class of functions having continuous $m$-th derivative on $\Omega$ \\
$\|x(t)\|_{L_2}$ & The $L_2$ norm of continuous-time function $x(t)$ while we use $\|x(t)\|$ for brevity \\
$\succ$        & A matrix $R\succ 0$ means that it is positive definite and $\succeq$ denotes positive semi-definite \\
$\otimes$ & Kronecker product: $\forall$ $A\in \mathbb{R}^{m\times n}$, $B\in \mathbb{R}^{p\times q}$,
$ A\otimes B\triangleq
\begin{bmatrix}
 a_{11}B &  \cdots & a_{1n}B \\
 \vdots &  \ddots & \vdots \\
a_{m1}B   & \cdots & a_{mn}B \\
\end{bmatrix}
$\\
$\odot$        & Hadamard product: $\forall$ $A,B\in \mathbb{R}^{m\times n}$, $ A\odot B\triangleq
\begin{bmatrix}
 a_{11}b_{11} &  \cdots & a_{1n}b_{1n} \\
 \vdots &  \ddots & \vdots \\
a_{m1}b_{m1}   & \cdots & a_{mn}b_{mn} \\
\end{bmatrix}
$
\\
$Z_i$          & The directly reachable regions from region $i$ except itself, i.e., $i \not\in Z_i$ \\
$n_{ij}(t)$    & Number of vehicles in region $i$ with destination to region $j$ at time $t$  \\
$n_{i}(t)$     & Accumulation or total number of vehicles in region ${i}$ at time $t$, and $n_{i}(t)=n_{ii}(t)+\sum\limits_{j\in Z_{i}} n_{ij}(t)$  \\
$q_{ij}(t)$    & Travel demand defined as a flow in which its origin is region $i$ and destination is region $j$ \\
$u_{ij}(t)$    & Perimeter controllers controlling the ratio of the transfer flow that transfers from region $i$ to region $j$ at time $t$\\
$G_{i}(n_{i}(t))$ & MFD that maps the network accumulation $n_{i}(t)$ to trip completion rate for region $i$ at time $t$ \\
$\alpha_n$     & The dimension of the accumulation state $n$ \\
$\alpha_u$     & the dimension of the perimeter control $u$ \\
$\alpha_q$     & the dimension of the travel demand $q$ \\
\hline
\end{tabular}
\end{threeparttable}
\end{table}

\begin{figure}[h]
  \centering
  \includegraphics[width=4in]{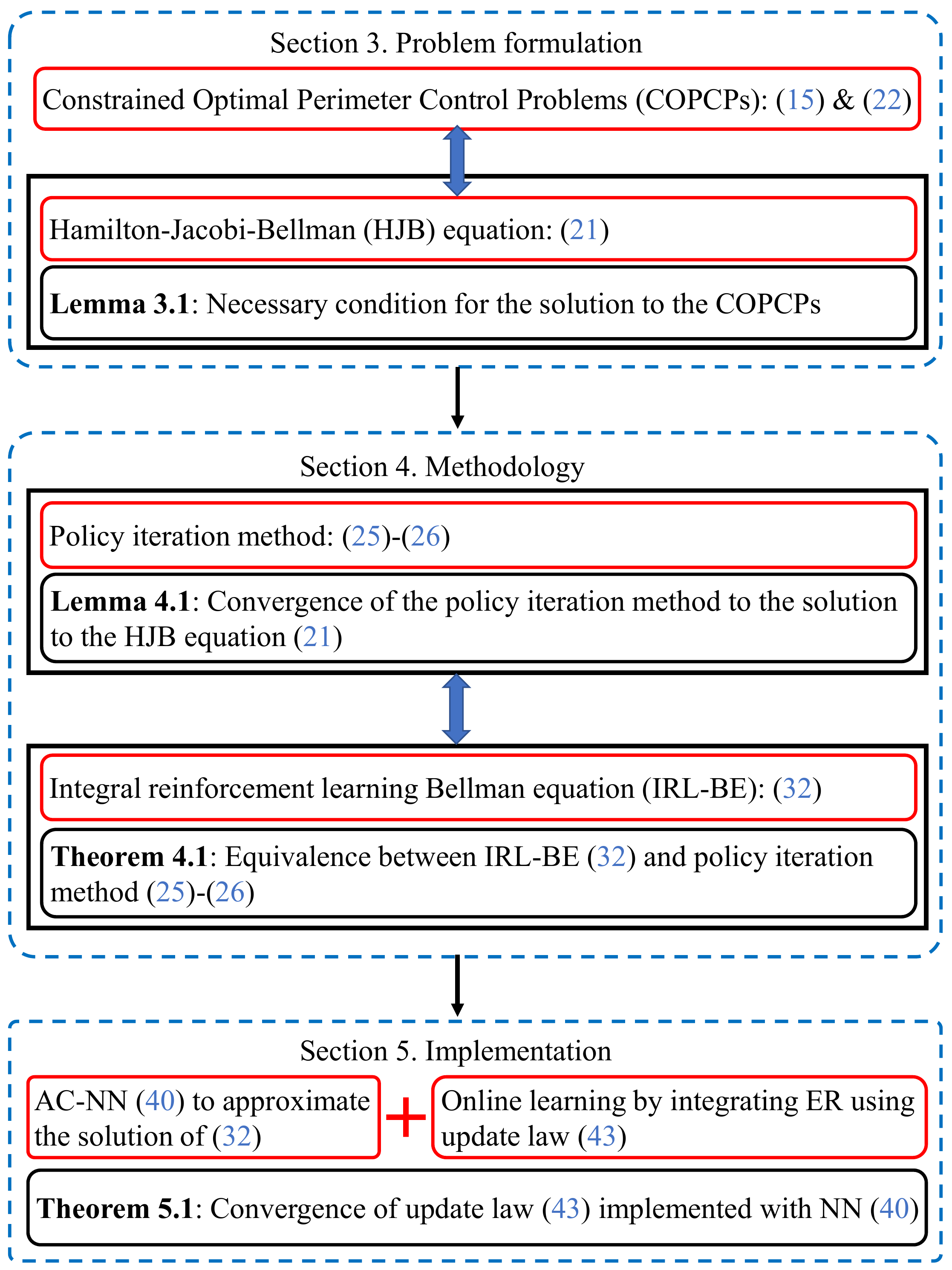}\\
  \caption{Gallery of this study}\label{gallery}
\end{figure}

For a better grasp of the logical structure of this paper, the flow for reasoning is depicted in \autoref{gallery}. Consider the MFD based system in the affine form as \eqref{eqmulti}. In the optimal regulation problem, the objective is to design an optimal perimeter control such that the accumulation state converges to the desired equilibrium by minimizing a cost functional defined by \eqref{UPobj}. This optimal cost function and perimeter control can be derived via solving an equivalent HJB equation \eqref{UpHJB}. Note that the solution of \eqref{UpHJB} may be intractable due to its strong nonlinearity. One of the most common methods to resolve this difficulty is the policy iteration method \eqref{eq8}-\eqref{eq9}. However, because of the heterogeneity of real-time data measurements and the lack of complete knowledge on the system dynamics, we cannot apply the policy iteration method directly in practice. Hence, an equivalent formulation of the policy iteration method, namely, the IRL Bellman equation given by \eqref{eq18} is established, which can adapt to the time-varying real-time data resolution and does not involve the system dynamics, i.e., ``model-free''. Finally, based on \eqref{eq18}, an online iterative learning scheme with experience replay adaptation law \eqref{ERupW} are established, which can be implemented with the AC-NN framework \eqref{NNe} to boost the computational efficiency and approximate the optimal perimeter controller $\tilde{u}^*$.

\section{Preliminaries: dynamic programming and integral reinforcement learning} \label{DPRL}

In this section, we outline how the HJB equation couples the performance functional, the nonlinear system dynamics, the IRL Bellman equation, and the neural network framework.

Consider the nonlinear system in the affine form  as
\begin{equation} \label{dynamics}
\begin{split}
  \dot {x}(t) & = f(x(t)) + g(x(t))u(t) \\
  y(t)  & =l(x(t))
\end{split}
\end{equation}
where $x(t)\in \mathbb {R}^{n}$, $u(t)\in \mathbb {R}^{m}$, and $y(t)\in \mathbb {R}^{p}$ represent the state, the control input, and the output of \eqref{dynamics}, respectively. We call $f(x(t))\in \mathbb {R}^{n}$ the drift dynamics, $g(x(t))\in \mathbb {R}^{n\times m}$  the input dynamics, and $l(x(t))\in \mathbb {R}^{p}$ the output dynamics, respectively. It is assumed that $f(0)=0$ and $f(x(t))+g(x(t))u(t)$ is locally Lipschitz and the system is stabilizable.

In the optimal regulation problem, the objective is to design an optimal control input such that the controlled state of (\ref{dynamics}) converges to the desired equilibrium by minimizing a cost functional defined as

\begin{equation*} {J}(x(0),u)=\int _{0}^\infty \mathcal{L}(x(t),u(t)) \textrm{d}t\equiv \int _{0}^\infty (N(x(t))+ U(u(t)))) \textrm{d}t \end{equation*}
where $N(x)\succeq 0$ and $U(u)\succeq 0$, with $\succeq$ denotes positive semi-definite. Generally, we choose $U(u)=u^T R u, \; R=R^{T}\succ 0$ and $R\in \mathbb{R}^{m\times m}$ for unconstrained control case with $\succ$ denotes positive definite.
However, for many real applications,  $u$ is saturated, i.e., $\mid u_{\xi} \mid \leq \lambda$, $\xi =1, \ldots, m$, where $\lambda >0$ is the performance limit of the actuator.   \cite{abu2008neurodynamic} proposed the following generalized non-quadratic functional to consider the effect of saturation on control input $u$.
\begin{equation*}
  U(u)=2\int^{u}_{0} \lambda \tanh^{-T}(s/\lambda)R\textrm{d}s,\ \lambda >0
\end{equation*}
  Without loss of generality, let $R=\textrm{diag}(\gamma_{1},\ldots,\gamma_{m})$ be a positive definite matrix of proper dimension.

\bdefinition
(\textsl{Admissible Control}) A control policy $\mu$ is admissible to \eqref{dynamics}, if $\mu(x)$ is continuous on $\Omega$, $\mu(0)=0$ and $\mu$ stabilizes the system \eqref{dynamics} on $\Omega$ with the value function \eqref{VF} being finite for $\forall x_{0}\in \Omega$.
\label{defAC}
\edefinition

The value function for an admissible control policy can be defined as
\begin{equation}
    {V}(x,u)=\int _{t}^\infty \mathcal{L}(x(\tau),u(\tau)) \textrm{d}\tau \equiv \int _{t}^\infty (N(x(\tau))+ U(u(\tau)))) \textrm{d}\tau
\label{VF}
\end{equation}
The Hamiltonian function is
\begin{equation*}
    {H}\left ({x,u,\frac { \partial V}{\partial x}}\right )=\mathcal{L}(x,u)+ \left(\frac { \partial V}{\partial x}\right)^{T} (f(x)+g(x) u)
\end{equation*}
Since the integrand of the performance functional does not depend on time explicitly and the terminal time is fixed (or infinite time) while \eqref{dynamics} is an autonomous dynamical system, the optimality is given by ${H}\left ({x,u,\frac { \partial V}{\partial x}}\right ) = 0$, i.e., the Bellman optimality equation
\begin{equation} \label{Bellman}
    \mathcal{L}(x,u^\star )+ \left(\frac { \partial V^\star }{\partial x}\right)^{T} (f(x)+g(x) u^\star )=0
\end{equation}
The optimal control can be obtained as
\begin{equation}
\label{ustar}
    u^\star = -\lambda \tanh\left (\frac{1}{2\lambda}R^{-1} g^{T}(x) \frac { \partial V^\star }{\partial x}\right )
\end{equation}

Since sensors collect data and transfer them to the controllers with prescribed time resolutions, we cannot apply the Bellman equation \eqref{Bellman} directly in practice. Moreover, \eqref{Bellman}  involves the exact system dynamics $f(x)$ and $g(x)$. To relax this requirement and to consider sensor data measurements, an  equivalent formulation of the the Bellman equation \eqref{Bellman} that does not involve the drift dynamics can be established
\begin{align*}
    {V}(x(t))=\int _{t}^{t+\Delta t}\!(N(x(\tau ))\!+\! U(u(\tau ))) \mathrm{d}\tau \!+\!{V}(x(t+\Delta t))
\end{align*}
for any time $t \geq 0$ and time interval $\Delta t>0$. $\Delta t$ is termed as the reinforcement interval, which can be adjusted in real-time according to the resolution of sensor data and the learning rate of the RL based algorithms. This equation is called IRL Bellman equation.  By iterating on the IRL Bellman equation and updating the control policy, we can obtain both the value function and the optimal control.

Given an admissible policy $u_0$, for $j=0,1, {\dots }$, given $u_j$, solve for the value $V_{j+1}(x)$ using the following IRL Bellman equation in iteration.
\begin{equation}
    {V}_{j+1}(x(t))=\int _{t}^{t+\Delta t}\big (N(x(\tau))+ U(u_{j}(\tau))\big ) \mathrm{d}\tau +{V}_{j+1}(x(t+\Delta t))
\label{onpolicyV}
\end{equation}
on convergence, set $V_{j+1}(x(t))=V_{j}(x(t))$. Update the control policy $u_{j+1}(x(t))$ using
\begin{equation}
    u_{j+1}(x(t))= -\lambda \tanh\left (\frac{1}{2\lambda}R^{-1} g^{T}(x(t)) \frac { \partial V_{j+1} }{\partial x}\right )
\label{onpolicy}
\end{equation}
\eqref{onpolicyV}-\eqref{onpolicy} are known as an on-policy RL algorithm\footnote{On-policy and off-policy RL algorithms are devised in the literature. Their essential difference lies in how the target policy and the behavior policy are implemented. The target policy is what we are learning about, i.e., the optimal control law or the solution to the HJB equation. The target policy can be regarded as the ideal optimal policy. The behavior policy generates the action and behavior, which can be regarded as the policy implemented. The target policy and the behavior policy are the same for on-policy RL algorithms while they are different for off-policy algorithms. Generally, similar to the decision process of human beings, the off-policy algorithms can learn the optimal policies but implement suboptimal policies.}.

To uniformly approximate the value function in \eqref{onpolicyV}, we can use the following neural-network-type structure.
\begin{equation*}
    \hat {V}(x) = \hat {W}_{c}^{T} \phi_{1}(x)
\end{equation*}
where $\phi _{1}(x): \mathbb {R}^{n}\rightarrow \mathbb {R}^{N}$ is the basis function vector and $N$ is the number of basis functions. With this value function approximation, its partial derivative $\frac{\partial \hat{V}}{\partial x}$ can be approximated accordingly. Using the above approximated value function, the constrained optimal control in \eqref{ustar} can be generated by
\begin{equation*}
   \hat{u} = -\lambda\tanh\left(\frac{1}{2\lambda}R^{-1}g^T(x)\hat{W}^T_c\frac{\partial \phi_{1}}{\partial x}\right)
\end{equation*}

Incorporating these approximations into the IRL Bellman equation yields
\begin{equation*}
    e(t)=\Delta \phi _{1}(x(t))^{T} \hat {W}_{c}+\int _{t-\Delta t}^{t}( N(x(\tau))+U(\hat{u}(\tau)))\textrm{d}\tau
\end{equation*}
where $\Delta \phi _{1}(x(t))=\phi _{1}(x(t))-\phi _{1}(x(t-\Delta t))$ and $e$ is the temporal difference (TD) error after using current approximated critic weight $\hat{W}_{c}$. To avoid the case that there are insufficient real-time data for updating the weights of the learning network and to use the  data in the history stack efficiently, we consider $\Delta \phi _{1}(x(t_{j}))$ as evaluated values of $\Delta \phi_{1}$ at the recorded time $t_j$. Then, we define the Bellman equation error (i.e., TD error) at the recorded time $t_j$ using the current critic weight estimation $\hat{W}_c$ as
\begin{equation*}
    e(t_{j})=\Delta \phi _{1}(t_{j})^{T} \hat {W}_{c}+\int _{t_{j}-\Delta t}^{t_{j}}( N(x(\tau ))+U(\hat{u}(\tau )))\textrm{d}\tau
\end{equation*}
Recent transition samples (historical data) are stored and repeatedly presented to the gradient-based update rule of  the weights of the learning network \eqref{updatingwe} so as to speed up the computation and to obtain an easy-to-check convergent condition for the IRL algorithm. This process is known as the experience replay (ER) technique. The weights of the learning network are updated via minimizing simultaneously the instantaneous TD error (the first part of \eqref{updatingwe} from real-time measurement) and the TD errors for the stored transition samples (the second part of \eqref{updatingwe}), which is given as
\begin{align}
    \dot {\hat {W}}_{c}=-\alpha _{c}\frac {\Delta \phi _{1}(x(t))}{(\Delta \phi _{1}(x(t))^{T}\Delta \phi _{1}(x(t))+1)^{2}}e(t)  -\,\alpha _{c}\sum _{j=1}^{l}\frac {\Delta \phi _{1}(x(t_{j}))}{(\Delta \phi_{1}(x(t_{j}))^{T}\Delta \phi_{1}(x(t_{j}))+1)^{2}}e(t_{j})
\label{updatingwe}
\end{align}
The optimal policy (\ref{onpolicy}) implemented by the on-policy IRL algorithms does not requires the knowledge of $f(x)$. However, it still relies on the input dynamics $g(x)$.
To get rid of $g(x)$, we may adopt an off-policy IRL algorithm that the control implemented (nearly optimal) can be different from the optimal control \eqref{onpolicy}. Towards this, we rewrite the affine dynamics as
\begin{equation}
    \dot x(t) = f(x(t)) \!+ \!g(x(t))u_{j}(t)\!+\!g(x(t))(u(t)\!-\!u_{j}(t))
\label{offdynamics}
\end{equation}
where $u_j(t)$ is the policy to be updated and $u(t)$ is the behavior policy that is actually implemented to the system dynamics to generate the data for learning.
Differentiating the value function $V(x)$ along the system trajectory (\ref{offdynamics}) and using \eqref{onpolicy} yields
\begin{align*}
    {\dot V_{j}}
    =\left ({\frac {\partial {V_{j}(x)}}{\partial x}}\right ){^{T}}(f + g {u_{j}}) + \left ({\frac {\partial {V_{j}(x)}}{\partial x}}\right )^{T}g(u_{j+1} - {u_{j}})
    =- N(x) - 2\varrho^{T}(u_{j+1}) R (u_{j+1}-u_{j})-2\int^{u_{j}}_{0}\varrho^{T}(s) R\textrm{d}s
\end{align*}
where $\varrho(s)=\lambda \tanh^{-1}(s/\lambda)$. Integrating  the above equation yields the off-policy IRL Bellman equation
\begin{align}
    &\hspace {-1.5pc}V_{j}(x(t+\Delta t))- V_{j}(x(t)) = \int _{t}^{t+\Delta t} \left(- N(x) - 2\varrho^{T}(u_{j+1}) R (u_{j+1}-u_{j})-2\int^{u_{j}}_{0}\varrho^{T}(s) R\textrm{d}s \right) \textrm{d}\tau
\label{offbell}
\end{align}
For an implemented control policy $u(t)$, the off-policy IRL Bellman equation (\ref{offbell}) can be solved for both value function $V_j$ and updated policy $u_{j+1}$ simultaneously without requiring any knowledge about the system dynamics.

\section{Problem statement} \label{Prob}

In this section, we first recapitulate the dynamics for a traffic network modeled by multi-region MFD systems. We then discuss the optimal perimeter control problem formulation.

\subsection{The multi-region MFD framework} \label{lrlo}

A heterogeneous urban network decomposed into $L$ ($L>1$) homogeneous subregions wherein each region admits a well-defined MFD is considered in line with \cite{haddad2015robust,ZHONG2018327}. Let the state vector be $n(t)\triangleq [n_{11}(t), \ldots, n_{ij}(t), \ldots, n_{LL}(t)]^{T}\in \mathbb{R}^{\alpha_{{n}}}$ and the travel demand vector be $q(t)\triangleq [q_{11}(t), \ldots, q_{ij}(t), \ldots, q_{LL}(t)]^{T}\in \mathbb{R}^{\alpha_{q}}$, respectively. The control vector is $u(t)\triangleq [u_{12}(t),\ldots,u_{ij}(t),\ldots,u_{Lj}(t)]^T\in \mathbb{R}^{\alpha_{u}}$, where $u_{ij}(t)$ controls the ratio of the transfer flow that transfer from region $i$ to $j$ at time $t$. Note that $n_i(t)$ and $u_{ij}(t)$ are subject to heterogeneous constraints as given by \eqref{htrconn}-\eqref{htrconu}. The dynamic flow conservation equations of the multi-region MFD system are then formulated as follows:
\begin{subequations}
\begin{align}
    \frac{\text{d}n_{ii}(t)}{\text{d}t} &=-\frac{n_{ii}(t)}{n_{i}(t)} G_{i}(n_{i}(t))+\sum\limits_{j\in Z_{i}} \frac{n_{ji}(t)}{n_{j}(t)}G_{j}(n_{j}(t)) u_{ji}(t)+q_{ii}(t)   \label{msvc1}\\
    \frac{\text{d}n_{ij}(t)}{\text{d}t}&=-\frac{n_{ij}(t)}{n_{i}(t)} G_{i}(n_{i}(t)) u_{ij}(t)+q_{ij}(t)    \label{msvc2}\\
    n_{i}(t)&=n_{ii}(t)+\sum\limits_{j\in Z_{i}} n_{ij}(t) \label{msvc3}
\end{align}
\end{subequations}
subject to
\begin{subequations}
\begin{align}
    & 0\le n_{i} (t)\le n^{jam}_{i} \label{htrconn}\\
    & 0\le u^{\min }_{ij} \le u_{ij} (t) \le u^{\max }_{ij} \le 1 \label{htrconu}
\end{align}
\end{subequations}
where $i=1, \ldots, L$ and $j\neq i$. The state dynamics \eqref{msvc1}-\eqref{msvc2} can be written in the following affine form \citep{SU2020102628}:
\begin{equation}\label{eq:oadyn}
  \dot{n}(t)=F(n(t))+S(n(t)) u(t)
\end{equation}

One significant traffic management purpose is to devise  perimeter control $u(t)$ to regulate the cross-boundary flows such that the network accumulations $n(t)$ can converge to the desired equilibrium $n^{*}$, i.e., set-point control \citep{zhong2018robust,ZHONG2018327}. The steady state $n^*$ and the corresponding control input $u^*$ can be solved from the  steady-state equations \citep{haddad2014robust,ZHONG2018327}:
\begin{subequations}
\begin{align}
    \frac{\mathrm{d}n^*_{ii}}{\mathrm{d}t} &= 0 =-\frac{n^*_{ii}}{\bar{n}_{i}} G_{i}(\bar{n}_{i})+\sum\limits_{j\in Z_{i}} \frac{n^*_{ji}}{\bar{n}_{j}}G_{j}(\bar{n}_{j}) u^*_{ji}+q^*_{ii}   \label{ss1}\\
    \frac{\mathrm{d}n^*_{ij}}{\mathrm{d}t} &= 0 =-\frac{n^*_{ij}}{\bar{n}_{i}} G_{i}(\bar{n}_{i}) u^*_{ij}+q^*_{ij}    \label{ss2}\\
    \bar{n}_{i} &=n^*_{ii}+\sum\limits_{j\in Z_{i}} n^*_{ij} \label{ss3}
\end{align}
\end{subequations}
subject to
\begin{equation*}
    0\le \bar{n}_{i} \le n^{jam}_{i},\hspace{1em}0\le u^{\min }_{ij} \le u^*_{ij} \le u^{\max }_{ij} \le 1
\end{equation*}
where $q^*_{ii}$ and $q^*_{ij}$ are nominal demand patterns.

It is a common practice to perform a coordinate transformation to reformulate the set-point control problem into a stabilization problem \citep{zhong2018robust,ZHONG2018327}. We define $\tilde{n}(t)=[\tilde{n}_1(t),\ldots, \tilde{n}_{\alpha_{n}}(t)]^T\in \mathbb{R}^{\alpha_n}$ and $\tilde{u}(t)=[\tilde{u}_1(t),\ldots, \tilde{u}_{\alpha_u}(t)]^T\in \mathbb{R}^{\alpha_u}$ as the new state vector and new control vector, respectively. $\tilde{n}=n-n^*$ denotes the difference between the actual accumulation and the desired steady-state accumulation, while $\tilde{u}=u-u^*$ is the difference between the actual control input and the steady-state control input. After the coordinate transformation, the multi-region MFD system \eqref{msvc1}-\eqref{msvc3} can be expressed by the following standard affine form:
\begin{equation}\label{eqmulti}
  \dot{\tilde{n}}(t)=\mathbf{F}(\tilde{n}(t))+\mathbf{S}(\tilde{n}(t)) \tilde{u}(t)
\end{equation}
Both the state vector and the control vector of system \eqref{eqmulti} are restricted into some compact sets say $\tilde{n}(t)\in \Omega\subset\mathbb{R}^{\alpha_{n}}$
and $\tilde{u}(t)\in\mathcal{U}\subset\mathbb{R}^{\alpha_u}$, where $\Omega$ and $\mathcal{U}$ are the universal sets of $\tilde{n}$ and $\tilde{u}$, respectively. $\mathbf{F}$ and $\mathbf{S}$ are unknown Lipschitz continuous nonlinear functions on $\Omega\subset \mathbb{R}^{\alpha_{n}}$ containing the origin.

In \ref{appenA}, we present the dynamics in the affine form for the two-region and the three-region MFD systems, which are widely investigated in the literature.

\subsection{Optimal perimeter control of multi-region MFD system}

Set-point control and minimizing the total time spent (TTS) are two main objectives considered in the optimal perimeter control problem of MFD systems. In this subsection, we present the formulation of constrained optimal perimeter control problem (COPCP) for multi-region MFD systems considering heterogeneous cross-boundary capacities.

As a special case, \cite{SU2020102628} showed that the set-point control problem of the two-region MFD system could be modeled as a constrained optimal control problem. We will extend the formulation of set-point constrained optimal perimeter control problem (S-COPCP) for the two-region MFD system to general multi-region MFD systems while considering the heterogeneous cross-boundary capacities. We will also derive the necessary condition for the S-COPCP of multi-region MFD systems. Next, we will present the COPCP for minimizing TTS (T-COPCP) of the multi-region MFD system and derive the optimal perimeter control law for the T-COPCP.

\vspace{6pt}
\noindent\textbf{Set-point COPCP (S-COPCP) of the Multi-region MFD System}
\vspace{6pt}

Consider the multi-region MFD system \eqref{eqmulti}, find the perimeter controller $\tilde{u}$ to minimize the following objective function:
\begin{align}\label{UPobj}
    & \min_{\tilde{u}} J(\tilde{n}_{0}) = \int^{\infty}_{0} \mathcal{L}(\tilde{n}(t), \tilde{u}(t)) \textrm{d}t \\
    & \mathbf{subject\ to}\ \eqref{eqmulti} \nonumber
\end{align}
where $\tilde{n}\in \Omega\subset\mathbb{R}^{\alpha_{n}}$ and $\tilde{u}\in\mathcal{U}\subset\mathbb{R}^{\alpha_u}$.

The utility function for the S-COPCP is given by
\begin{equation}\label{utilfunL}
  \mathcal{L}(\tilde{n}(t),\tilde{u}(t))\triangleq N(\tilde{n}(t))+U(\tilde{u}(t))
\end{equation}
where $N(\tilde{n})$ represents the cumulative error between the system state and the desired equilibrium, and $U(\tilde{u})$ is the required control effort for unconstrained control case. Generally,
$N(\tilde{n})\triangleq \tilde{n}^TQ\tilde{n}\succeq 0$ with $Q\in \mathbb{R}^{\alpha_n\times \alpha_n}$ and $Q\succ 0$, and  $U(\tilde{u})\triangleq \tilde{u}^TR\tilde{u}\succeq 0$ with $R\in \mathbb{R}^{\alpha_u\times\alpha_u}$ and $R\succ 0$.

\begin{figure}
  \centering
  \includegraphics[width=2.5in]{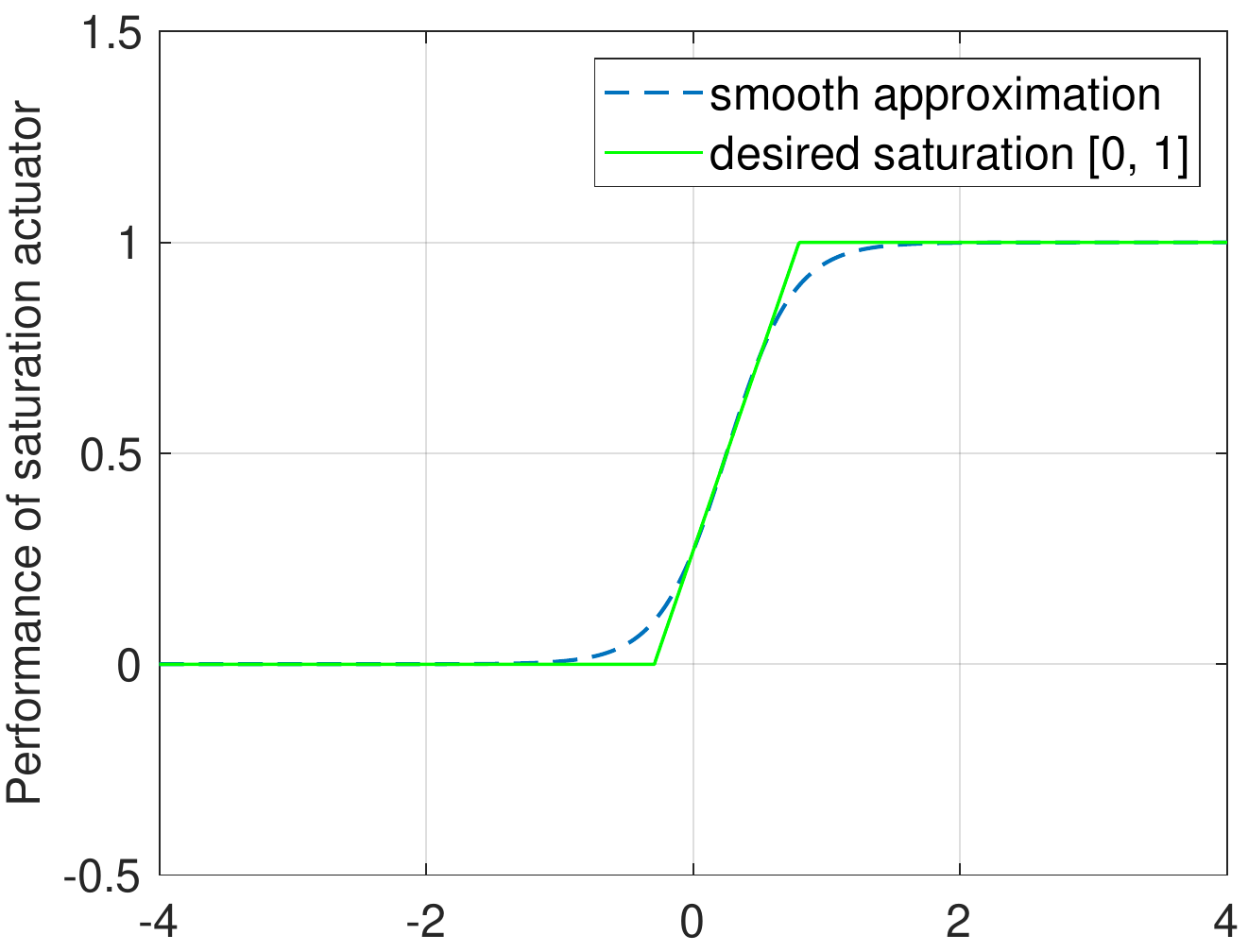}\\
  \caption{Performance of the proposed saturation actuator}\label{pfm_sac}
\end{figure}

Without loss of generality, let $R=\mathrm{diag}(\gamma_1,\ldots,\gamma_{\alpha_u})\in \mathbb{R}^{\alpha_u\times\alpha_u}$ with $\gamma_{k_u}>0$, $k_u=1,\ldots,\alpha_u$. To handle the heterogeneous cross-boundary capacities \eqref{htrconu} in the perimeter controller design, i.e., $\tilde{u}^{\min}_{k_u}\leq \tilde{u}_{k_u}\leq \tilde{u}^{\max}_{k_u}$, inspired by \cite{abu2006nonlinear} and \cite{lyshevski1998optimal}, for each $\tilde{u}_{k_u}$ we define the following function:
\begin{equation*}
  U_{k_u}(\tilde{u}_{k_u})=2\underline{v}_{k_u}\gamma_{k_u}\int^{\tilde{u}_{k_u}}_{\overline{v}_{k_u}}\tanh^{-1}\left(\frac{v_{k_u}-\overline{v}_{k_u}}{\underline{v}_{k_u}}\right) \mathrm{d}v_{k_u}
\end{equation*}
where $\overline{v}_{k_u}=\frac{\tilde{u}^{\max}_{k_u}+\tilde{u}^{\min}_{k_u}}{2},\ \underline{v}_{k_u}=\frac{\tilde{u}^{\max}_{k_u}-\tilde{u}^{\min}_{k_u}}{2}$.
Based on the features of inverse hyperbolic tangent function, $U_{k_u}(\tilde{u}_{k_u})$ can be regarded as a penalty function which limits the input $\tilde{u}_{k_u}$ to $(\tilde{u}^{\min}_{k_u},\tilde{u}^{\max}_{k_u})$. \autoref{pfm_sac} shows that the saturation actuator (blue dotted line) developed by the proposed penalty function can well approximate the non-smooth control constraint (green solid line) in a smooth manner.
Thus, $U(\tilde{u})$ is defined as
\begin{align}\label{conipt}
 U(\tilde{u}) & = \sum^{\alpha_{u}}_{k_u=1}U_{k_u}(\tilde{u}_{k_u}) = \sum^{\alpha_{u}}_{k_u=1} 2\underline{v}_{k_u}\gamma_{k_u}\int^{\tilde{u}_{k_u}}_{\overline{v}_{k_u}}\tanh^{-1}\left(\frac{v_{k_u}-\overline{v}_{k_u}}{\underline{v}_{k_u}}\right) \mathrm{d}v_{k_u}
              = 2\underline{v}^TR\int^{\tilde{u}}_{\overline{v}} \tanh^{-1}\left(\frac{1}{\underline{v}}\odot(v-\overline{v})\right)\mathrm{d}v
\end{align}
where $\overline{v}\triangleq [\overline{v}_1,\ldots,\overline{v}_{\alpha_u}]^T\in \mathbb{R}^{\alpha_u}$, $\underline{v}\triangleq [\underline{v}_1,\ldots,\underline{v}_{\alpha_u}]^T\in \mathbb{R}^{\alpha_u}$.

The value function $V: \mathbb{R}^{\alpha_{n}}\rightarrow \mathbb{R}$ is defined as
\begin{align}\label{valfun}
  V(\tilde{n}(t)) &= \int^{\infty}_{t} \mathcal{L}(\tilde{n}(\tau), \tilde{u}(\tau)) \textrm{d}\tau \equiv\int^{\infty}_{t} (N(\tilde{n}(\tau))+U(\tilde{u}(\tau))) \mathrm{d}\tau
                  = \int^{\infty}_t \left( \tilde{n}^T(\tau)Q\tilde{n}(\tau) + 2\underline{v}^TR\int^{\tilde{u}(\tau)}_{\overline{v}}\tanh^{-1}\left(\frac{1}{\underline{v}}\odot (v-\overline{v})\right)\mathrm{d}v \right) \mathrm{d}\tau
\end{align}
Following the development in \autoref{DPRL}, we can obtain the following Bellman equation
\begin{equation}\label{vbelleq}
  \mathcal{L}(\tilde{n}(t),\tilde{u}(t)) + \left(\frac{\partial V}{\partial \tilde{n}}\right)^T(\mathbf{F}(\tilde{n})+\mathbf{S}(\tilde{n})\tilde{u})  =0
\end{equation}

Now we can present the necessary condition for the solution to the S-COPCP of the multi-region MFD system.

\blemma\label{lem1}
Suppose that $V^*$ is the optimal value function for the S-COPCP of the multi-region MFD system. It follows that
\begin{enumerate}[1)]
  \item the constrained optimal perimeter control is given by
  \begin{equation}\label{optconvec}
    \tilde{u}^*=-\underline{v}\odot\tanh(D^*)+\overline{v},\ \mathrm{with}\ D^*=\frac{1}{2\underline{v}}\odot\left(R^{-1}\mathbf{S}^T\frac{\partial V^*}{\partial \tilde{n}}\right)
  \end{equation}
  where $D^*=[D^*_1,\ldots,D^*_{\alpha_u}]^T\in \mathbb{R}^{\alpha_u}$ is the unconstrained optimal control input;
  \item the necessary condition for the solution to the S-COPCP, i.e., $(V^*,D^*)$ should satisfy the following equation:
  \begin{equation}\label{UpHJB}
    0= \tilde{n}^TQ\tilde{n}+\left(\frac{\partial V^*}{\partial \tilde{n}}\right)^T\mathbf{F}+\left(\frac{\partial V^*}{\partial \tilde{n}}\right)^T\mathbf{S}\overline{v} + \underline{v}^{2T}R\ln(\mathbf{1}_{\alpha_u}-\tanh^2(D^*))
  \end{equation}
  where $\mathbf{1}_{\alpha_u}\in \mathbb{R}^{\alpha_u}$ is a column vector with each element equal to $1$.
\end{enumerate}
\elemma
The proof of \autoref{lem1} is presented in \ref{appenB}. Note that \eqref{UpHJB} is the HJB equation for the S-COPCP of the multi-region MFD system. To find the optimal feedback control policy $\tilde{u}^*$ that minimizes \eqref{UPobj}, it is necessary to solve the HJB equation \eqref{UpHJB} for the value function $V^{*}$ and unconstrained control $D^{*}$, and then substitute them into \eqref{optconvec}. However, the HJB equation \eqref{UpHJB} is extremely difficult to solve due to its strong nonlinearity. In the subsequent sections, a data-driven online algorithm will be presented to find an approximate solution to \eqref{UpHJB} without requiring the system dynamics.

\vspace{6pt}
\noindent\textbf{MinTTS COPCP (T-COPCP) of the Multi-region MFD System}
\vspace{6pt}

Another commonly adopted perimeter control objective is to minimize the total time spent (TTS) during the simulation period:
\begin{align}\label{eq:vftts}
  & \min_{u}\bar{J}(n_{0}) = \int^{t_f}_{0} \left(\sum^{L}_{i=1}n_{i}(t)+\bar{\lambda}\|u(t)\|\right)\mathrm{d}t \\
  & \mathbf{subject\ to}\  \eqref{eq:oadyn} \nonumber
\end{align}
where $n$ and $u$ are constrained by \eqref{htrconn}-\eqref{htrconu}. The last term of the value function \eqref{eq:vftts} is to damp oscillation of the control input, where $\bar{\lambda}$ is a positive constant to adjust the weight of the norm. Different from S-COPCP, the formulation of T-COPCP does not require coordinate transformation.

The Hamiltonian function can be formulated as:
\begin{equation}\label{eq:ttshf}
  \bar{H}\left(n,u,\frac{\partial \bar{V}}{\partial n}\right) = \sum^{L}_{i=1}n_{i}(t)+\bar{\lambda}\|u(t)\| + \left(\frac{\partial \bar{V}}{\partial n}\right)^T\cdot (F(n)+S(n)u)
\end{equation}
where $\bar{V}(n(t))=\int^{t_f}_{t} \left(\sum^{L}_{i=1}n_{i}(\tau)+\bar{\lambda}\|u(\tau)\|\right)\mathrm{d}\tau$.

Similar to the deduction of \autoref{lem1}, the corresponding constrained optimal control law is
\begin{equation}\label{eq:ttsocc}
  u^* = -\underline{v}'\odot \tanh(\bar{D}^*)+\overline{v}',\ \mathrm{with}\ \bar{D}^*=\frac{1}{\underline{v}'}\odot\left(\frac{1}{2\bar{\lambda}}S^T\frac{\partial \bar{V}^*}{\partial n}\right)
\end{equation}
where $\overline{v}'\triangleq [\overline{v}'_1,\ldots,\overline{v}'_{\alpha_u}]^T\in \mathbb{R}^{\alpha_u}$, $\underline{v}'\triangleq [\underline{v}'_1,\ldots,\underline{v}'_{\alpha_u}]^T\in \mathbb{R}^{\alpha_u}$ with $\overline{v}'_{k_u}=\frac{u^{\max}_{k_u}+u^{\min}_{k_u}}{2},\ \underline{v}'_{k_u}=\frac{u^{\max}_{k_u}-u^{\min}_{k_u}}{2}$.

Note that \eqref{eq:oadyn} and \eqref{eqmulti} are in the same affine form. The theoretical results developed for system \eqref{eqmulti} (regarding S-COPCP) can be applied to system \eqref{eq:oadyn} (regarding T-COPCP).

\section{Data-driven IRL based adaptive optimal perimeter control} \label{metdlg}

Parallel to the development in \autoref{DPRL}, to relax the requirement of system knowledge and consider sensor data measurements, we will establish an equivalent formulation of the HJB equation \eqref{UpHJB} that does not involve the system dynamics. Towards this, in this section a recapitulation of the policy iteration method for solving \eqref{UpHJB} will be presented. Based on the policy iteration method, a data-driven model-free adaptive optimal perimeter controller, which considers the heterogeneous discrete-time sensor data, is developed through the lens of the integral reinforcement learning (IRL).

Note that it is difficult to give an analytical solution to \eqref{UpHJB} due to the strong nonlinearity. Policy iteration is one of the most common methods to resolve this difficulty. The policy iteration method considering heterogeneous cross-boundary capacities is as follows:

\begin{enumerate}
  \item (Policy evaluation) Given an initial admissible control policy $\tilde{u}^{0}(\tilde{n})$, find $V^{k}(\tilde{n})$ successively approximated by solving the following equation with $V^{k}(0)=0$
      \begin{equation}\label{eq8}
        \mathcal{L}(\tilde{n}, \tilde{u}^{k})+\left(\frac{\partial V^{k+1}}{\partial \tilde{n}}\right)^{T} (\mathbf{F}+\mathbf{S} \tilde{u}^{k})=0, k=0,1,\ldots
      \end{equation}
  \item (Policy improvement) Update the control policy simultaneously by
      \begin{equation}\label{eq9}
        \tilde{u}^{k+1}(\tilde{n})=-\underline{v}\odot\tanh(D^{k+1})+\overline{v}, D^{k+1}=\frac{1}{2\underline{v}}\odot\left(R^{-1} \mathbf{S}^{T} \frac{\partial V^{k+1}}{\partial \tilde{n}}\right)
      \end{equation}
\end{enumerate}
where $k$ is the iterative index. The policy evaluation is implemented to update the iterative value function that satisfies the Bellman equation \eqref{vbelleq}. Then based on value iteration, the policy improvement is implemented to obtain the iterative control law sequence that minimizes the total cost in each period. From the policy improvement, we can always find another control law sequence that is better, or at least no worse. The following lemma demonstrates the convergence of $V^k$ and $\tilde{u}^k$ (i.e., $D^k$) by iterating \eqref{eq8}-\eqref{eq9} to the optimal value function $V^*$ and optimal perimeter control $\tilde{u}^*$ (i.e., $D^*$).

\blemma \label{thm1}
Let $V^{k}(\tilde{n})\in C^1(\Omega)$ on $\Omega$ where $V^{k}(\tilde{n})\geq 0$, $V^{k}(0)=0$ and $\tilde{u}^{k}(\tilde{n})$ is admissible to \eqref{eqmulti}, $k=0,1,\ldots$. If $(V^{k+1}(\tilde{n}), \tilde{u}^{k}(\tilde{n}))$ and $(V^{k+2}(\tilde{n}), \tilde{u}^{k+1}(\tilde{n}))$ both satisfy \eqref{vbelleq} with the boundary condition $V^{k+1}(0)=0$, $V^{k+2}(0)=0$, then
\begin{enumerate}[1)]
  \item the obtained control policies $\tilde{u}^{k+1}(\tilde{n})$ in \eqref{eq9} are admissible for \eqref{eqmulti} on $\Omega$;
  \item $V^{*}(\tilde{n})\leq V^{k+2}(\tilde{n})\leq V^{k+1}(\tilde{n}), \forall \tilde{n}\in \Omega$;
  \item $\lim_{k\rightarrow \infty} V^{k}(\tilde{n})=V^{*}(\tilde{n})$;
  \item $\lim_{k\rightarrow \infty} \tilde{u}^{k}(\tilde{n})=\tilde{u}^{*}(\tilde{n})$.
\end{enumerate}
\elemma

We present the proof of \autoref{thm1} in \ref{appenC}.

\autoref{thm1} indicates that using the policy iteration method, $(V^{k},\tilde{u}^{k})$ (i.e., $(V^{k},D^{k})$) can approximate the optimal solution $(V^*,\tilde{u}^*)$ (i.e., $(V^*,D^*)$) to the HJB equation \eqref{UpHJB}. However, \eqref{eq8}-\eqref{eq9} requires identification of the MFD dynamics $\mathbf{F}$ and $\mathbf{S}$. To enable a data-driven method without requiring calibration of the MFD dynamics \eqref{eqmulti}, the main idea is to get rid of the system dynamics in the HJB equation \eqref{UpHJB}.
We can adopt an off-policy IRL algorithm that the control implemented can be different from the optimal control \eqref{eq9}. Towards this, we rewrite the traffic dynamics \eqref{eqmulti} as
\begin{equation}\label{ofpdyna}
    \dot{\tilde{n}}=\mathbf{F}(\tilde{n})+\mathbf{S}(\tilde{n}) \tilde{u}^k+\mathbf{S}(\tilde{n}) (\tilde{u}-\tilde{u}^k)
\end{equation}
where $\tilde{u}^k$ is the policy to be updated and $\tilde{u}$ is the behavior policy that is actually implemented to the system dynamics to generate the data for learning.

\bremark
On-policy and off-policy are two important RL methods. The policy iteration \eqref{eq8}-\eqref{eq9} can be regarded as a class of on-policy methods. When using the on-policy methods, the learned control policy should be applied to generate data simultaneously even before it converges. Although the on-policy methods can provide nearly unbiased estimates of the policy gradient, they (e.g., Sarsa) are usually data-intensive and their learning process is time-consuming. Furthermore, the data usage of on-policy learning methods is low because the samples generated previously would be discarded along with each policy changes. Thus, the implementation of on-policy learning method is generally difficult.  Unlike the on-policy learning, the off-policy learning evaluates the target policy when executing other behavior policies.

There are several practical reasons that the implemented control can differ from the optimal control to be learned.
As discussed in \cite{ZHONG2018327}, the traffic managers may have difficulties in calibrating a detailed functional form and its steady state for the time-varying travel demand and the MFD dynamics. Thus, they may not be able to implement the model-based optimal control (that is a `miracle' to the manager) in the learning process. On the other hand, the traffic managers definitely have a preferable network condition (or state) for management purposes. The implemented control $\tilde{u}$ can be arbitrary enables the traffic managers to enforce their preference as a `priori' for traffic management purposes.  This can be regarded as a superiority of the proposed off-policy IRL-based learning algorithms. This implemented control can stimulate the network dynamics so that the learning algorithms can observe the evolution of traffic states and the network performance to adjust the adaptive optimal control iteratively.
\eremark

Now we derive the IRL Bellman equation that does not involve the system dynamics $\mathbf{F}$ and $\mathbf{S}$, i.e., ``model-free''. The time derivative of $V^{k+1}(\tilde{n}(t))$ for the $\{k+1\}$-th iteration equals
\begin{equation}\label{eq17a}
    \frac{\textrm{d}V^{k+1}}{\textrm{d}t}=\left(\frac{\partial V^{k+1}}{\partial \tilde{n}}\right)^{T} (\mathbf{F}+\mathbf{S} \tilde{u})
\end{equation}
Subtracting \eqref{eq8} from \eqref{eq17a}, we have
\begin{align}\label{itndV}
  \frac{\mathrm{d} V^{k+1}}{\mathrm{d} t} & = \left(\frac{\partial V^{k+1}}{\partial \tilde{n}}\right)^T(\mathbf{F}+\mathbf{S}\tilde{u}) - \left(\frac{\partial V^{k+1}}{\partial \tilde{n}}\right)^T(\mathbf{F}+\mathbf{S}\tilde{u}^k) - \mathcal{L}(\tilde{n},\tilde{u}^k)
    = \left(\frac{\partial V^{k+1}}{\partial \tilde{n}}\right)^T\mathbf{S}(\tilde{u}-\tilde{u}^k) - \mathcal{L}(\tilde{n},\tilde{u}^k)
\end{align}
From the second equation of \eqref{eq9}, we have
\begin{align}\label{tsfDk}
    & 2\underline{v}\odot D^{k+1} = R^{-1}\mathbf{S}^T\frac{\partial V^{k+1}}{\partial \tilde{n}} \nonumber \\
  \Rightarrow & \left(\frac{\partial V^{k+1}}{\partial \tilde{n}}\right)^T\mathbf{S}R^{-1} = (2\underline{v}\odot D^{k+1})^T \nonumber \\
  \Rightarrow & \left(\frac{\partial V^{k+1}}{\partial \tilde{n}}\right)^T\mathbf{S} = 2(\underline{v}\odot D^{k+1})^TR
\end{align}
Substituting \eqref{tsfDk} into \eqref{itndV} yields
\begin{equation}\label{eq17}
    \frac{\textrm{d}V^{k+1}}{\textrm{d}t}  = 2(\underline{v}\odot D^{k+1})^{T} R (\tilde{u}-\tilde{u}^{k})-\mathcal{L}(\tilde{n},\tilde{u}^{k})
\end{equation}
Integrating both sides of \eqref{eq17} on the interval $[t, t+\Delta t]$, we obtain
\begin{align}
    & V^{k+1}(\tilde{n}(t+\Delta t))-V^{k+1}(\tilde{n}(t)) = \int^{t+\Delta t}_{t}2(\underline{v}\odot D^{k+1})^{T} R (\tilde{u}-\tilde{u}^{k})\textrm{d}\tau - \int^{t+\Delta t}_{t}\mathcal{L}(\tilde{n},\tilde{u}^{k})\textrm{d}\tau \nonumber\\
    \Rightarrow &  V^{k+1}(\tilde{n}(t)) = \int^{t+\Delta t}_{t}\mathcal{L}(\tilde{n},\tilde{u}^{k})\textrm{d}\tau - \int^{t+\Delta t}_{t}2(\underline{v}\odot D^{k+1})^{T} R (\tilde{u}-\tilde{u}^{k})\textrm{d}\tau + V^{k+1}(\tilde{n}(t+\Delta t))\label{eq18}
\end{align}
for any time $t \geq 0$ and time interval $\Delta t>0$. As introduced in \autoref{DPRL}, $\Delta t$ is termed as the reinforcement interval. There is a trade-off between the learning rate and the reinforcement interval. It is found by \cite{modares2014flow} that the larger the reinforcement interval $\Delta t$ is, the smaller the learning rate should be chosen.

\eqref{eq18} is called IRL Bellman equation, which no longer involves the model information of the traffic dynamics. Thus, solving \eqref{eq18} instead of the HJB equation \eqref{UpHJB}, we can obtain a data-driven IRL based adaptive optimal perimeter controller, which is ``model-free''.

Note that the convergence of the iteration sequence $\{(V^{k+1},D^{k+1})\}$ by using \eqref{eq8}-\eqref{eq9} to the optimality has been checked by \autoref{thm1}. Hence, we only need to justify the equivalence between the policy iterative equations \eqref{eq8}-\eqref{eq9} and the IRL Bellman equation \eqref{eq18}, whereby the convergence and optimality of the IRL approach can also be derived.

\btheorem \label{thm2}
The IRL Bellman equation \eqref{eq18} gives the same solution to the value function as the Bellman equation \eqref{eq8} and the same updated control policy as \eqref{eq9}.
\etheorem

\bproof
The proof of \autoref{thm2} is divided into two fold.

1) First, we prove that \eqref{eq8}-\eqref{eq9} $\Rightarrow$ \eqref{eq18}. Provided that $(V^{k+1},D^{k+1})$ is the solution of the policy iterative equations \eqref{eq8}-\eqref{eq9}, from the derivation of \eqref{eq18}, one can easily deduce that $(V^{k+1},D^{k+1})$ is the solution of \eqref{eq18}.

2) Next, we prove that \eqref{eq18} $\Rightarrow$ \eqref{eq8}-\eqref{eq9}. Provided that $(V^{k+1},D^{k+1})$ is the solution of the IRL Bellman equation \eqref{eq18} and that $D^{k+1}=\frac{1}{2\underline{v}}\odot\left(R^{-1} \mathbf{S}^{T} \frac{\partial V^{k+1}}{\partial \tilde{n}}\right)$.

Dividing both sides of \eqref{eq18} by $\Delta t$ and taking limit results in
\begin{align}\label{ndVk}
    & \lim_{\Delta t\rightarrow 0}\frac{V^{k+1}(\tilde{n}(t+\Delta t))-V^{k+1}(\tilde{n}(t))}{\Delta t} = \lim_{\Delta t\rightarrow 0} \frac{\int^{t+\Delta t}_{t}2(\underline{v}\odot D^{k+1})^{T} R (\tilde{u}-\tilde{u}^{k})\textrm{d}\tau - \int^{t+\Delta t}_{t}\mathcal{L}(\tilde{n},\tilde{u}^{k})\textrm{d}\tau}{\Delta t} \nonumber \\
  \Rightarrow & \frac{\mathrm{d} V^{k+1}}{\mathrm{d}t} = 2(\underline{v}\odot D^{k+1})^{T} R (\tilde{u}-\tilde{u}^{k}) - \mathcal{L}(\tilde{n},\tilde{u}^{k})
\end{align}
Substituting $D^{k+1}=\frac{1}{2\underline{v}}\odot\left(R^{-1} \mathbf{S}^{T} \frac{\partial V^{k+1}}{\partial \tilde{n}}\right)$ into \eqref{ndVk}, we have
\begin{equation}\label{nitndV}
  \frac{\mathrm{d} V^{k+1}}{\mathrm{d}t} = \left(\frac{\partial V^{k+1}}{\partial \tilde{n}}\right)^T\mathbf{S}(\tilde{u}-\tilde{u}^k) - \mathcal{L}(\tilde{n},\tilde{u}^k)
\end{equation}
Combining \eqref{ofpdyna} and \eqref{nitndV}, it follows that
\begin{align}\label{eqBE}
  \frac{\mathrm{d} V^{k+1}}{\mathrm{d}t} & = \left(\frac{\partial V^{k+1}}{\partial \tilde{n}}\right)^T\mathbf{S}(\tilde{u}-\tilde{u}^k) - \mathcal{L}(\tilde{n},\tilde{u}^k) + \left(\frac{\partial V^{k+1}}{\partial \tilde{n}}\right)^T\mathbf{F} - \left(\frac{\partial V^{k+1}}{\partial \tilde{n}}\right)^T\mathbf{F} \nonumber \\
    & = \left(\frac{\partial V^{k+1}}{\partial \tilde{n}}\right)^T(\mathbf{F}+\mathbf{S}\tilde{u}) - \left(\frac{\partial V^{k+1}}{\partial \tilde{n}}\right)^T(\mathbf{F}+\mathbf{S}\tilde{u}^k) - \mathcal{L}(\tilde{n},\tilde{u}^k) \nonumber \\
  \Rightarrow & \mathcal{L}(\tilde{n},\tilde{u}^k) + \left(\frac{\partial V^{k+1}}{\partial \tilde{n}}\right)^T(\mathbf{F}+\mathbf{S}\tilde{u}^k) = \left(\frac{\partial V^{k+1}}{\partial \tilde{n}}\right)^T(\mathbf{F}+\mathbf{S}\tilde{u}) - \frac{\mathrm{d} V^{k+1}}{\mathrm{d}t}
\end{align}
From \eqref{eq17a} we have the right side of \eqref{eqBE} equals to $0$. Hence,
\begin{equation}\label{feqBE}
  \mathcal{L}(\tilde{n}, \tilde{u}^{k})+\left(\frac{\partial V^{k+1}}{\partial \tilde{n}}\right)^{T} (\mathbf{F}+\mathbf{S} \tilde{u}^{k})=0
\end{equation}
\eqref{feqBE} is the same as \eqref{eq8}. This completes the proof.
\eproof

Note that we have proven the equivalence between the Bellman equation and the IRL Bellman equation, which does not involve the traffic dynamics (model-free). By iterating $V^k$ on the IRL Bellman equation and updating the control policy $D^k$ (i.e., $\tilde{u}^k$), we can approach both the optimal value function $V^*$ and the optimal perimeter control $\tilde{u}^*$. In the subsequent section, we will develop an online learning approach via the IRL Bellman equation \eqref{eq18} to approximate the optimal value function and perimeter controller.

\section{Online learning by integrating experience replay}\label{ImpmNN}

For the implementation of the conventional off-line learning based RL approach, sufficient historical data must be collected beforehand and the collected data set would be used repeatedly during the learning process. This implies that only recurrent traffic conditions can be well handled by the conventional off-line learning based RL approach. To adapt to new or unseen data samples and possible changes of traffic conditions, an online iterative learning approach based on the IRL is proposed in this section. Employing off-policy methods, the proposed online learning approach can be integrated with the ER technique to reduce the requirement on real-time data samples and simultaneously reduce the computational burden.

\begin{figure}[!h]
\centering
\includegraphics[width=3.1in]{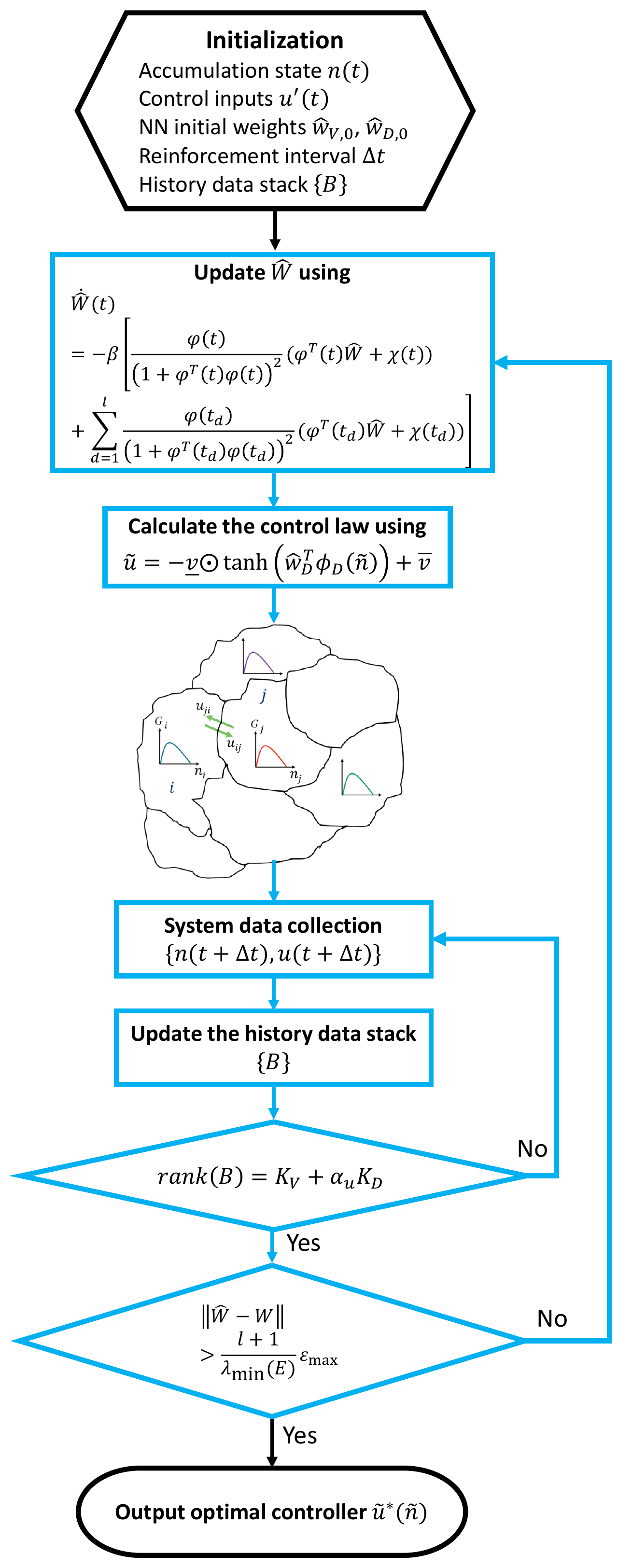}\\
\caption{The online iterative learning algorithm}\label{onalgrthm}
\end{figure}

The online (incorporated with ER) learning method is constructed via the actor-critic (AC) neural network (NN) framework. The neural networks can learn the unknown macroscopic traffic dynamics and achieve the adaptive optimal perimeter control with the IRL. The critic (i.e., policy evaluation) NN and the actor (i.e., policy improvement) NN are tuned sequentially. The flow chart of the proposed IRL algorithm is shown in \autoref{onalgrthm}. The algorithm starts by evaluating the cost of a given initial admissible control policy and then uses this information to obtain a new and improved control policy that generates a lower associated cost than the previous one does. These two steps of policy evaluation and policy improvement are repeated until the actual policy remains unchanged after the policy improvement step, whereby the convergence to the optimal controller is achieved.
The convergence and stability analysis are evidenced via the Lyapunov theory.

Accordingly, at any time $t>\Delta t$ with reinforcement interval $\Delta t>0$, given that $\tilde{u}'$ is an admissible control, the IRL Bellman equation \eqref{eq18} can thus be rewritten as follows
\begin{equation}\label{intval}
  V(\tilde{n}(t-\Delta t)) = \int^{t}_{t-\Delta t}\mathcal{L}(\tilde{n}(\tau),\tilde{u}'(\tau))\mathrm{d}\tau - \int^{t}_{t-\Delta t}2(\underline{v}\odot D(\tau))^TR(\tilde{u}-\tilde{u}'(\tau))\mathrm{d}\tau + V(\tilde{n}(t))
\end{equation}

We utilize an AC-NN framework to approximate the value function and the control policy (i.e., the solution of \eqref{intval}) simultaneously:
\begin{equation}\label{oapproVD}
  V(\tilde{n}) = w^T_V \phi_V (\tilde{n})+\varepsilon_V (\tilde{n}),\quad D(\tilde{n}) = w^T_D \phi_D (\tilde{n})+\varepsilon_D (\tilde{n})
\end{equation}
where $\phi_V: \mathbb{R}^{\alpha_{n}}\rightarrow \mathbb{R}^{K_V}$, $\phi_D: \mathbb{R}^{\alpha_{n}}\rightarrow \mathbb{R}^{K_D}$ are vectors of linearly independent activation functions, $w^T_V\in \mathbb{R}^{K_V}$, $w^T_D\in \mathbb{R}^{K_D\times \alpha_u}$ are the NN weights of appropriate dimensions, $\varepsilon_V (\tilde{n})$ and $\varepsilon_D (\tilde{n})$ are the approximation errors of the critic NN and the actor NN, respectively.

Using the approximations \eqref{oapproVD} in \eqref{intval} and considering $\varepsilon_V=0$, $\varepsilon_D=0$ for the ideal weights $w_V$ and $w_D$, one has
\begin{equation}\label{bdapper}
  \varepsilon_B(t)\triangleq \int^{t}_{t-\Delta t}\mathcal{L}(\tilde{n}(\tau),\tilde{u}'(\tau))\mathrm{d}\tau - \int^{t}_{t-\Delta t}2(\underline{v}\odot w^T_D \phi_D (\tilde{n}(\tau)))^TR(\tilde{u}-\tilde{u}'(\tau))\mathrm{d}\tau + w^T_V(\phi_V (\tilde{n}(t))-\phi_V (\tilde{n}(t-\Delta t)))
\end{equation}
where $\varepsilon_B(t)$ is the Bellman equation error at time $t$. $\varepsilon_B$ is assumed to be bounded on the compact set $\Omega$ given the ideal weights $w_V$ and $w_D$ \citep{modares2014flow}. That is, there exists a bound $\varepsilon_{\max}$ such that $\|\varepsilon_B\|\leq \varepsilon_{\max}$.

Note that the ideal weights $w_V$ and $w_D$ that provide the best approximate solution for \eqref{bdapper} are unknown. Hence, the estimations of value function and control policy are given by
\begin{equation}\label{NNe}
  \hat{V}(\tilde{n}) = \hat{w}^T_V \phi_V (\tilde{n}),\quad \hat{D}(\tilde{n}) = \hat{w}^T_D \phi_D (\tilde{n})
\end{equation}
where $\hat{w}_V\in \mathbb{R}^{K_V}$, $\hat{w}_D\in \mathbb{R}^{K_D\times \alpha_u}$ are estimations of $w_V$ and $w_D$, respectively. These estimations are usually learned from training data.

Using \eqref{NNe} in \eqref{intval}, the approximation error of the IRL Bellman equation, i.e., the TD error, at time $t$ is given by
\begin{equation}\label{rsderr}
\begin{split}
  e(t) & = \hat{V}(\tilde{n}(t))-\hat{V}(\tilde{n}(t-\Delta t)) - \int^{t}_{t-\Delta t}2(\underline{v}\odot \hat{D}(\tau))^TR(\tilde{u}-\tilde{u}'(\tau))\mathrm{d}\tau + \int^t_{t-\Delta t}\mathcal{L}(\tilde{n}(\tau),\tilde{u}'(\tau))\mathrm{d}\tau \\
    & = \phi^T_{V}(\tilde{n}(t))\hat{w}_{V}-\phi^T_{V}(\tilde{n}(t-\Delta t))\hat{w}_{V} - \int^t_{t-\Delta t}2(\underline{v}\odot (\hat{w}^T_D\phi_D(\tilde{n}(\tau))))^TR(\tilde{u}+\underline{v}\odot\tanh(\hat{w}^T_D\phi_D(\tilde{n}(\tau)))-\overline{v})\mathrm{d}\tau \\
    & \ \ + \int^t_{t-\Delta t}\left(\tilde{n}^T(\tau)Q\tilde{n}(\tau) + 2\underline{v}^TR\int^{-\underline{v}\odot\tanh(\hat{w}^T_D\phi_D(\tilde{n}(\tau)))+\overline{v}}_{\overline{v}}\tanh^{-1}\left(\frac{1}{\underline{v}}\odot(v-\overline{v})\right)\mathrm{d}v \right)\mathrm{d}\tau
\end{split}
\end{equation}
Let $\hat{W}=[\hat{w}^T_V, vec^T(\hat{w}_D)]^T\in \mathbb{R}^{K_V+\alpha_u K_D}$ be the estimated weight of the AC-NNs, where $vec(\hat{w}_D)\in \mathbb{R}^{\alpha_u K_D}$ is the vectorization of matrix $\hat{w}_{D}\in \mathbb{R}^{K_D\times\alpha_u}$. Thus, \eqref{rsderr} can be rewritten as
\begin{equation}\label{resd}
  e(t)=\varphi^T(\tilde{n}(t),\tilde{u}(t))\hat{W}+\chi(\tilde{n}(t))
\end{equation}
where
\begin{align*}
   \varphi(\tilde{n}(t),\tilde{u}(t))&=\left[
     \begin{array}{c}
       \phi_V(\tilde{n}(t))-\phi_V(\tilde{n}(t-\Delta t)) \\
       \int^t_{t-\Delta t} (2\underline{v}\odot R(\tilde{u}+\underline{v}\odot\tanh(\hat{w}^T_D\phi_D(\tilde{n}(\tau)))-\overline{v}))\otimes\phi_D(\tilde{n}(\tau))\mathrm{d}\tau \\
     \end{array}
  \right] \\
  \chi(\tilde{n}(t))&=\int^t_{t-\Delta t}\mathcal{L}(\tilde{n}(\tau),\tilde{u}'(\tau))\mathrm{d}\tau = \int^t_{t-\Delta t}\left(\tilde{n}^T(\tau)Q\tilde{n}(\tau) + 2\underline{v}^TR\int^{-\underline{v}\odot\tanh(\hat{w}^T_D\phi_D(\tilde{n}(\tau)))+\overline{v}}_{\overline{v}}\tanh^{-1}\left(\frac{1}{\underline{v}}\odot(v-\overline{v})\right)\mathrm{d}v \right)\mathrm{d}\tau
\end{align*}

To enable online learning, we use the gradient-descent method to update the estimated AC-NN weights.  Both real-time data and historical data are used to estimate the weights of the NNs to guarantee the data richness and efficiency.

As discussed in \autoref{DPRL}, the ER technique can be integrated with the IRL algorithm to speed up the computation. Based on the generalized least-squares (GLS) principle, we aim to update the estimated weight vector $\hat{W}$ to minimize $\|e(t)\|+\sum^l_{d=1}\|e(t_d)\|$, where the first part denotes the instantaneous TD error and the second part denotes the TD errors for the stored transition samples. In order to ensure the existence of the solution, we need the following assumption.

\basm \label{condt}
Define $B=[\varphi(t_1), \ldots, \varphi(t_l)]$ as a matrix of the stored data, where $l$ is the number of samples stored in the history stack. There are as many linearly independent elements as the number of corresponding NN's hidden neurons for the stored data matrix $B$ such that $\mathrm{rank}(B)=K_V+\alpha_u K_D$.
\easm

This rank condition is to verify the richness of the stored data, i.e., whether it is sufficient to solve the GLS problem and to guarantee the convergence to a near-optimal control \citep{modares2014flow}. Based on \eqref{resd}, for the online iterative learning, the gradient-based adaptation law with ER is given by
\begin{align}\label{ERupW}
  \dot{\hat{W}}(t) & = -\beta\left( \frac{\varphi(t)}{(1+\varphi^T(t)\varphi(t))^2}e(t) + \sum^l_{d=1}\frac{\varphi(t_d)}{(1+\varphi^T(t_d)\varphi(t_d))^2}e(t_d) \right) \nonumber \\
    & = -\beta\left( \frac{\varphi(t)}{(1+\varphi^T(t)\varphi(t))^2}(\varphi^T(t)\hat{W}+\chi(t))+ \sum^l_{d=1}\frac{\varphi(t_d)}{(1+\varphi^T(t_d)\varphi(t_d))^2} (\varphi^T(t_d)\hat{W}+\chi(t_d)) \right)
\end{align}
where $\beta >0$ is the learning rate, $t$ is the current time and the index $d$ refers to the $d$-th sample data ($d=1,\ldots,l$) stored in the history stack $B$. In \eqref{ERupW}, the first term is a gradient-descent update law for minimizing $\|e(t)\|$, while the second term attempts to minimize $\sum^l_{d=1}\|e(t_d)\|$.

Denote the optimal value of weight by $W=[{w}^T_V, vec^T({w}_D)]^T$ and recall that the estimated weight is defined by $\hat{W}=[\hat{w}^T_V, vec^T(\hat{w}_D)]^T$. The following theorem demonstrates the convergence of the weight estimation error of AC-NNs, $\tilde{W}(t)=W-\hat{W}(t)$, using Lyapunov method.

\btheorem \label{lem42}
If the stored data $B$ for AC-NNs \eqref{NNe} with the ER adaptation law \eqref{ERupW} satisfy \autoref{condt},
\begin{enumerate}[1)]
  \item for bounded $\varepsilon_B$, the weight estimation error $\tilde{W}(t)=W-\hat{W}(t)$ converges exponentially to the residual set $R_s=\{\tilde{W}\ |\ \|\tilde{W}(t)\|\leq c\cdot\varepsilon_{\max}\}$, where $c>0$ is a constant;
  \item the system state $\tilde{n}$ is asymptotically stable.
\end{enumerate}
\etheorem

\bproof
1) Based on \eqref{bdapper}, \eqref{rsderr}, \eqref{resd} and $\tilde{W}(t)=W-\hat{W}(t)$, the TD errors for he current time $t$ and the recorded time $t_d$ can be rewritten respectively as
\begin{subequations}
\begin{align}
  & e(t)=\varphi^T(t)\hat{W}+\chi(t) = \varphi^T(t)W-\varphi^T(t)\tilde{W}+\chi(t) = -\varphi^T(t)\tilde{W}+(\chi(t)+\varphi^T(t)W)=-\varphi^T(t)\tilde{W}+\varepsilon_B(t) \label{crtTD}
  \\
  & e(t_d)=\varphi^T(t_d)\hat{W}+\chi(t_d) = \varphi^T(t_d)W-\varphi^T(t_d)\tilde{W}+\chi(t_d) = -\varphi^T(t_d)\tilde{W}+(\chi(t_d)+\varphi^T(t_d)W)=-\varphi^T(t_d)\tilde{W}+\varepsilon_B(t_d)
  \label{rctTD}
\end{align}
\end{subequations}
From $\tilde{W}(t)=W-\hat{W}(t)$, one has $\dot{\tilde{W}}(t)=-\dot{\hat{W}}(t)$. Substituting \eqref{crtTD}-\eqref{rctTD} into \eqref{ERupW} and denoting $\bar{\varphi}=\varphi/(1+\varphi^T\varphi)$ and $m=1+\varphi^T\varphi$, we can obtain
\begin{align}
  \dot{\tilde{W}}(t) & = \beta\left(\frac{\varphi(t)}{(1+\varphi^T(t)\varphi(t))^2}e(t) + \sum^l_{d=1}\frac{\varphi(t_d)}{(1+\varphi^T(t_d)\varphi(t_d))^2}e(t_d)\right) \nonumber \\
    & = \beta\left(\frac{\bar{\varphi}(t)}{m(t)}(-\varphi^T(t)\tilde{W}+\varepsilon_B(t)) + \sum^l_{d=1}\frac{\bar{\varphi}(t_d)}{m(t_d)}(-\varphi^T(t_d)\tilde{W}+\varepsilon_B(t_d))\right) \nonumber \\
    & = -\beta\left(\frac{\bar{\varphi}(t)}{m(t)}\varphi^T(t)+\sum^l_{d=1}\frac{\bar{\varphi}(t_d)}{m(t_d)}\varphi^T(t_d)\right)\tilde{W} + \beta\left(\frac{\bar{\varphi}(t)}{m(t)}\varepsilon_B(t)+\sum^l_{d=1}\frac{\bar{\varphi}(t_d)}{m(t_d)}\varepsilon_B(t_d)\right) \nonumber \\
    & = -\beta\left(\bar{\varphi}(t)\bar{\varphi}^T(t)+\sum^l_{d=1}\bar{\varphi}(t_d)\bar{\varphi}^T(t_d)\right)\tilde{W} + \beta\bar{\varepsilon}_B \label{dWtu}
\end{align}
where $\bar{\varepsilon}_B=\frac{\bar{\varphi}(t)}{m(t)}\varepsilon_B(t)+\sum^l_{d=1}\frac{\bar{\varphi}(t_d)}{m(t_d)}\varepsilon_B(t_d)$.

Now we choose the Lyapunov function as
\begin{equation}\label{lpnfun}
  L=\frac{1}{2\beta}\tilde{W}^T(t)\tilde{W}(t)
\end{equation}
Differentiating \eqref{lpnfun} along the trajectories of \eqref{dWtu}, one has
\begin{equation}\label{lpntdf}
\begin{split}
  \dot{L} & =\frac{1}{\beta}\tilde{W}^T\dot{\tilde{W}} \\
    & = \frac{1}{\beta}\tilde{W}^T\cdot \left(-\beta\left(\bar{\varphi}(t)\bar{\varphi}^T(t)+\sum^l_{d=1}\bar{\varphi}(t_d)\bar{\varphi}^T(t_d)\right)\tilde{W} + \beta\bar{\varepsilon}_B\right) \\
    & = -\tilde{W}^T\left(\bar{\varphi}(t)\bar{\varphi}^T(t)+\sum^l_{d=1}\bar{\varphi}(t_d)\bar{\varphi}^T(t_d)\right)\tilde{W} + \tilde{W}^T\bar{\varepsilon}_B
\end{split}
\end{equation}

If \autoref{condt} is satisfied, then $\bar{\varphi}(t)\bar{\varphi}^T(t)+\sum^l_{d=1}\bar{\varphi}(t_d)\bar{\varphi}^T(t_d)>0$. Suppose that $\varepsilon_B$ is bounded by $\varepsilon_{\max}$, i.e., $\|\varepsilon_B\|\leq \varepsilon_{\max}$, $\dot{L}$ is negative definite provided that
\begin{equation}\label{condtW}
  \|\tilde{W}(t)\|>\frac{l+1}{\lambda_{\min} (E)}\varepsilon_{\max} = c\cdot \varepsilon_{\max}
\end{equation}
where $c=\frac{l+1}{\lambda_{\min} (E)}>0$ and $\lambda_{\min}(E)$ is the minimum eigenvalue of $E$ with $E=\bar{\varphi}(t)\bar{\varphi}^T(t)+\sum^l_{d=1}\bar{\varphi}(t_d)\bar{\varphi}^T(t_d)$. Hence, the weight estimation error $\tilde{W}$ converges exponentially to the residual set $R_s=\{\tilde{W}\ |\ \|\tilde{W}(t)\|\leq c\cdot\varepsilon_{\max}\}$.

2) For system \eqref{eqmulti}, define Lyapunov function candidate as \eqref{valfun}. Take the time derivative of $V$ and we can obtain
\begin{equation*}
  \dot{V} = -\mathcal{L}(\tilde{n},\tilde{u}) = -N(\tilde{n})-U(\tilde{u})
\end{equation*}

Recall that $N(\tilde{n})$ and $U(\tilde{u})$ are positive definite functions. Then we have $V(\tilde{n}(t))\geq 0$, $\dot{V}\leq 0$ and $V(\tilde{n}(t))=0$ if and only if $\tilde{n}=0$, i.e., $n=n^*$. That is, $V(\tilde{n})$ is a Lyapunov function. The closed-loop system is thus asymptotically stable. This completes the proof.
\eproof

\autoref{lem42} indicates that using the gradient-based adaptation law with ER \eqref{ERupW}, the AC-NN framework \eqref{NNe} can approximate the optimal value function $V^*(\tilde{n})$ and perimeter control policy $\tilde{u}^*(\tilde{n})$. The value function \eqref{valfun} is proven to be a Lyapunov function for the MFD dynamics. Hence, the initial accumulation state $n_0$ can be asymptotically stabilized by the obtained perimeter controller at the desired steady state $n^*$.

\section{Numerical experiments} \label{Simul}

\subsection{Settings of the test environment}

\begin{figure}[!htb]
\centering
\subfigure[The two-region MFD network topology]{
\begin{minipage}[b]{0.4\linewidth}\label{twotopo}
\centering
\includegraphics[width=2.0in]{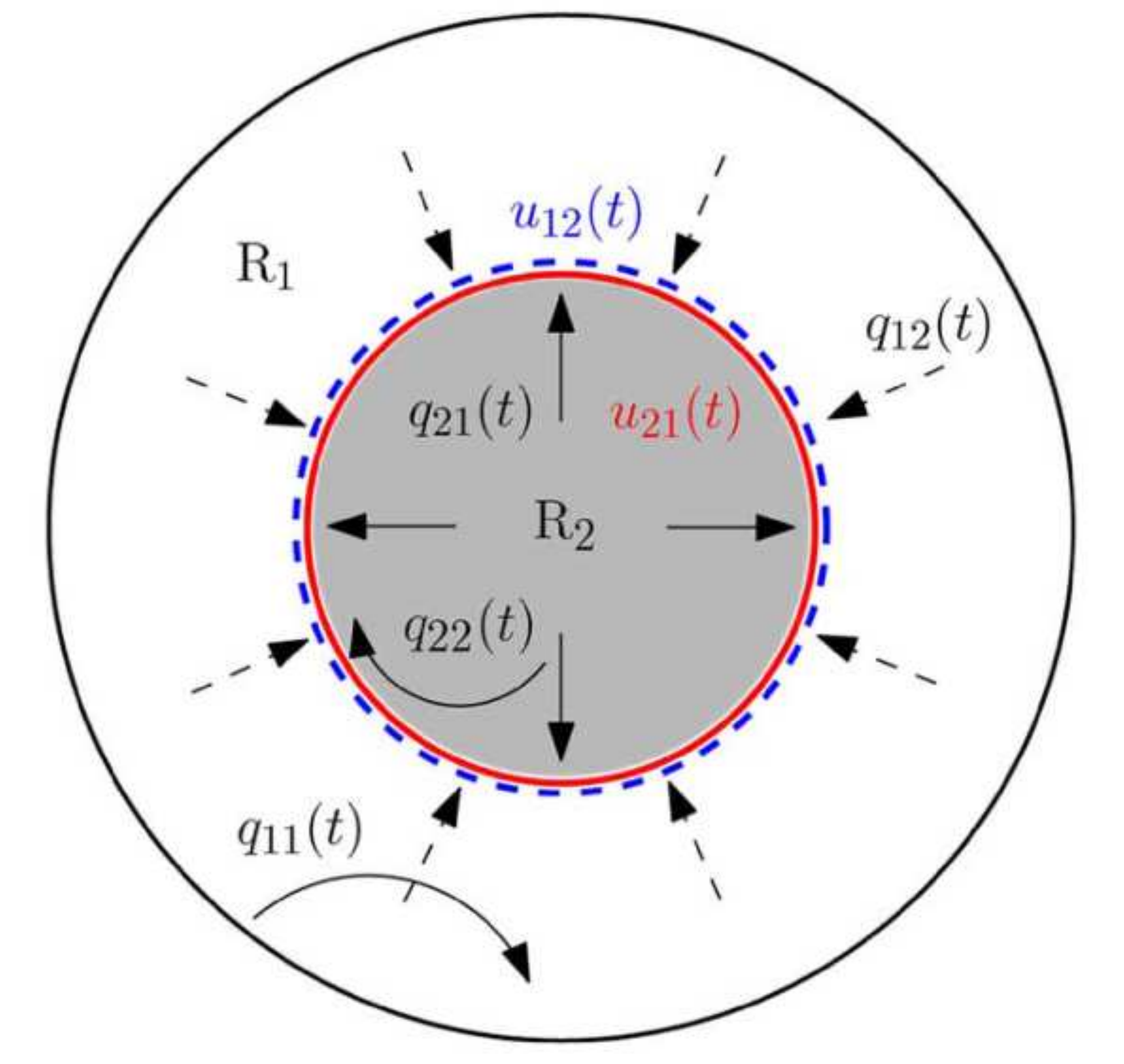}
\end{minipage}
}
\subfigure[The three-region MFD network topology]{
\begin{minipage}[b]{0.4\linewidth}\label{trtopo}
\centering
\includegraphics[width=2.0in]{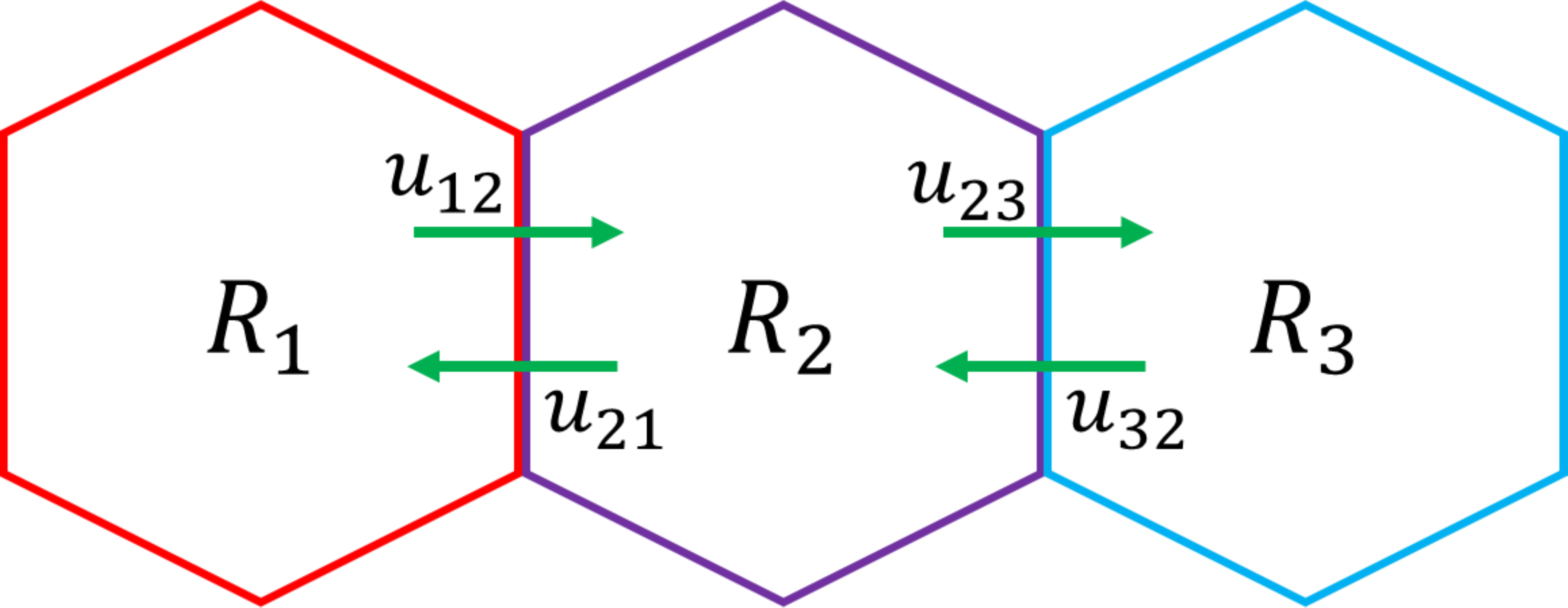}
\end{minipage}
}
\centering
\caption{Network topologies}\label{nettopo}
\end{figure}

To test the performance of the proposed method, two scenarios with different purposes and settings are simulated (see \autoref{egset}). The network topologies used in these numerical examples are shown in \autoref{nettopo}. For set-point control objective, the two-region MFD system \citep{haddad2015robust} with constant demand pattern is considered in Scenario 1 for demonstration of the convergent speed and computation efficiency of the proposed method. For min TTS control objective, a three-region MFD system as in \cite{ZHONG2018327} with time-varying travel demand is considered in Scenario 2. The robustness and adaptiveness of the proposed method are validated by conducting experiments under various demand patterns. The subregion MFD functions of all the examples are assumed to be the same. The true, but unknown MFD functions and the parameters are given in \autoref{egset} to generate the I/O data for learning the traffic dynamics only. Note that they are not involved in the controller design. For the examples in Scenario 1, the cost functions $N(\tilde{n})=\tilde{n}^TQ\tilde{n}$ and $U(\tilde{u})$ is defined by \eqref{conipt}, where $Q=10^{-2}\cdot \mathcal{I}_{\alpha_n}$, $R=\mathcal{I}_{\alpha_u}$ with $\mathcal{I}_{x}$ denoting the identity matrix of dimension $x$. For the experiment in Scenario 2, the objective function is defined by \eqref{eq:vftts}, where $\bar{\lambda}=1$.

\begin{table}[!htb]
\caption{Scenario description}\label{egset}
\centering
\begin{threeparttable}
\begin{tabular}{lllll}
\hline
         & Network                  & Demand                   & Controllers    &  MFD parameters \\
\hline
1-A      & \multirow{3}*{2-region}  & \multirow{3}*{constant}  & IRL, N-DP      & $G_{i}(n_{i})=$ \\
1-B      &                          &                          & IRL (various reinforcement intervals) &  \multirow{2}*{$\frac{1.4877\cdot 10^{-7}n^{3}_{i}-2.9815\cdot 10^{-3}n^{2}_{i}+15.0912 n_{i}}{3600}$ (veh/s)} \\
1-C      &                          &                          & IRL, MPC       &   \\
\cline{1-4}
2        &               3-region   & time-varying             & IRL, MPC       & $n_{i}^{jam}=10000$ (veh), $n_{i}^{cr}=3392$ (veh) \\
\hline
\end{tabular}
\end{threeparttable}
\end{table}

The stabilizing control law by \cite{haddad2015robust} embedded with a sequence of randomly generated deviations is adopted to initialize the online learning algorithm.
The AC-NN framework is employed for approximating the optimal value function and control policy in all examples. Let $\phi_{V}\in \mathbb{R}^{K_{V}}$ and $\phi_{D}\in \mathbb{R}^{K_{D}}$ denote the activation functions of the online learning approach.

For Scenario 1, inspired by \cite{abu2005nearly}, we adopt an AC-NN framework with 84 critic NN hidden neurons and 4 actor NN hidden neurons, i.e., $K_{V}=84$ and $K_{D}=4$. Suppose $x=[x_{1},x_{2},x_{3},x_{4}]^T$, the activation function of critic NN is
\begin{equation}\label{v84}
  \phi^{p_V}_{V}(x)= x^{i}_{1}x^{j}_{2}x^{m}_{3}x^{n}_{4}
\end{equation}
where $i+j+m+n=6$ and $p_V=1,\ldots,84$, and the activation function of actor NN is
\begin{equation}\label{v4}
  \phi^{p_D}_{D}(x)= x_{p_D}
\end{equation}
where $p_D=1,\ldots,4$.

For Scenario 2, we set $K_{V}=210$ and $K_{D}=7$. Suppose $x=[x_{1},x_{2},x_{3},x_{4},x_{5},x_{6},x_{7}]^T$, the activation function of critic NN is
\begin{equation}\label{v210}
  \phi^{p_V}_{V}(x)= x^{i}_{k_1}x^{j}_{k_2}x^{m}_{k_3}x^{n}_{k_4}
\end{equation}
where $i+j+m+n=6$, $p_V=1,\ldots,210$ and $k_1,k_2,k_3,k_4\in\{1,\ldots,7\}$, and the activation function of actor NN is
\begin{equation}\label{v7}
  \phi^{p_D}_{D}= x_{p_D}
\end{equation}
where $p_D=1,\ldots,7$.

The sample size and replay buffer (history data stack) size for AC-NN updates in all the examples are 250 and 1000, respectively. The computer processor is Intel Core i7-9850 CPU \@2.60 GHz, and the simulation platform is MATLAB R2022a.

\subsection{Set-point control}

In this subsection, we apply the proposed IRL based online iterative learning approach to the two-region network with constant travel demand. A two-region network as shown in \autoref{twotopo} is considered in this scenario. As explained, the objective of set-point perimeter control is to regulate the network traffic state to the desired stable equilibrium. Comparison in terms of control performance and computational efficiency is made between the proposed IRL approach and other existing controllers, e.g., the state-of-the-art MPC by \cite{geroliminis2013optimal} and the neuro-dynamic programming (N-DP) method by \cite{SU2020102628}. Note that MPC is a model-based controller, and that N-DP requires partial information of the MFD system, while the IRL does not rely on any knowledge of the traffic dynamics.

\vspace{6pt}
\noindent \textbf{Scenario 1-A: Comparison between the IRL and the N-DP approaches}
\vspace{6pt}

In Scenario 1-A, we present a comparison between the proposed off-policy learning based IRL approach and the on-policy learning based N-DP approach by \cite{SU2020102628}.
In line with \cite{haddad2015robust}, $\bar{n}=[3000,3000]^T$ (veh), which is close to the critical accumulation, is chosen as the desired equilibrium. In addition, the demand pattern is set to be constant as $q = [1.6, 1.6, 1.6, 1.6]^T$ (veh/s). Thus, the steady-state accumulation for each direction and the corresponding control inputs as solved from the steady-state equations \eqref{ss1}-\eqref{ss3} are $n^* = [1538.9, 1461.1, 1461.1, 1538.9]^T$ (veh) and $u^*= [0.5267, 0.5267]^T$. The initial regional accumulations are set to be $[1800, 3100]^T $ (veh) with OD-specific initial accumulations being $n_{11}(0)=540$ (veh), $n_{12}(0)=1260$ (veh), $n_{21}=2170$ (veh), $n_{22}(0)=930$ (veh).

Note that N-DP requires input data with high resolution to solve the HJB equation for the optimal controller. For a fair comparison, the first case is that the sample time interval (and thus reinforcement interval) and the control update step are set as 1 second for both methods. To consider more practical situations, in the second case and third case, we set the sample time interval and the control update step to 15 seconds \citep{haddad2015robust} and 30 seconds, respectively. Sensitivity analysis of the reinforcement interval for the IRL is presented in Scenario 1-B.

\begin{figure}[!htb]
\subfigure[State evolutions with $\Delta t=1s$]{
\begin{minipage}[t]{0.33\linewidth}\label{com_state1}
\centering
\includegraphics[width=2.1in]{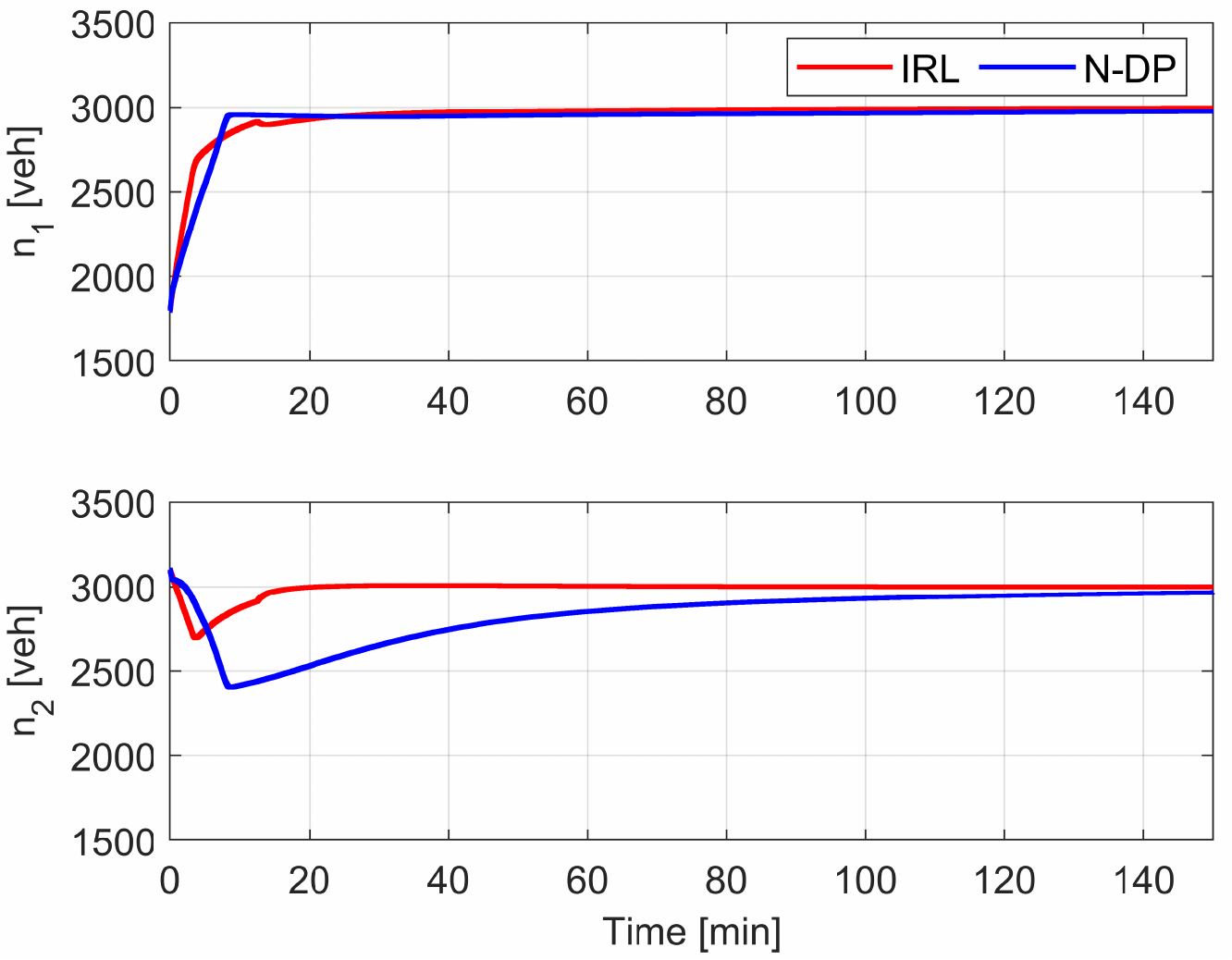}
\end{minipage}
}%
\subfigure[State evolutions with $\Delta t=15s$]{
\begin{minipage}[t]{0.33\linewidth}\label{com_state15}
\centering
\includegraphics[width=2.1in]{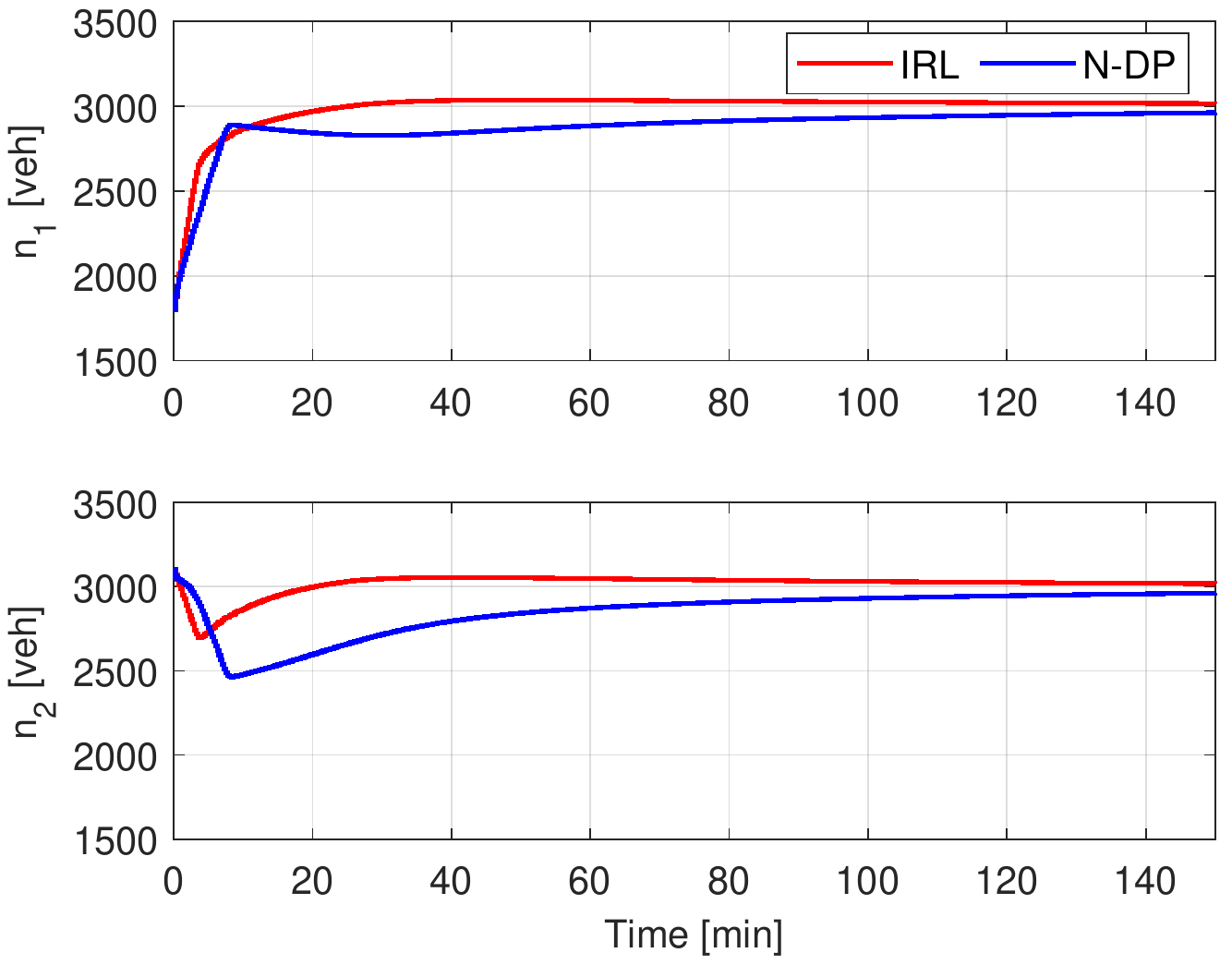}
\end{minipage}
}%
\subfigure[State evolutions with $\Delta t=30s$]{
\begin{minipage}[t]{0.33\linewidth}\label{com_state30}
\centering
\includegraphics[width=2.1in]{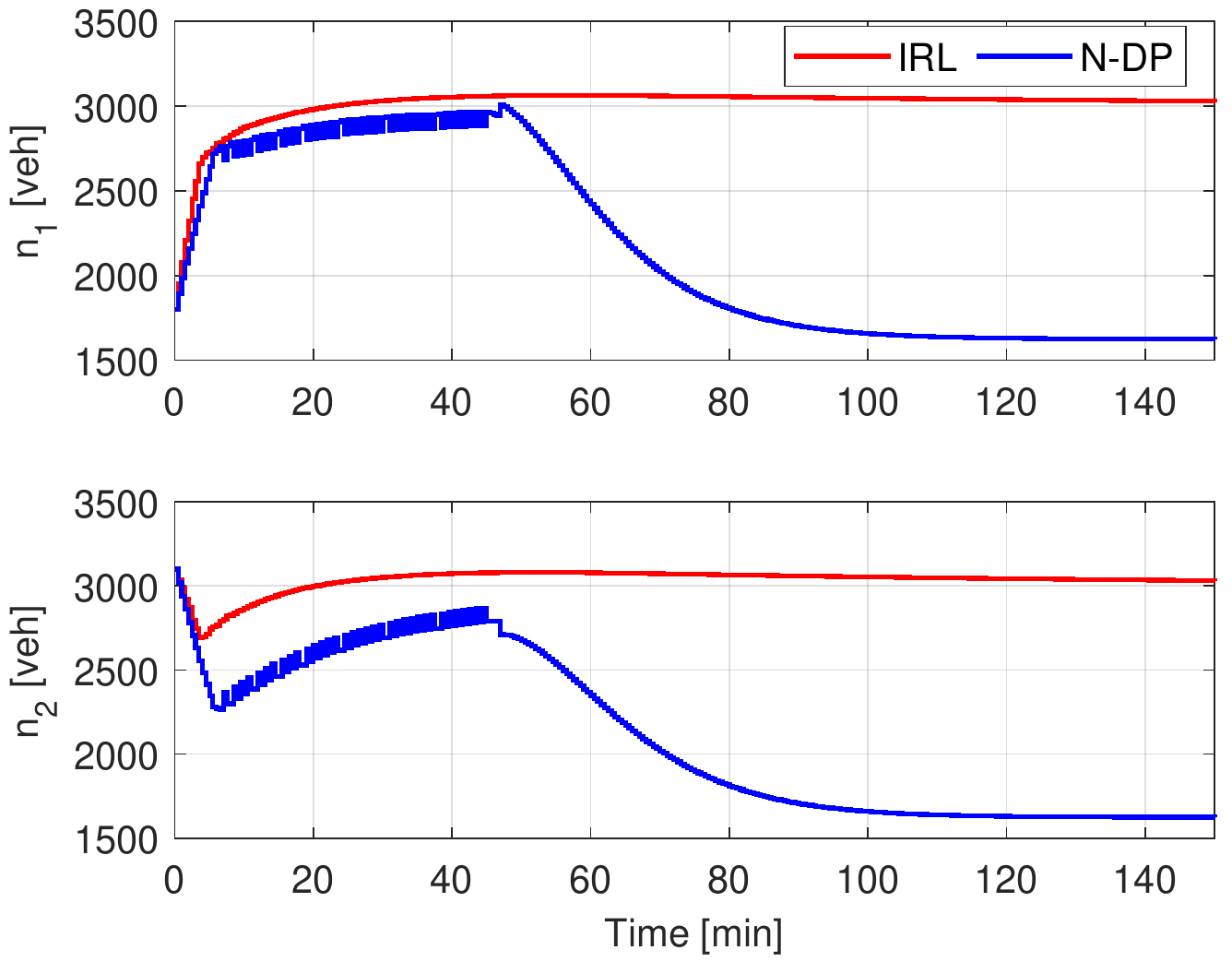}
\end{minipage}
}%
\centering
\caption{Simulation results of Scenario 1-A}\label{sce1b_c1}
\end{figure}

As shown in \autoref{com_state1}, when $\Delta t=1$, both the IRL and the N-DP can regulate the accumulation states to the desired equilibrium $[3000, 3000]^T$ (veh) in an asymptotic manner. Specifically, the N-DP controller achieves a shorter settling time\footnote{The settling time is the time required for the dynamics to reach and stay within a small range of certain percentage (usually 5\% or 2\%) of the desired steady state \citep[see Fig. 3.23 and Chapter 3.4.3 in][]{franklin2015feedback}. In our case, the settling time is the time required for $\tilde{n}_i(t)$ to reach and stay within 2\% of the steady state $\bar{n}_i$.} than the IRL approach for accumulation state $n_1(t)$, while the IRL is much better than the N-DP in the settling time for $n_2(t)$. Note that the N-DP control algorithm has been well-trained in an off-line manner before it is applied. However, only by interacting with the environment and learning the macroscopic traffic dynamics online, the proposed IRL approach can achieve settling times of around 20 minutes for both $n_1$ and $n_2$. Besides, $n_2(t)$ has experienced an overshoot to around $2500$ (veh) applying the N-DP based controller, while the overshoot induced by the IRL approach is much smaller. Moreover, applying the N-DP, the increase of the reinforcement interval slows down the convergent speed of accumulation states (see \autoref{com_state15}) or even cannot regulate them to the desired equilibria (see \autoref{com_state30}). However, the IRL approach can stabilize the accumulations at the desired steady states in all the cases within around 20 minutes. These results indicate that the IRL approach can achieve decent convergence and stability of the accumulation states under different $\Delta t$, while the control performance of N-DP deteriorates as $\Delta t$ increases.

There are also differences in the computational complexity and data usage efficiency. Note that $20000$ samples are used for off-line training in each iteration (40 iterations in total) for the N-DP approach, whereas only $250$ samples are used in each iteration for the IRL approach. This is the advantage of integrating the ER technique with IRL, i.e., fast convergence of the iterative learning process can be guaranteed. Take the first case as an example, because of the reduction of samples used for each iteration, the total computation time of the IRL approach is less than $8$ seconds while that of the N-DP approach is more than $2$ minutes. Different from the conventional RL methods (e.g., N-DP) which are usually trained off-line and data intensive, the improvement in data usage efficiency by integrating IRL with ER indicates the real-time applicability of the proposed IRL based perimeter control schemes.

\vspace{6pt}
\noindent \textbf{Scenario 1-B: Sensitivity analysis of the reinforcement interval}
\vspace{6pt}

\begin{figure}[!htb]
\centering
\subfigure[State $n_{1}$ evolution]{
\begin{minipage}[t]{0.4\linewidth}\label{n1delT}
\centering
\includegraphics[width=2.5in]{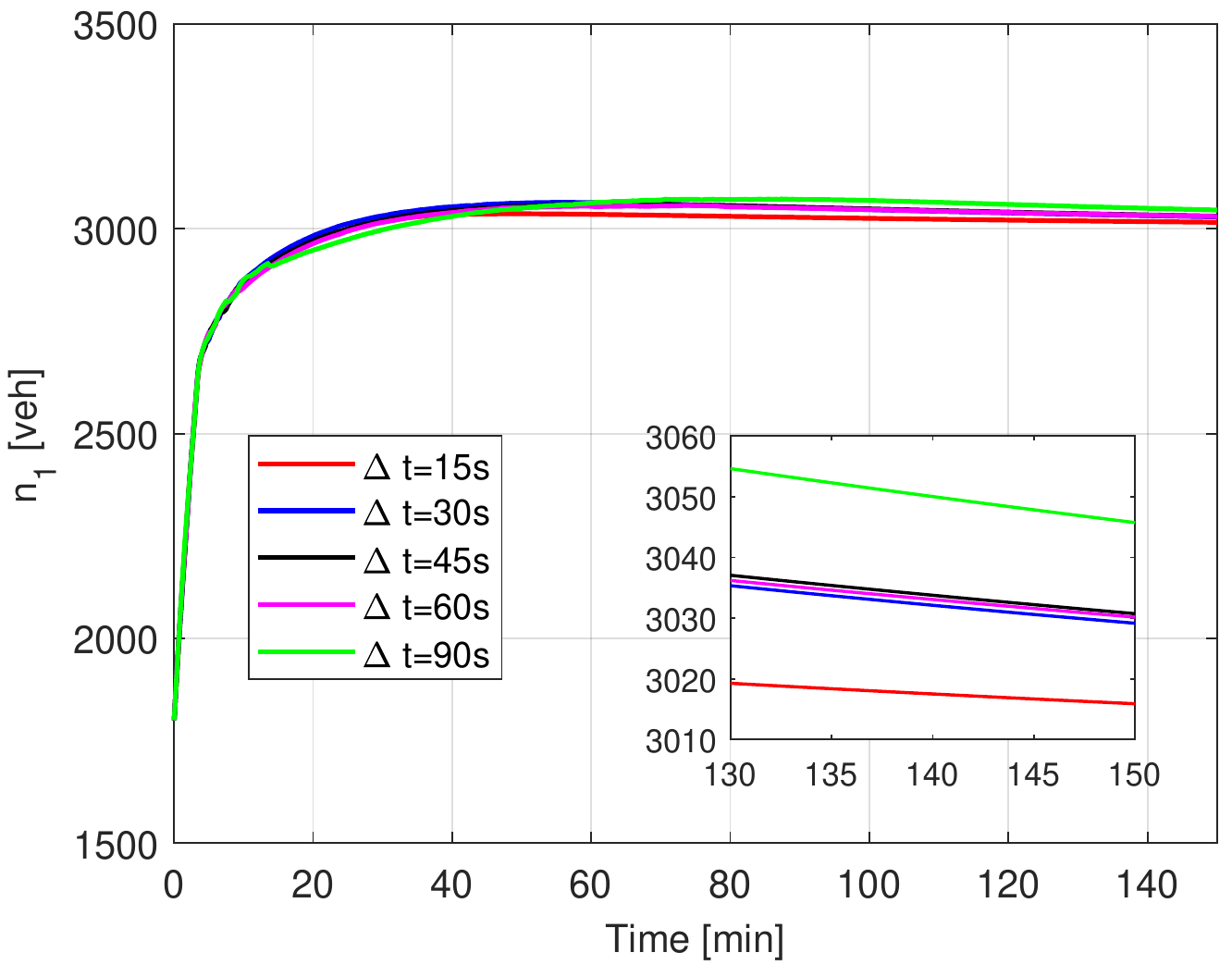}
\end{minipage}
}
\subfigure[State $n_{2}$ evolution]{
\begin{minipage}[t]{0.4\linewidth}\label{n2delT}
\centering
\includegraphics[width=2.5in]{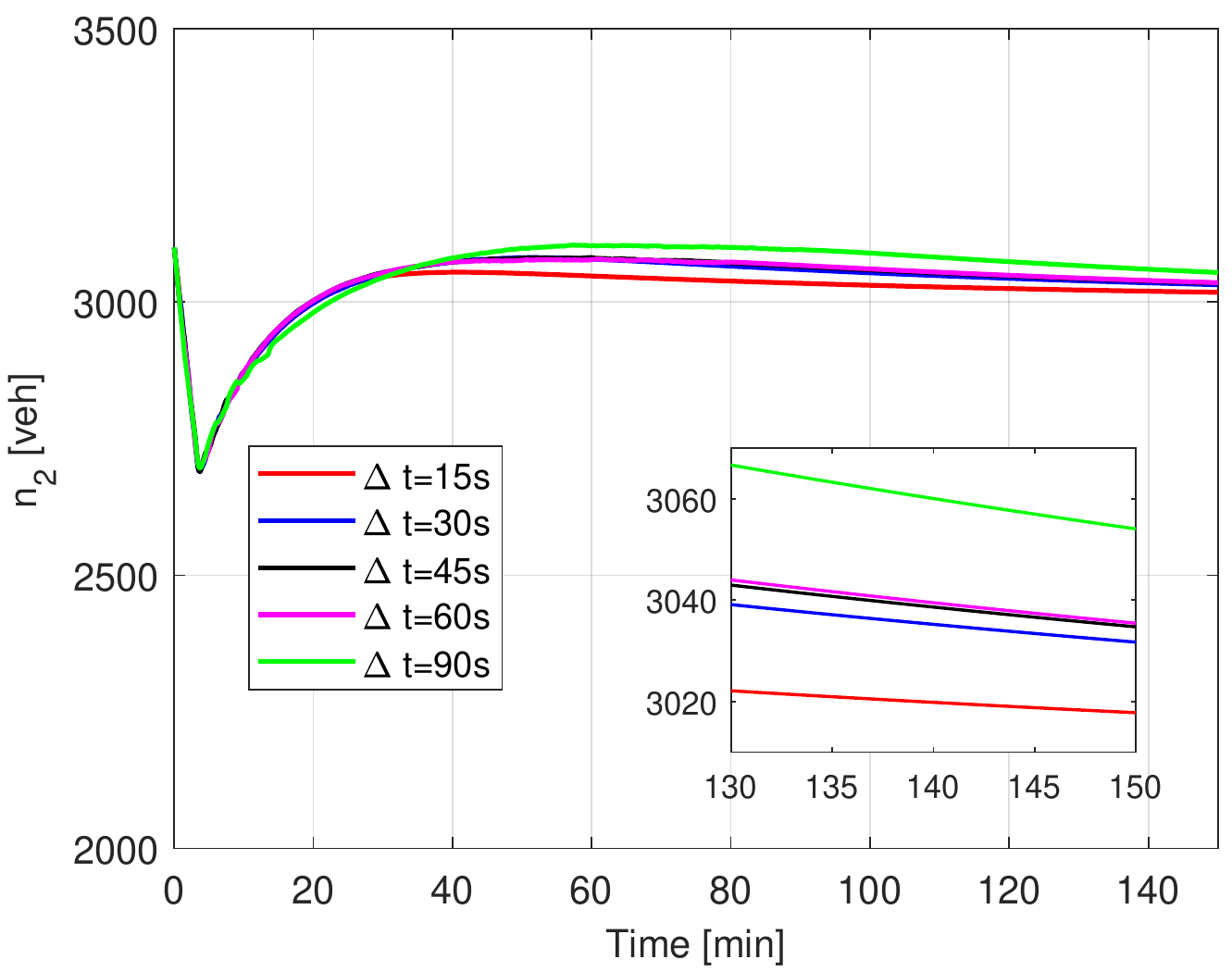}
\end{minipage}
}
\\
\subfigure[Control $u_{12}$ evolution]{
\begin{minipage}[t]{0.4\linewidth}\label{u12delT}
\centering
\includegraphics[width=2.5in]{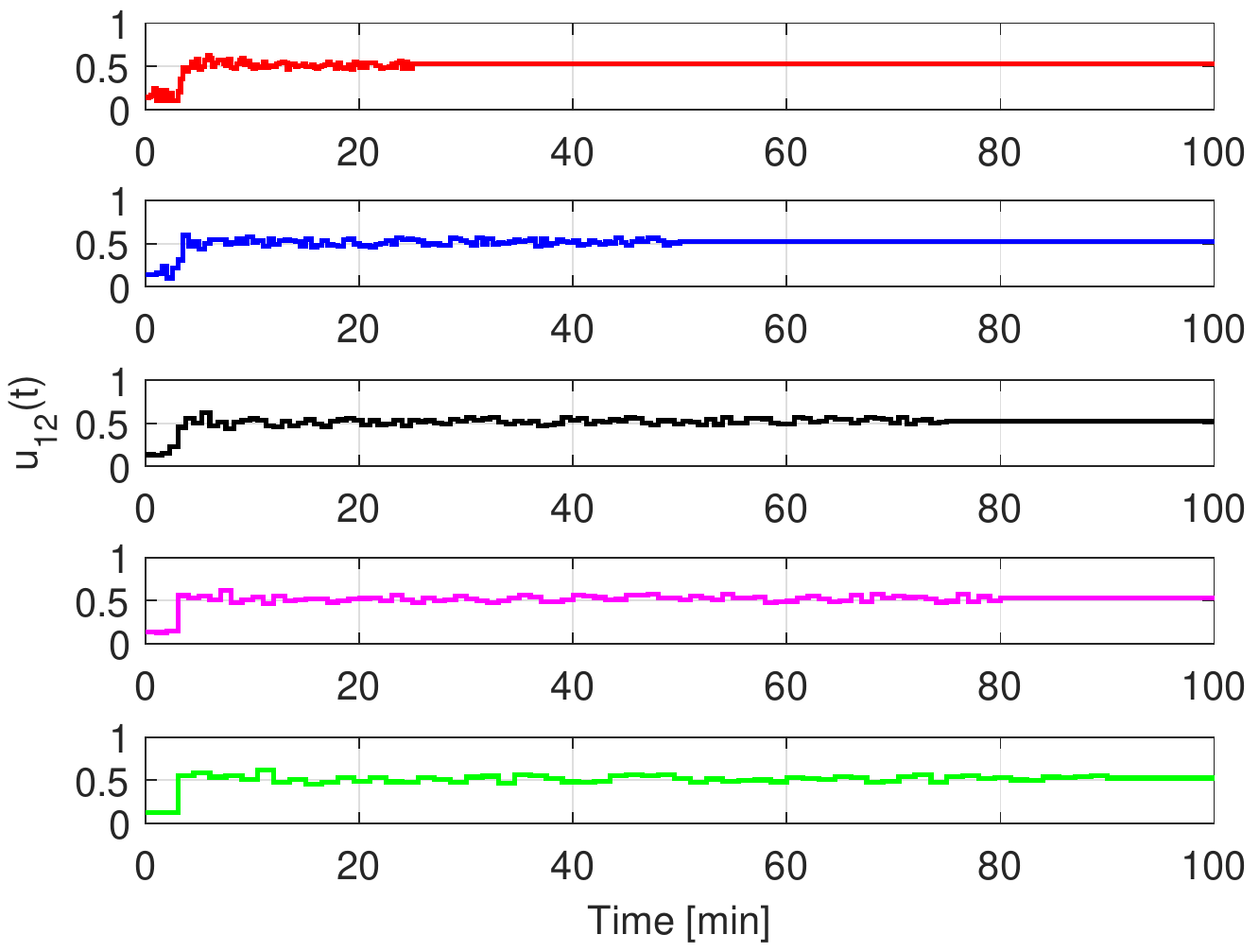}
\end{minipage}
}
\subfigure[Control $u_{21}$ evolution]{
\begin{minipage}[t]{0.4\linewidth}\label{u21delT}
\centering
\includegraphics[width=2.5in]{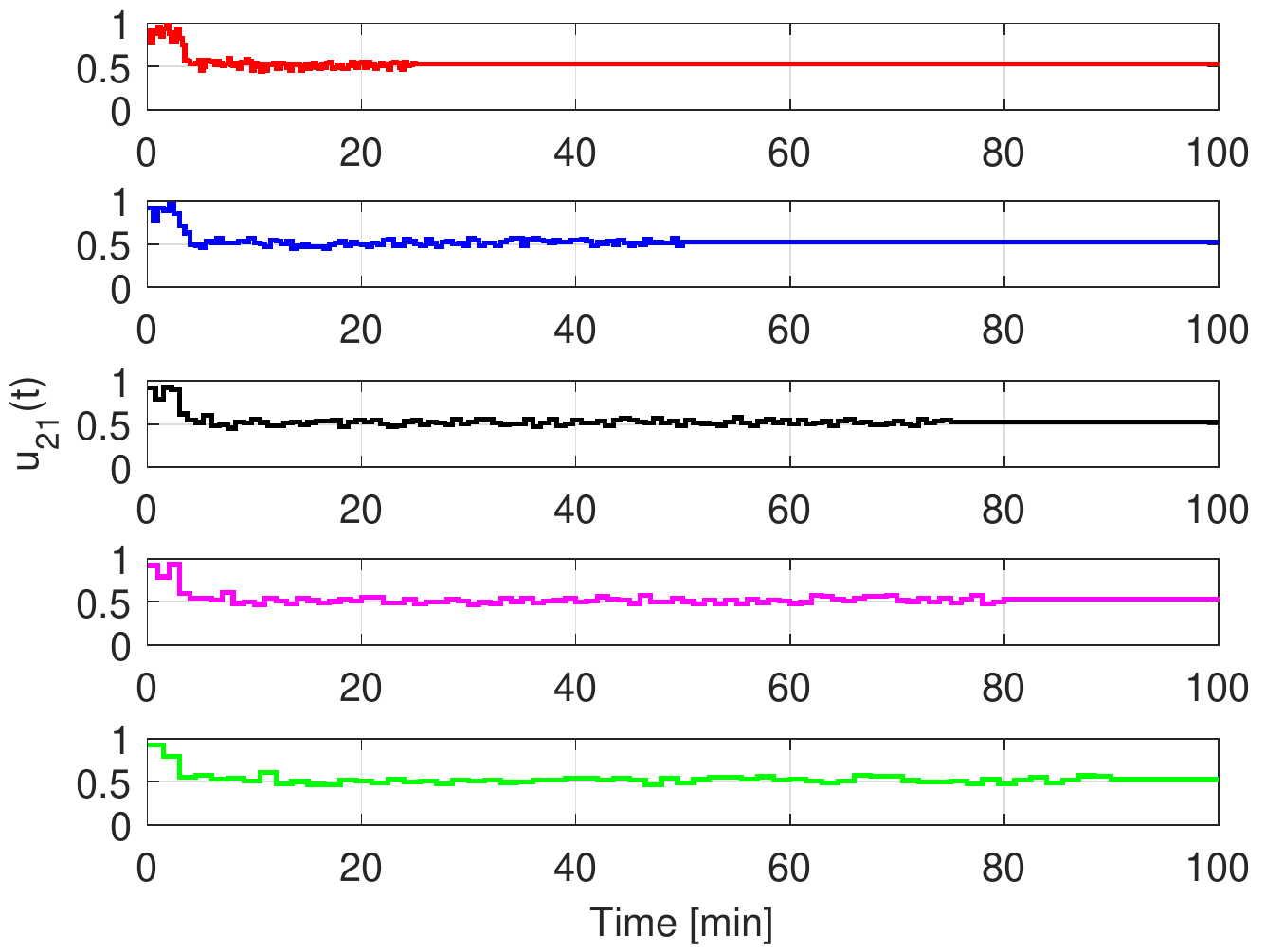}
\end{minipage}
}
\centering
\caption{Simulation results of Scenario 1-B with reinforcement intervals equal to control update steps}
\end{figure}

In Scenario 1-B, sensitivity analysis of the reinforcement interval $\Delta t$ for the proposed IRL method is performed using the network and settings of Scenario 1-A. We assume that the data resolution of traffic sensors is identical to the reinforcement interval so that one data sample is collected in a reinforcement interval. The larger $\Delta t$ is, the lower frequency the sensors upload traffic data to the management center. For instance, $\Delta t = 60$s means the sensors upload data every 1 minute. Note that the larger the reinforcement interval is, the smaller learning rate that could be chosen to achieve the AC-NN weights convergence \citep{modares2014flow}. In this example, each learning rate $\beta\in \{0.01, 0.007, 0.005, 0.003, 0.0001, 0.00007\}$ is chosen respectively for each reinforcement interval $\Delta t\in\{15s, 20s, 30s,$ $45s, 60s, 90s\}$.

Regarding the nature of perimeter control actuation approaches, e.g., traffic signal controls which can be changed only with a new traffic signal cycle, the control update step cannot be smaller than the sample time interval (i.e., the reinforcement interval). Therefore, the sensitivity analysis of reinforcement interval for the IRL algorithm is divided into the following two folds.

The first case is that the control update intervals are equal to the tested reinforcement intervals. \autoref{n1delT} and \autoref{n2delT} present the accumulation trajectories $n_{1}(t)$ and $n_{2}(t)$ over time under different $\Delta t$, respectively, while the control input evolutions are illustrated by \autoref{u12delT} and \autoref{u21delT}. When $\Delta t=15$s, the results show that both the initial states and controls converge very fast to the desired equilibrium. As $\Delta t$ increases, the convergent speed of the perimeter control gain decreases. This slows down the convergent speed of the accumulation states. This is because the algorithm has to wait longer to collect new data to update the weights of the AC-NNs, which also results in less frequent updates of the control inputs. However, \autoref{n1delT} and \autoref{n2delT} indicate that the variation of $\Delta t$ does not significantly influence the convergence of the accumulation states. That is to say, the proposed online learning approach is robust to the variation of real-time data resolution.

\begin{figure}[!htb]
\centering
\subfigure[State $n_{1}$ evolution]{
\begin{minipage}[t]{0.4\linewidth}\label{fcn1delT}
\centering
\includegraphics[width=2.5in]{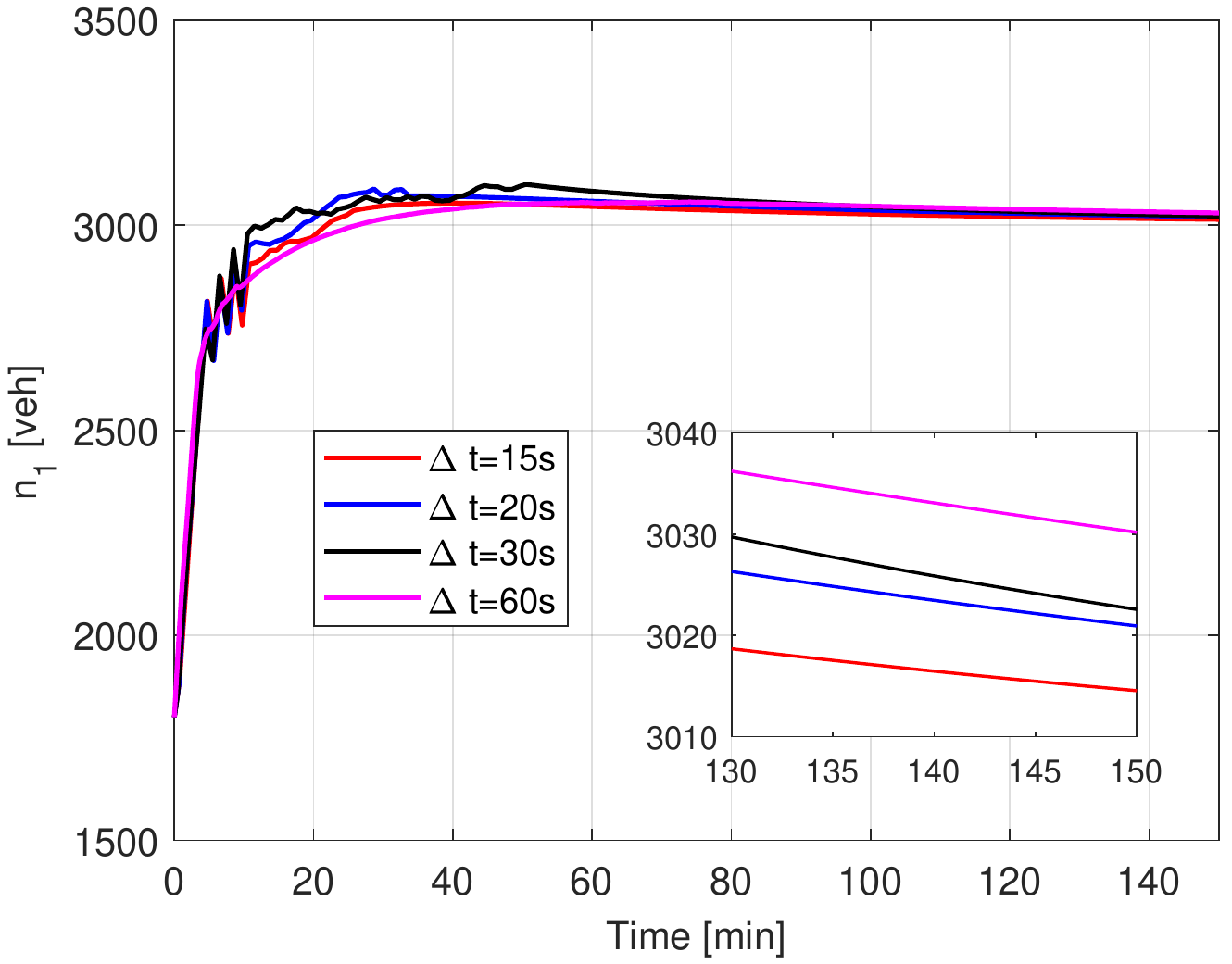}
\end{minipage}
}
\subfigure[State $n_{2}$ evolution]{
\begin{minipage}[t]{0.4\linewidth}\label{fcn2delT}
\centering
\includegraphics[width=2.5in]{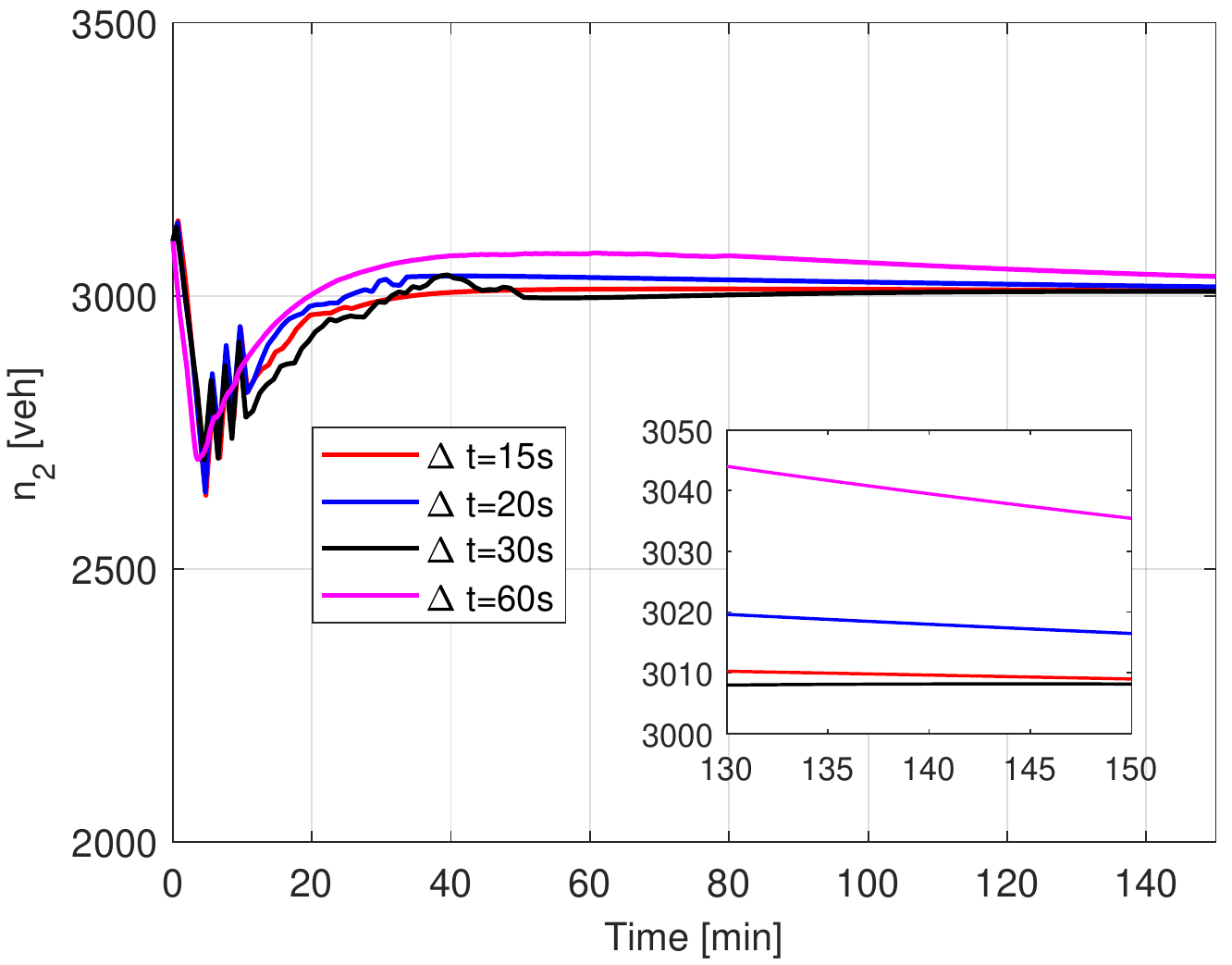}
\end{minipage}
}
\centering
\caption{Simulation results of Scenario 1-B with fixed control update steps}\label{fig:fcrtcom}
\end{figure}

Next, we fix the control update steps to 60 seconds while vary the reinforcement intervals in $\Delta t\in\{15s, 20s, 30s,$ $60s\}$. \autoref{fcn1delT}-\autoref{fcn2delT} shows that the accumulation states can converge to the desired steady states in around 30 minutes. We can also observe that as $\Delta t$ increases, both the convergent speeds of the perimeter control gain and the accumulation states decrease. The IRL based perimeter controller with $\Delta t=15s$ still achieves the shortest settling time. In the early stage of the training and implementation of the IRL controllers, the larger difference between $\Delta t$ and the control update step, the stronger oscillation of the accumulation state occurs. However, with the control update steps fixed, the variation of $\Delta t$ still does not significantly affect the convergence of the accumulation states.

These results imply the feasibility of online tuning of the reinforcement interval to adapt to heterogeneous real-time sensor data resolution without affecting the system stability. This is a key advantage of the proposed IRL based online learning algorithm over the traditional RL based methods.

\vspace{6pt}
\noindent \textbf{Scenario 1-C: Comparison between the IRL and the MPC approaches}
\vspace{6pt}

Scenario 1-C adopts the same settings as Scenario 1-A except the initial condition, the set-point value and the reinforcement interval. Unlike Scenario 1-A corresponding to a mild traffic condition where all regions are regulated in an uncongested regime (i.e., below the critical accumulation), the initial accumulation state values are set to far exceed the critical accumulations, i.e., $[4300,3700]^T$ (veh) with $n_{11}(0)=430$ (veh), $n_{12}(0)=3870$ (veh), $n_{21}=370$ (veh), $n_{22}(0)=3330$ (veh). Besides, in Scenario 1-C, both regions are regulated around set points in the congested regimes, e.g., $\bar{n}=[4000,4000]^T$ (veh). Performance comparison is conducted between the IRL and the state-of-the-art MPC method, where for both the controllers the sample time interval and the control update step are set as 60 seconds. For the MPC controller, the prediction horizon is set to be 30 (i.e., 30 minutes simulation time).

\begin{figure}[!htb]
\subfigure[State $n_1$ evolution]{
\begin{minipage}[t]{0.4\linewidth}\label{fig:sp4000state_n1}
\centering
\includegraphics[width=2.5in]{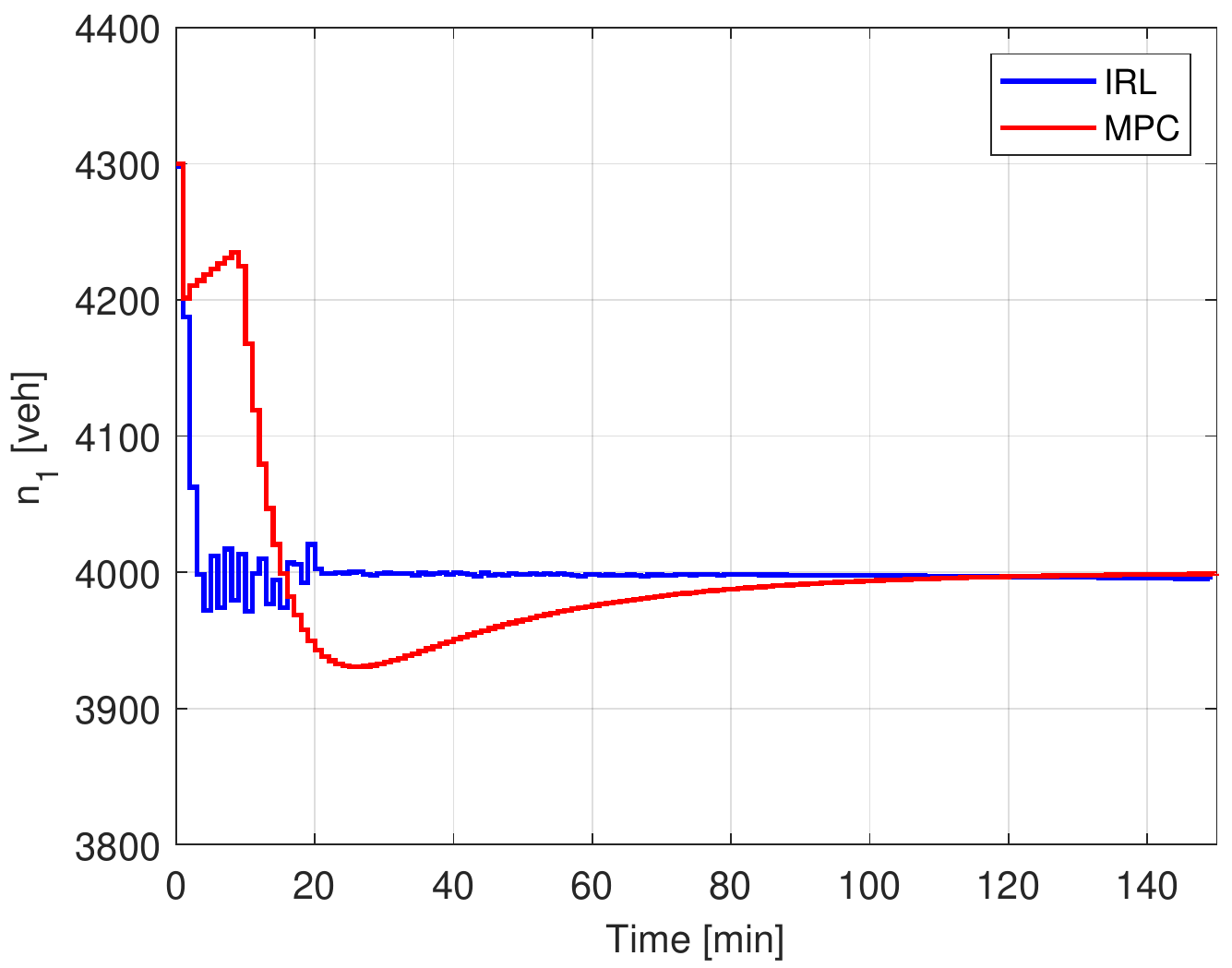}
\end{minipage}
}
\subfigure[State $n_2$ evolution]{
\begin{minipage}[t]{0.4\linewidth}\label{fig:sp4000state_n2}
\centering
\includegraphics[width=2.5in]{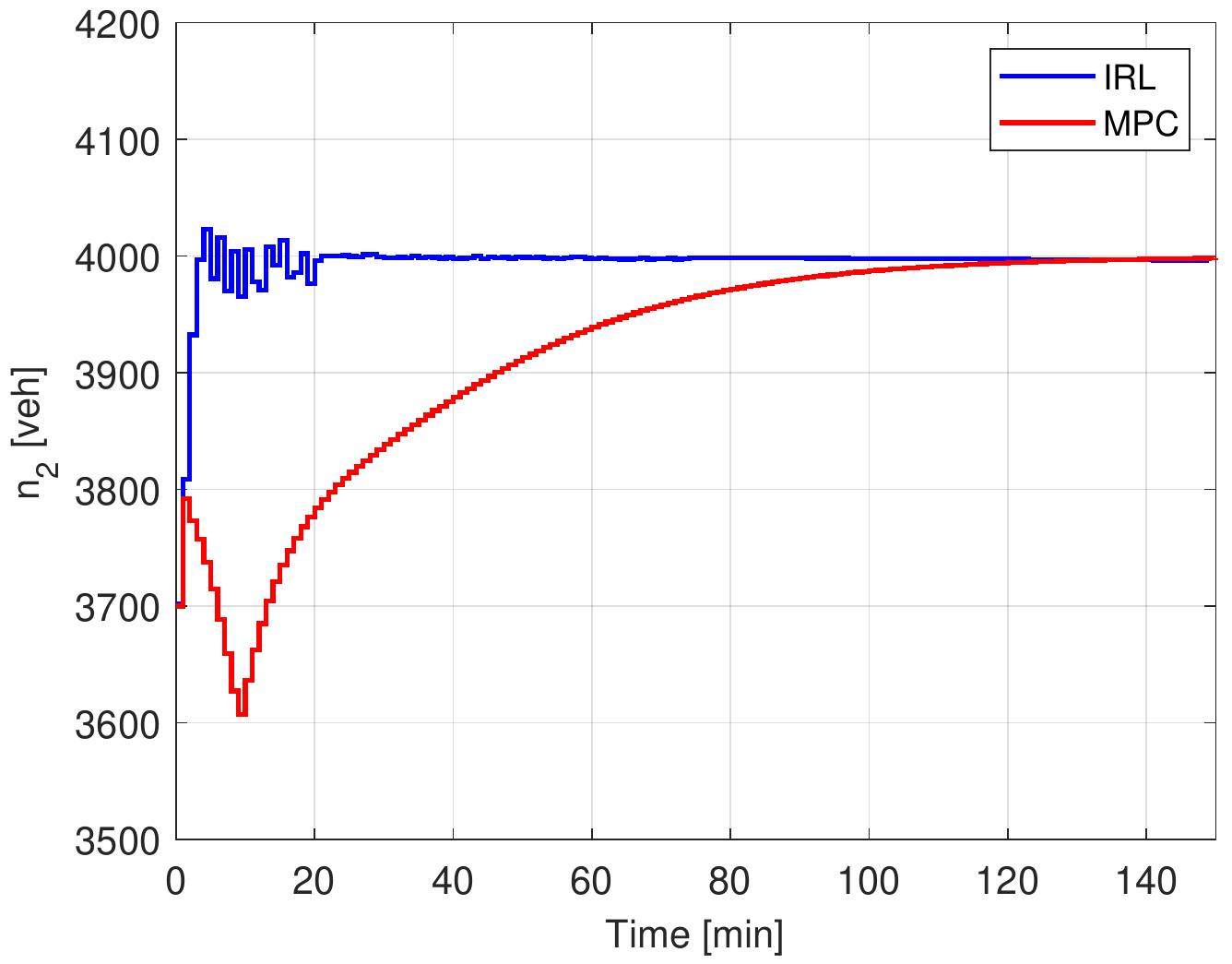}
\end{minipage}
}
\centering
\caption{Simulation results of Scenario 1-C}\label{fig:cspIRLMPC}
\end{figure}

\autoref{fig:cspIRLMPC} shows that both the IRL and the MPC controllers can stabilize the accumulation states at the desired equilibrium. Regulated by the IRL controller, both $n_1$ and $n_2$ converge very fast to the steady states, while regulated by the MPC controller, one can observe small overshoots of the accumulation states. The settling time and the average CPU time per control update step\footnote{The CPU time is defined as the average computation time per control update step. They were measured by the tic and toc functions of MATLAB R2022a. We present the average value of 10 tests for each controller.} of different control schemes are reported in \autoref{comptwo}. Despite no model knowledge available, the IRL approach can achieve a 20-minute settling time, which indicates that the IRL can have a decent control performance in a congested traffic situation. The CPU times per control update step of both methods are extremely small, which are far less than the 60-second control update step. These results imply the real-time applicability of the proposed model-free IRL approach.

\begin{table}[!htb]
    \caption{Summary of settling time and computation time for Scenario 1-C}\label{comptwo}
    \centering
    \begin{threeparttable}
\begin{tabular}{lccc}
  \hline
                                                     & State    & IRL                & MPC       \\
  \hline
  \multirow{2}*{Settling time  (min)}                & $n_{1}$  & $\approx$ 22       & $\approx$ 29   \\
                                                     & $n_{2}$  & $\approx$ 21       & $\approx$ 53    \\
                                                              \cline{2-4}
  \hline
  CPU time per step  (sec)                           &   --   &  6.57$\times 10^{-6}$  & 1.79$\times 10^{-2}$  \\
  \hline
\end{tabular}
\end{threeparttable}
\end{table}

\subsection{TTS minimization}

In this subsection, the objective function is related to minimizing the total time spent (TTS) for the urban network subject to uncertainties in travel demands. We apply the proposed IRL based perimeter controller to the three-region network shown by \autoref{trtopo} with a time-varying demand pattern. The perimeter controller is subject to heterogeneous cross-boundary capacities, i.e.,
\begin{equation*}
0.1\leq u_{12}\leq 0.7,\; 0.3\leq u_{21}\leq 1,\; 0.4\leq u_{23}\leq 1,\; 0.2\leq u_{32}\leq 0.9
\end{equation*}

\vspace{6pt}
\noindent \textbf{Scenario 2: Three-region MFD system with uncertain time-varying travel demand}
\vspace{6pt}

\begin{figure}[!htb]
\subfigure[Nominal demand]{
\begin{minipage}[t]{0.33\linewidth}\label{fig:basedm}
\centering
\includegraphics[width=2.1in]{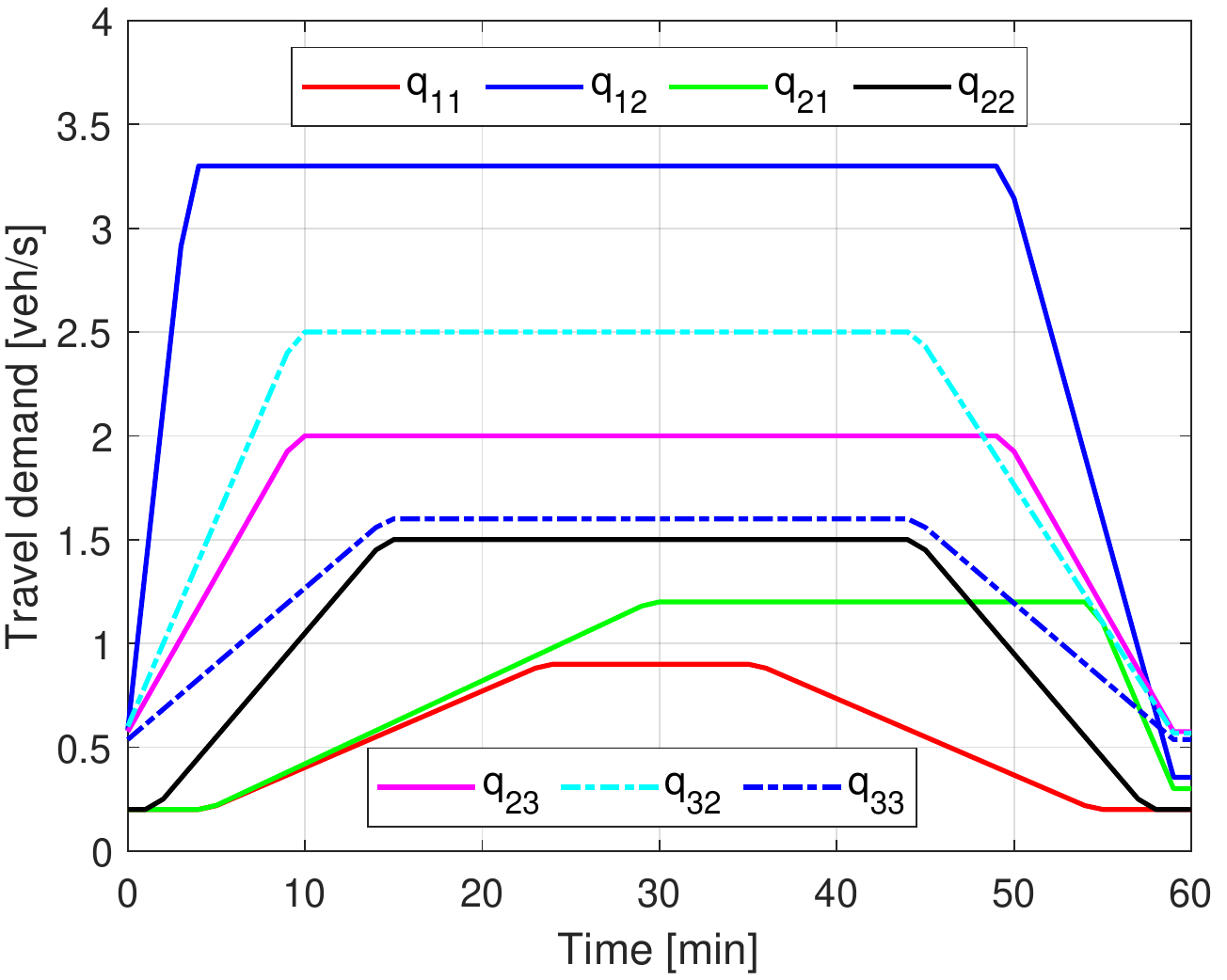}
\end{minipage}
}%
\subfigure[Noisy demand]{
\begin{minipage}[t]{0.33\linewidth}\label{fig:noisydm}
\centering
\includegraphics[width=2.1in]{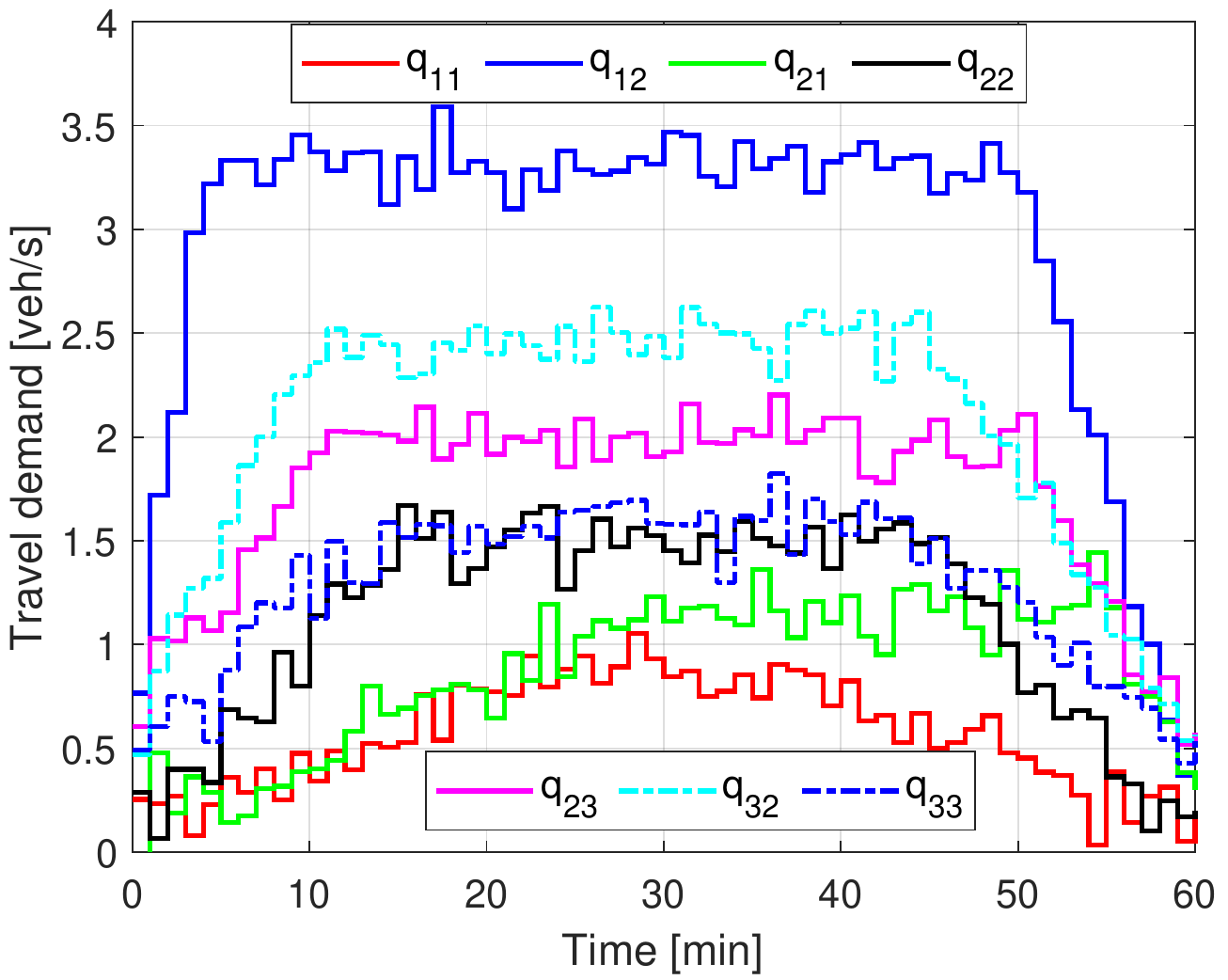}
\end{minipage}
}%
\subfigure[Abrupt changes of demand]{
\begin{minipage}[t]{0.33\linewidth}\label{fig:shchdm}
\centering
\includegraphics[width=2.1in]{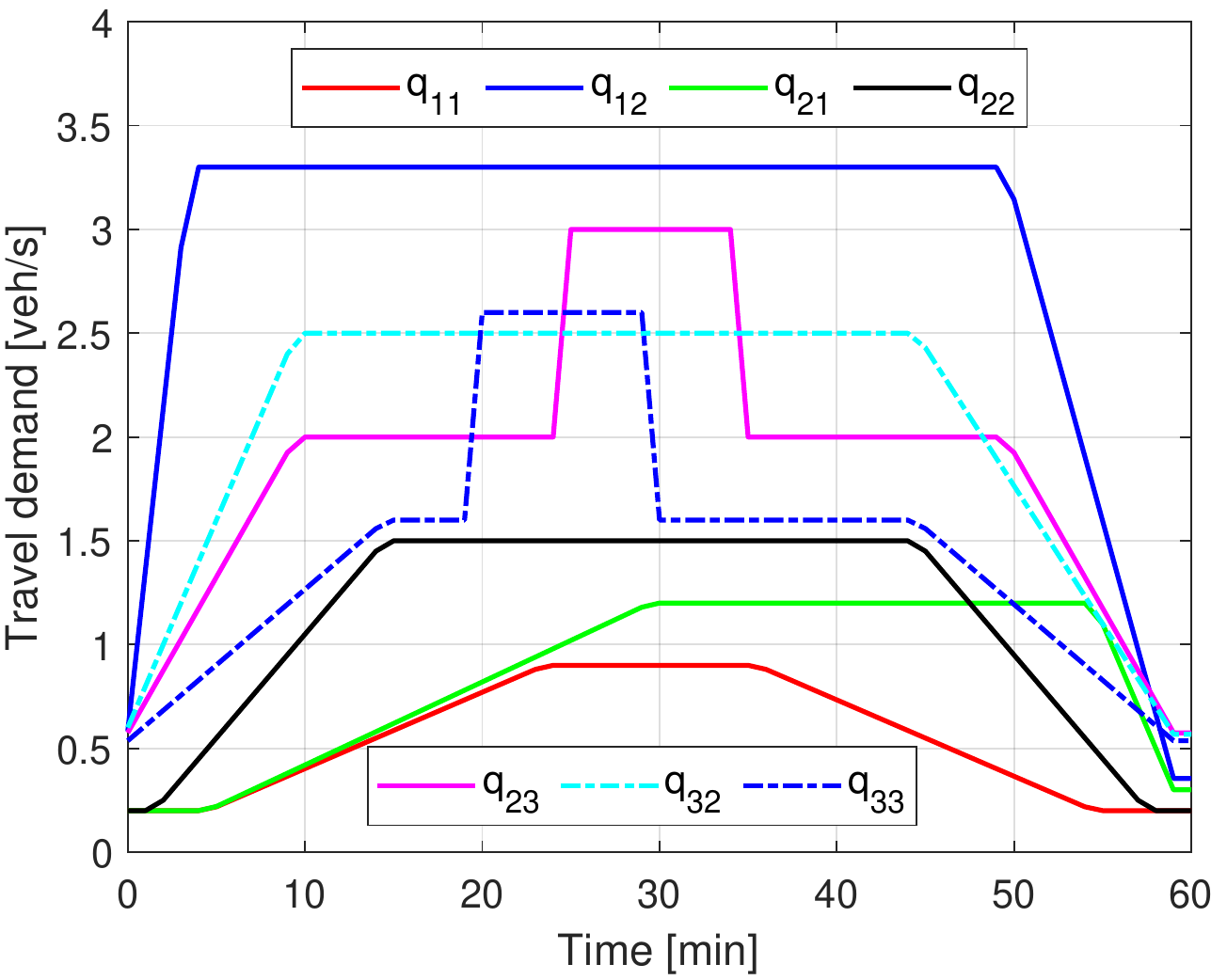}
\end{minipage}
}%
\centering\\
\subfigure[States under nominal demand]{
\begin{minipage}[t]{0.33\linewidth}\label{fig:nmstate}
\centering
\includegraphics[width=2.1in]{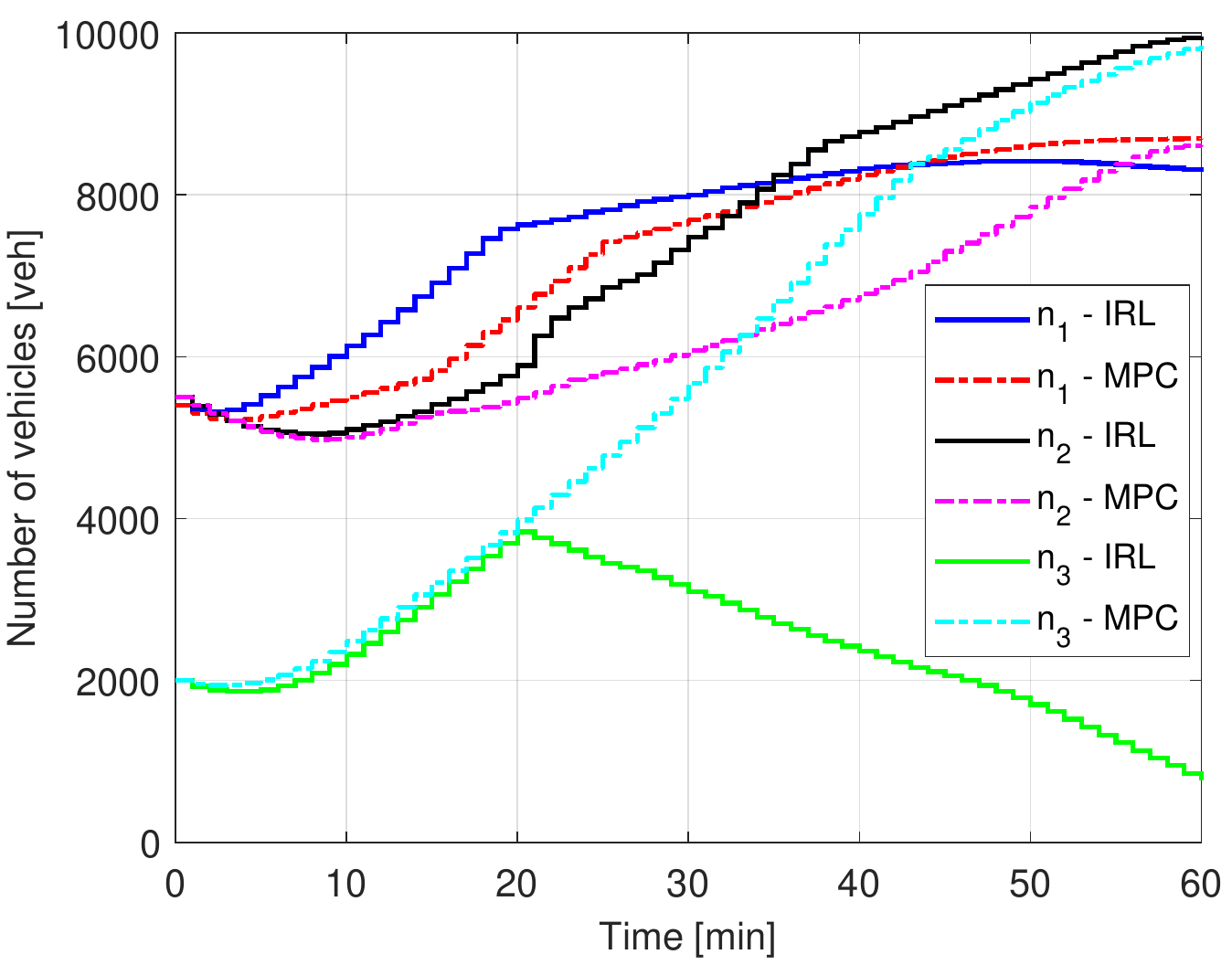}
\end{minipage}
}%
\subfigure[States under noisy demand]{
\begin{minipage}[t]{0.33\linewidth}\label{fig:nsstate}
\centering
\includegraphics[width=2.1in]{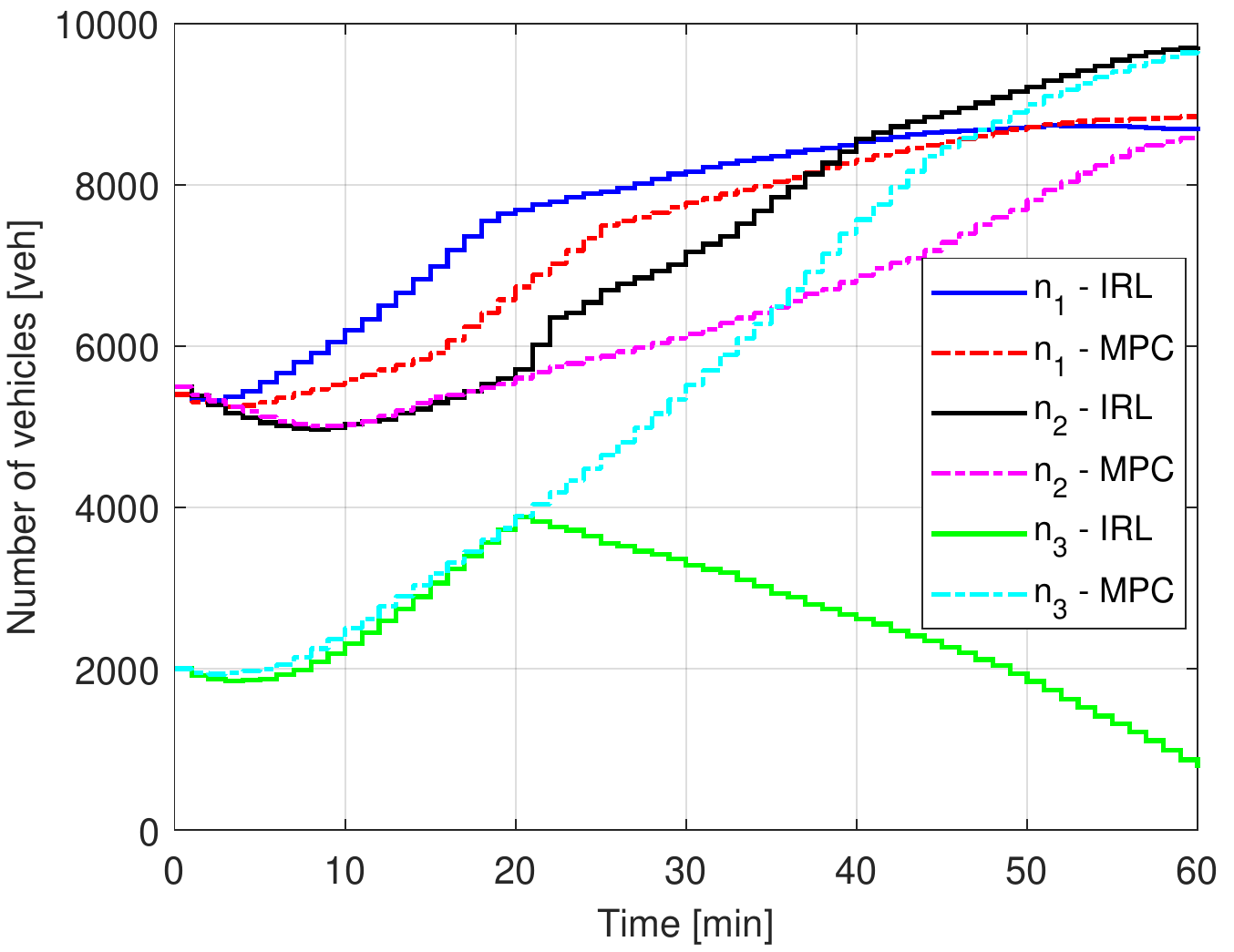}
\end{minipage}
}%
\subfigure[States under abrupt-change demand]{
\begin{minipage}[t]{0.33\linewidth}\label{fig:scstate}
\centering
\includegraphics[width=2.1in]{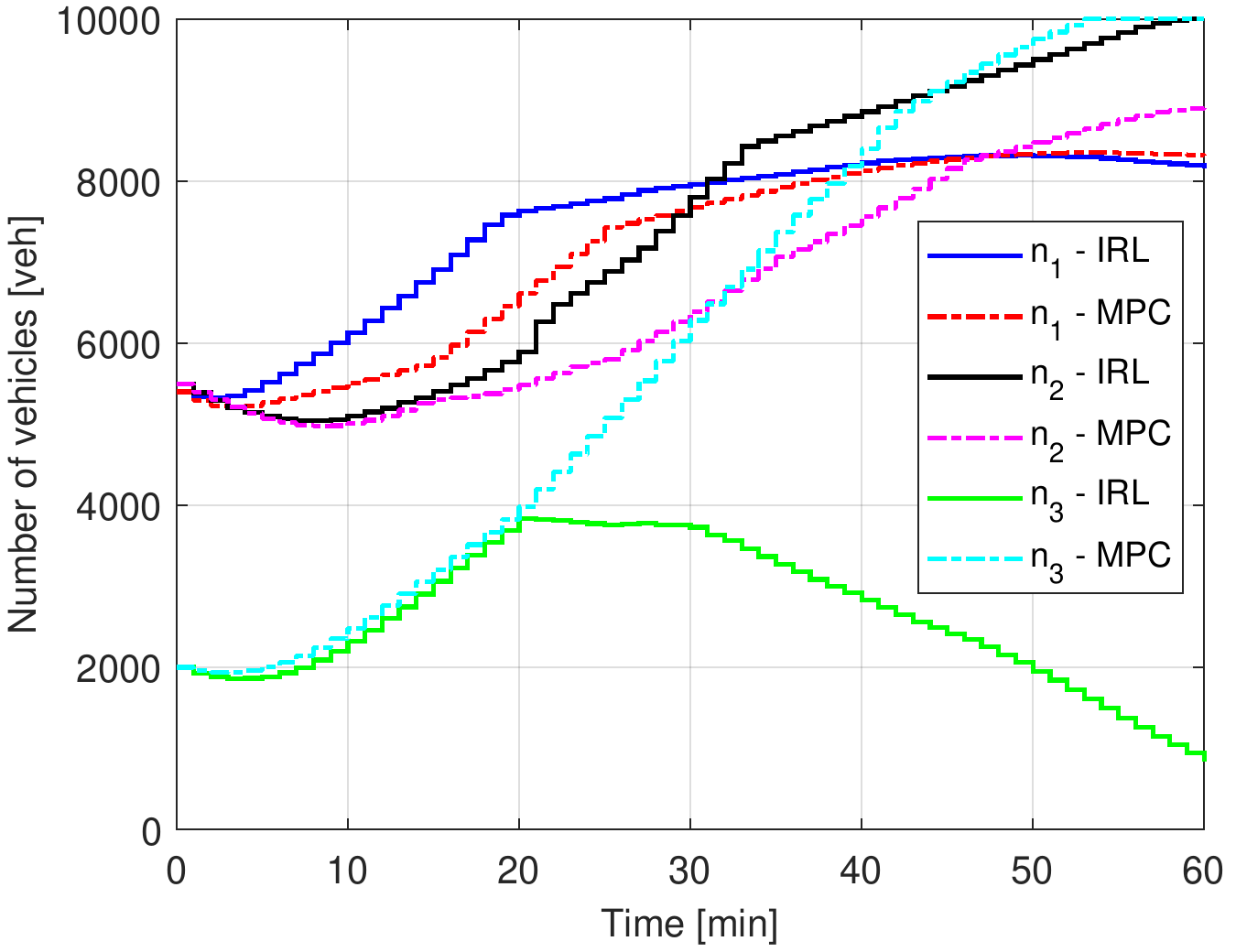}
\end{minipage}
}%
\centering\\
\subfigure[Controls under nominal demand]{
\begin{minipage}[t]{0.33\linewidth}\label{fig:nmctrl}
\centering
\includegraphics[width=2.1in]{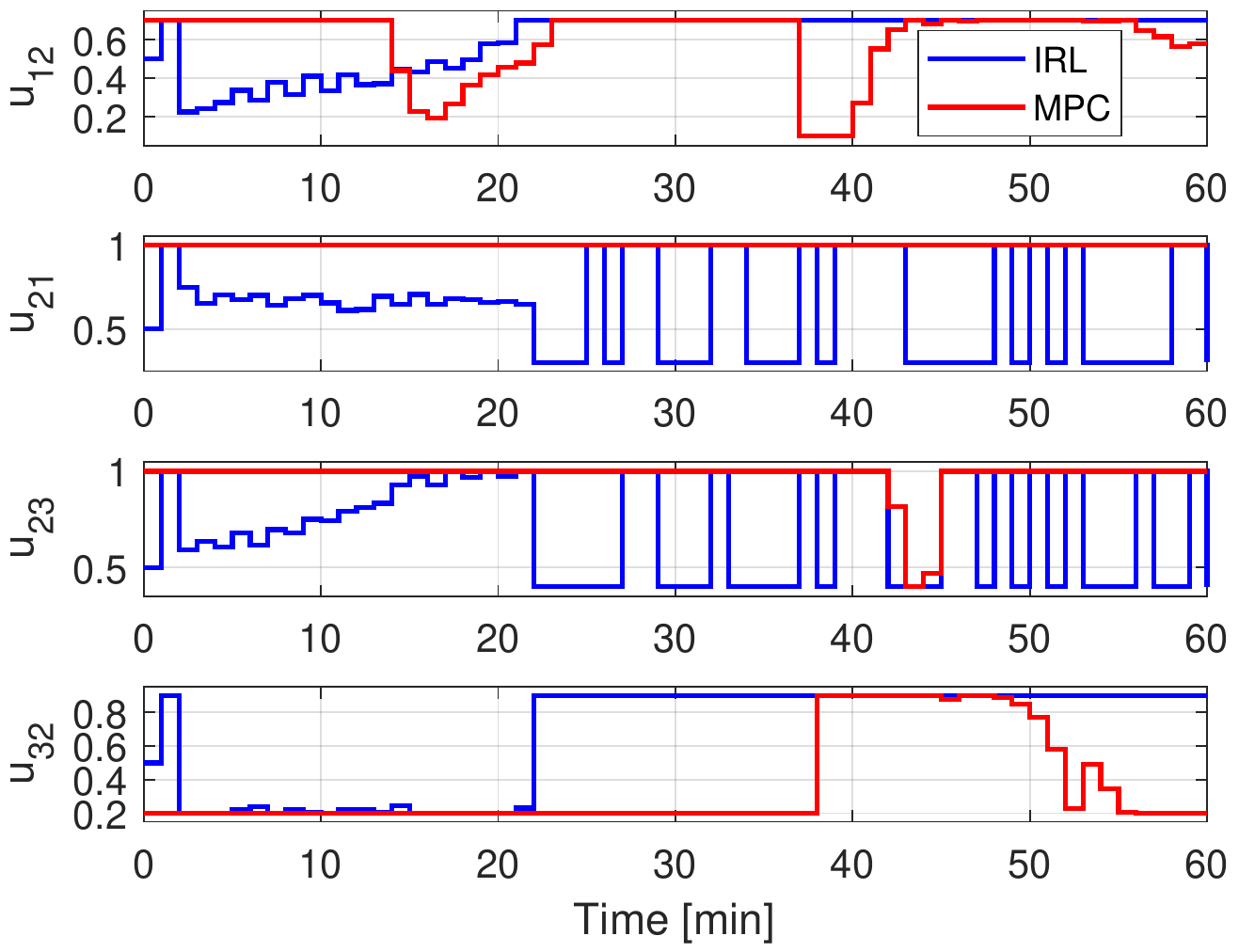}
\end{minipage}
}%
\subfigure[Controls under noisy demand]{
\begin{minipage}[t]{0.33\linewidth}\label{fig:nsctrl}
\centering
\includegraphics[width=2.1in]{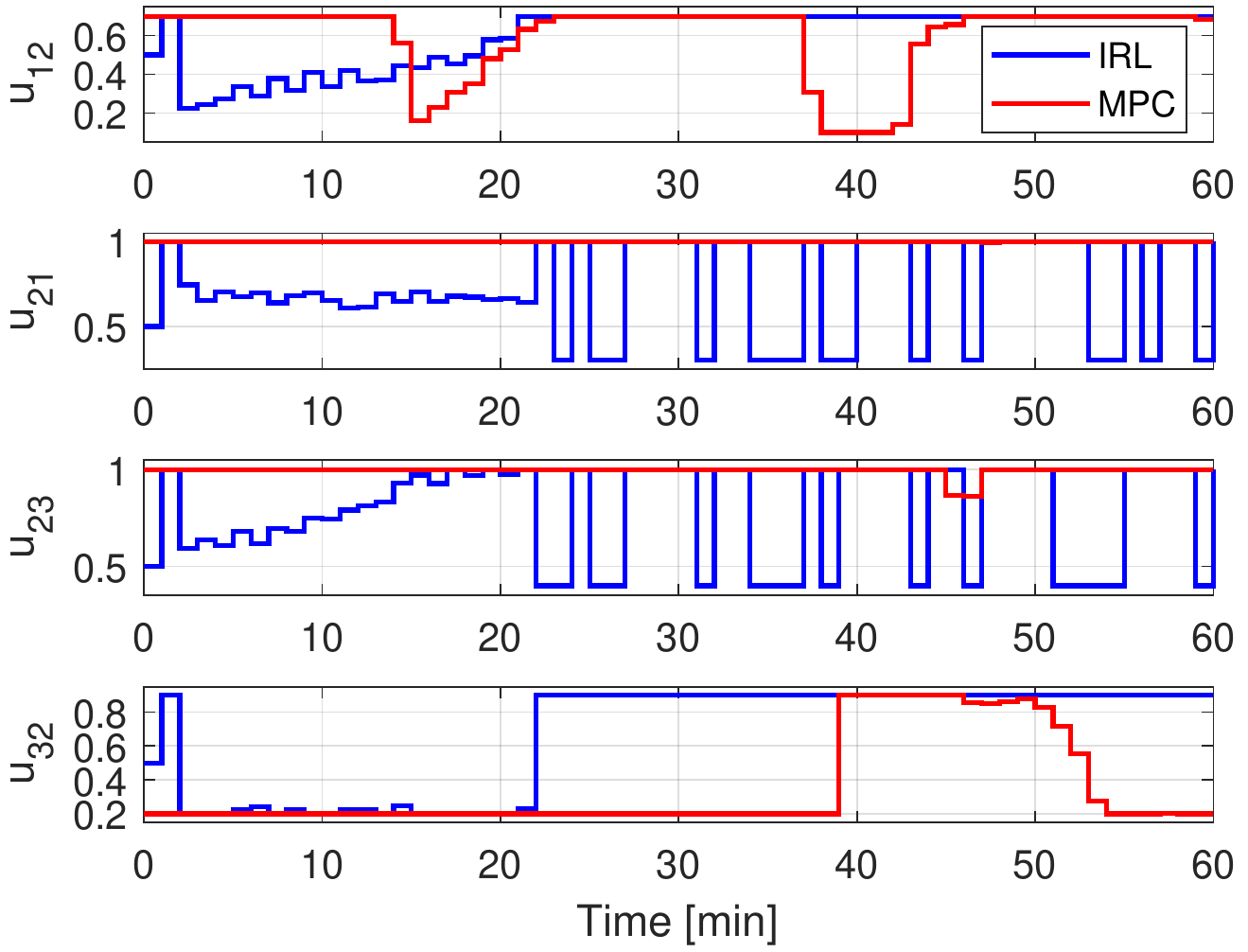}
\end{minipage}
}%
\subfigure[Controls under abrupt-change demand]{
\begin{minipage}[t]{0.33\linewidth}\label{fig:scctrl}
\centering
\includegraphics[width=2.1in]{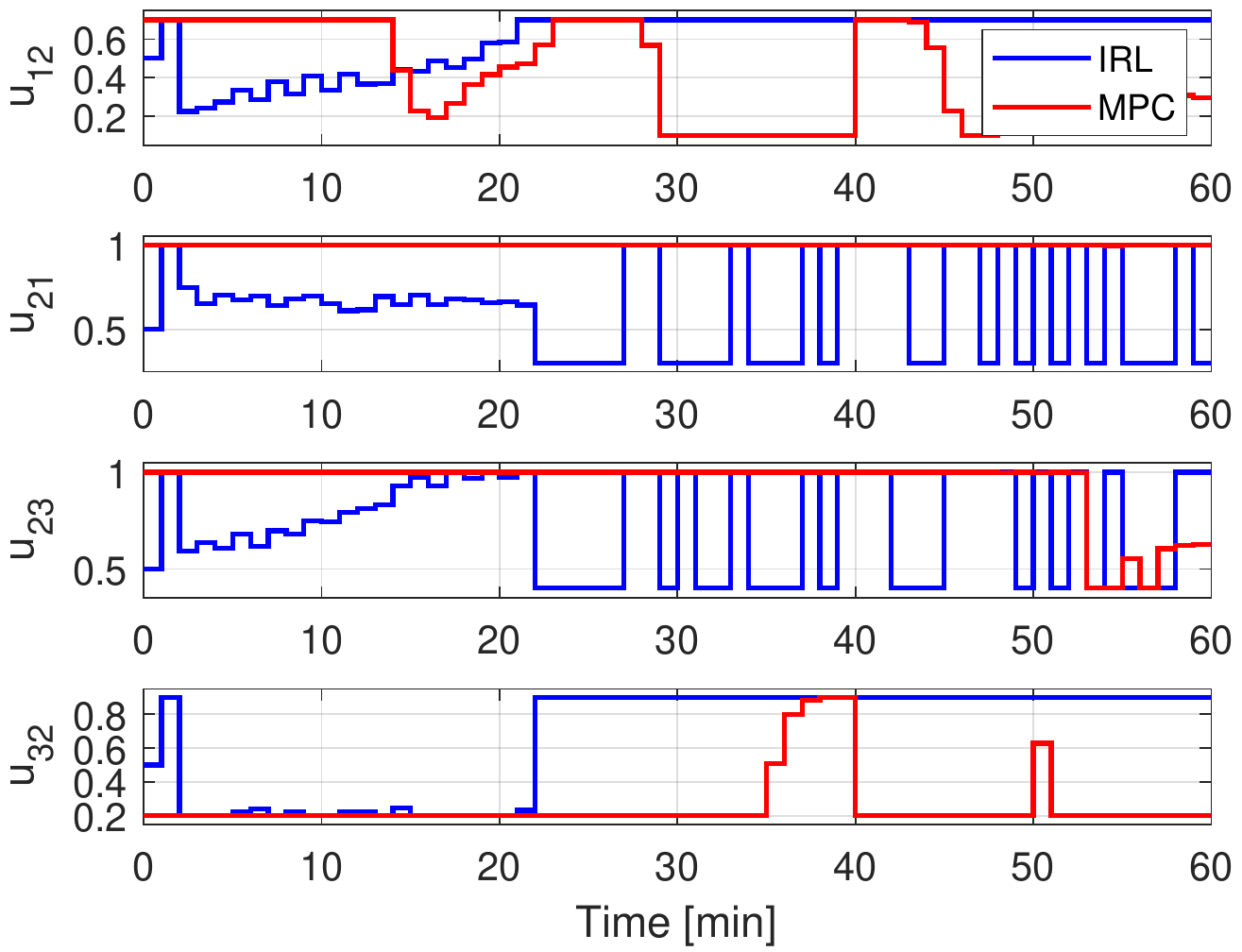}
\end{minipage}
}%
\centering\\
\subfigure[TTS under nominal demand]{
\begin{minipage}[t]{0.33\linewidth}\label{fig:nmTTS}
\centering
\includegraphics[width=2.1in]{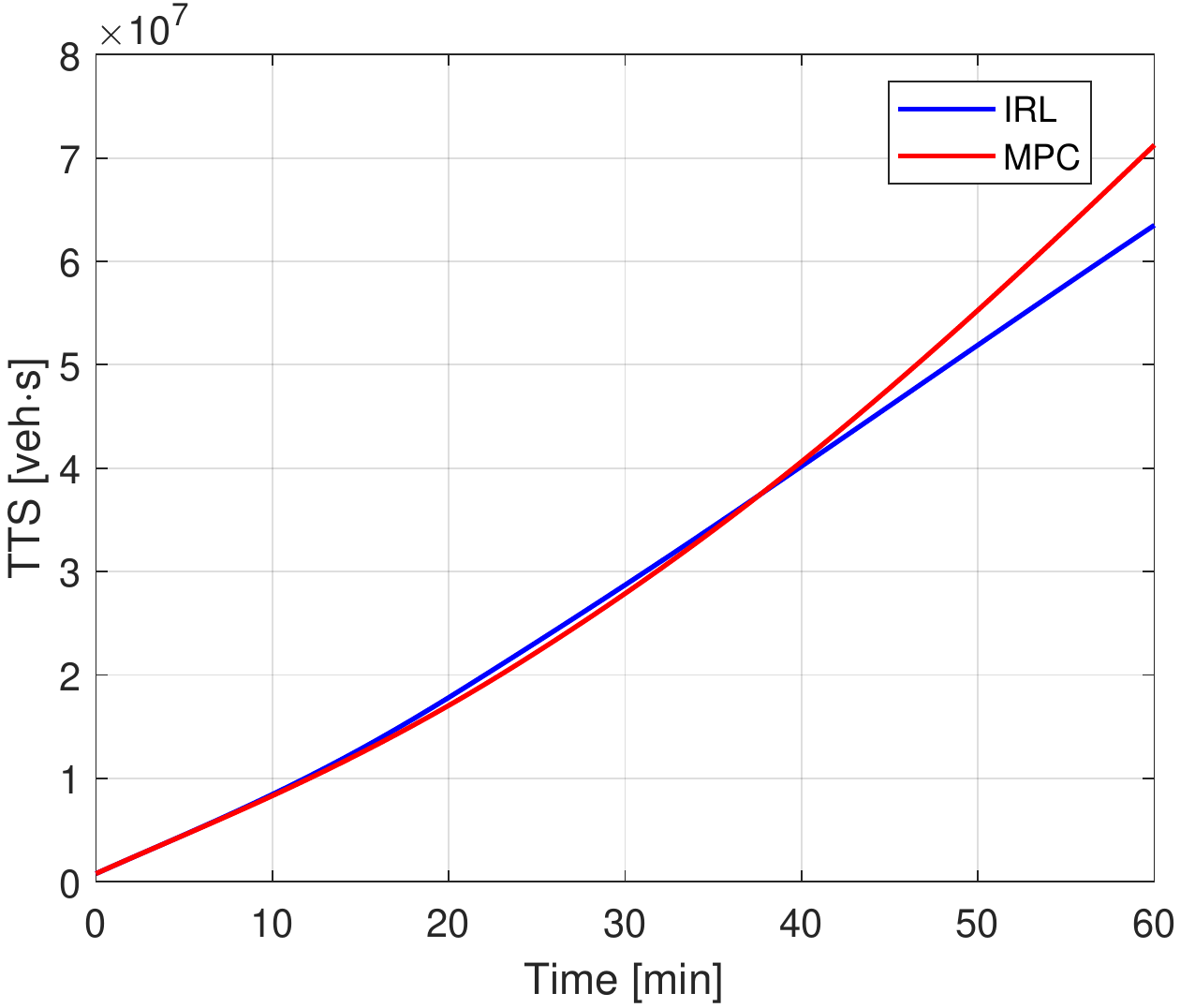}
\end{minipage}
}%
\subfigure[TTS under noisy demand]{
\begin{minipage}[t]{0.33\linewidth}\label{fig:nsTTS}
\centering
\includegraphics[width=2.1in]{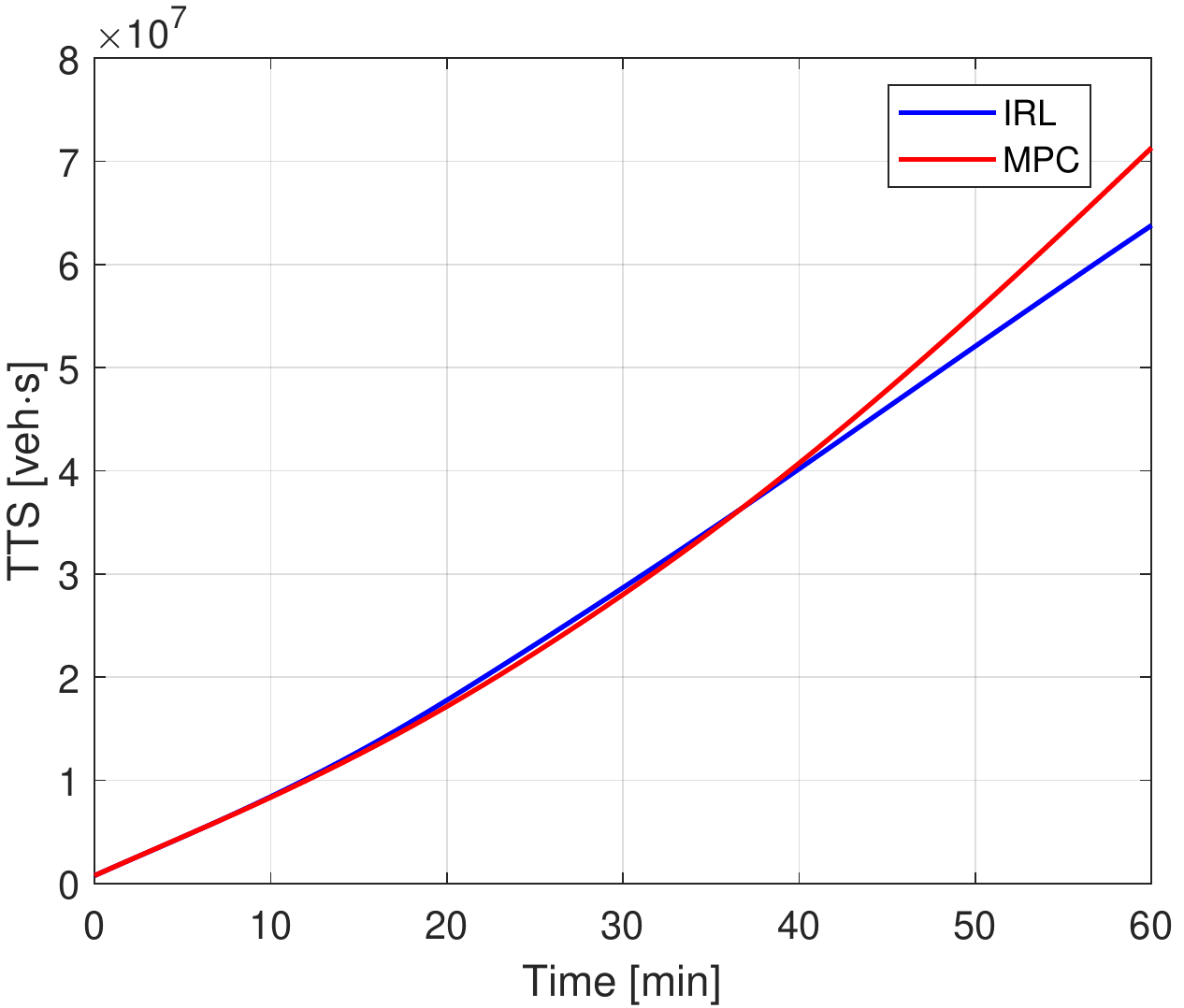}
\end{minipage}
}%
\subfigure[TTS under abrupt-change demand]{
\begin{minipage}[t]{0.33\linewidth}\label{fig:scTTS}
\centering
\includegraphics[width=2.1in]{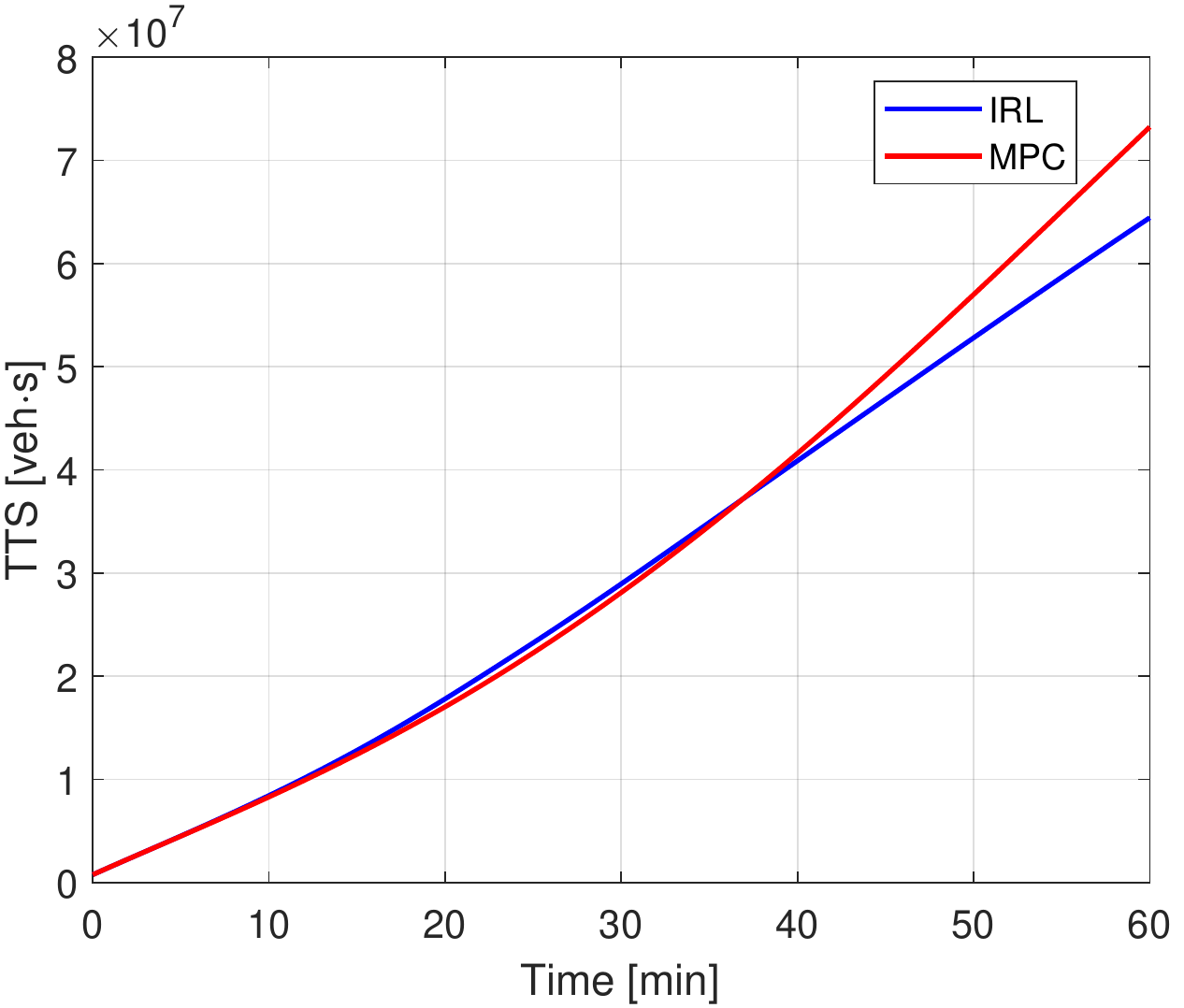}
\end{minipage}
}%
\centering
\caption{Simulation results of Scenario 2}\label{fig:num_TTS}
\end{figure}

In this scenario, a time-varying demand pattern is used to mimic a scenario of peak-hour traffic with congestion onset, stationary congestion and congestion dissolving processes. Comparisons between the IRL approach and MPC are made under three different travel demand patterns, i.e., 1) the nominal deterministic travel demand pattern (\autoref{fig:basedm}), 2) the nominal demand pattern subject to external disturbances (\autoref{fig:noisydm}), and 3) the travel demand pattern subject to an abrupt change during the stationary congestion period (\autoref{fig:shchdm}). The initial accumulation state is set as $n(0)=[5400, 5500, 2000]^T$ (veh). Namely, $n_1$ and $n_2$ are initiated in a very congested state while $n_3$ in an uncongested initial state. For both the IRL and the MPC, the control update interval is set as 60 seconds.

The results are given in \autoref{fig:num_TTS}, where the evolution of regional accumulations (\autoref{fig:nmstate}, \autoref{fig:nsstate} and \autoref{fig:scstate} for the nominal, noisy and abrupt-change demand cases, respectively) are shown alongside graphs of control input evolutions (\autoref{fig:nmctrl}, \autoref{fig:nsctrl} and \autoref{fig:scctrl}) and TTS evolutions (\autoref{fig:nmTTS}, \autoref{fig:nsTTS} and \autoref{fig:scTTS}), for the IRL and the MPC schemes. The achieved TTS and the average CPU time per control update step of different control schemes are summarized in \autoref{compthree}.

\autoref{fig:nmstate}, \autoref{fig:nsstate} and \autoref{fig:scstate} illustrate that in all the demand cases, Region 1 and 2 are congested at the beginning while Region 3 is uncongested, and all the regional accumulation states experience increase during the early stage due to the increase in inflow demands. It is noteworthy that the congestion in Region 3 regulated by the IRL starts to dissipate after 20 minutes, while the accumulation state of Region 3 regulated by MPC continues being increasingly congested.
As observed from \autoref{fig:nmTTS}, \autoref{fig:nsTTS} and \autoref{fig:scTTS}, the IRL is superior to the MPC in minimizing TTS for the whole network. Based on \autoref{compthree}, in the abrupt-change demand case, the IRL achieves a $12\%$ decrease in TTS over MPC, while the same performance metrics are $11\%$ and $10\%$ respectively for the nominal and noisy demand cases.
These results demonstrate that the proposed IRL based control strategy can well learn and adapt to the dynamic nature of the travel demand and hence guarantee the robustness of the traffic dynamics.

\begin{table}[!htb]
    \caption{Summary of TTS and computation time for Scenario 2}\label{compthree}
    \centering
    \begin{threeparttable}
\begin{tabular}{lcccccc}
  \hline
  \multirow{2}*{ }                                            & \multicolumn{2}{c}{nominal demand} & \multicolumn{2}{c}{noisy demand}  & \multicolumn{2}{c}{abrupt-change demand} \\
                                                              \cline{2-7}
                                                              &  IRL     &  MPC           &  IRL        &  MPC       &  IRL      &   MPC        \\
  \hline
  TTS ($\times 10^7$ veh$\cdot$sec)                           &  6.35    &  7.12          &  6.38       &  7.13      &  6.44     &   7.32       \\
  CPU time per step (sec)                                     & $1.07\times 10^{-5}$ & $5.85\times 10^{-1}$ & $1.14\times 10^{-5}$ & $5.93\times 10^{-1}$ & $1.24\times 10^{-5}$ & $5.99\times 10^{-1}$  \\
  \hline
\end{tabular}
\end{threeparttable}
\end{table}

To close the discussion, the numerical results indicate that the proposed IRL based adaptive perimeter controller can not only stabilize the network accumulation states at the desired equilibrium, but also achieve improvement in min TTS compared to the state-of-the-art MPC scheme. These results demonstrate the effectiveness and efficiency of IRL under a variety of data resolutions. In addition, the proposed approach has been examined under various traffic conditions and demand patterns, which implies a promising application of IRL for macroscopic traffic control. The perimeter control in essence is a type of gating control actualized on the boundaries to regulate the cross-boundary traffic flows between different regions. Such kinds of perimeter control are deployed in many metropolises such as Guangzhou and Hong Kong utilizing the existing infrastructure. For example, such perimeter control has been implemented on the cross-Zhujiang-river bridges connecting two busy business districts to manage the peak-hour traffic. Similar perimeter control strategies are also implemented on existing infrastructures in Hong Kong, such as the Hung Hom Cross Harbor Tunnel connecting Hong Kong Island and Kowloon area. For a detailed discussion on the potential applications of the perimeter control, readers may refer to \cite{zhong2018robust,ZHONG2018327}.

\section{Microscopic simulation}\label{SimEx}

To further demonstrate the applicability of the proposed IRL approach to perimeter control of MFD based networks, a microscopic simulation example is presented in this section.
Both training of the proposed IRL algorithm and its performance evaluation are conducted using SUMO as the environment \citep{SUMO2018}. The simulation and calculation are implemented in Python 3.6 and MATLAB R2022a.

\begin{figure}[!htb]
  \centering
  \includegraphics[width=6.4in]{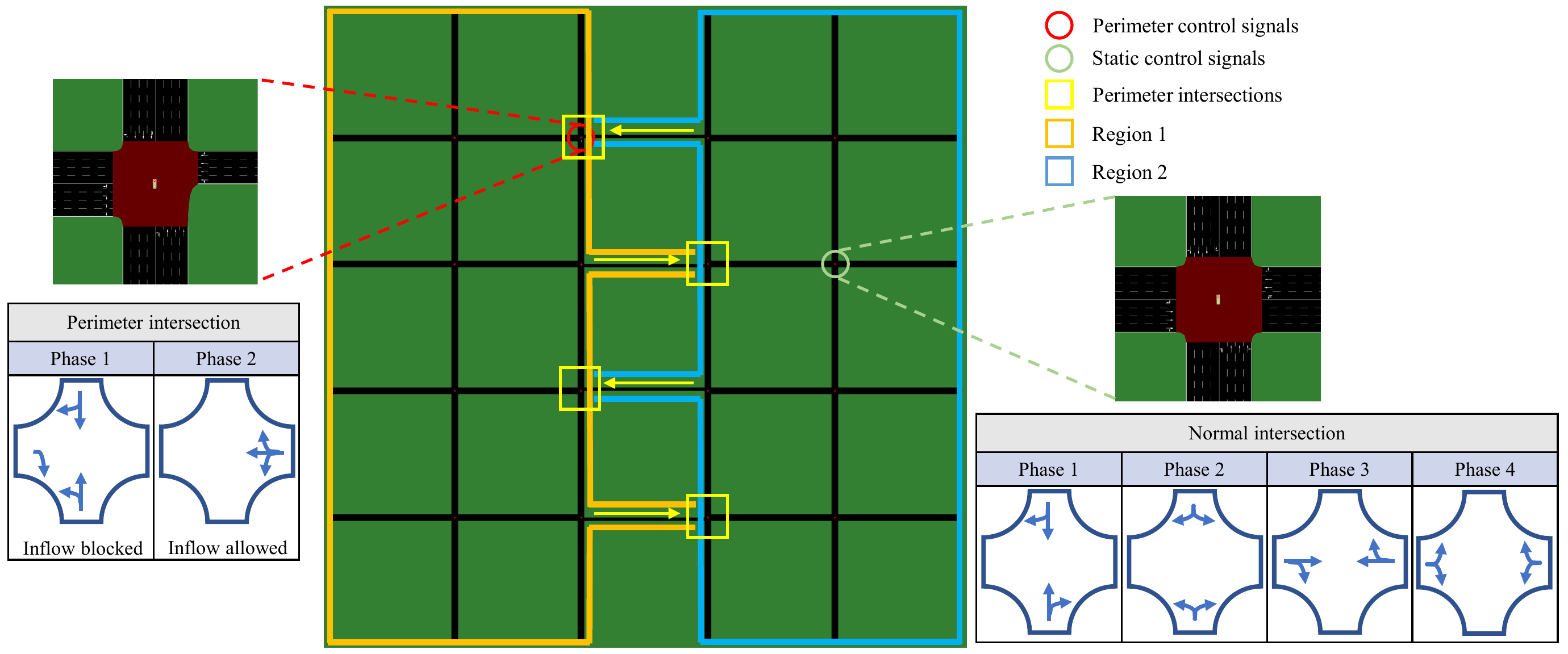}\\
  \caption{The simulated grid network}\label{fig:simtopo}
\end{figure}

The microscopic simulation is carried out using a grid road network as depicted in \autoref{fig:simtopo}, which comprises 2 regions (regions 1 and 2 surrounded by orange lines and blue lines, respectively), 16 signalized intersections (12 normal intersections applying an identical static signal plan and 4 perimeter intersections applying the IRL based signal plans for perimeter control actuation) and 76 links. All links are 500 meters long and comprise 4 lanes. There are 4 special links connecting the two network regions as marked alongside yellow arrows. Note that these 4 links are unidirectional and that their end nodes are signalized intersections working as the perimeter controllers (as marked with yellow rectangles). For the perimeter intersections, a two-phased signal with a 120-second cycle time is adopted (see \autoref{fig:simtopo}). Both the sample time interval and the control update step are equal to the perimeter control signal cycle time. Perimeter control inputs are implemented by changing the green duration ratios of the corresponding perimeter intersections. Let $GR_{ij}(k)$ denotes the green duration ratio of phase 2 at the $k$-th ($k=1,\ldots,90$) control update step for actualizing the computed result of $u_{ij}(k)$ and $GR_{ij}(k)=u_{ij}(k)$. The calculation of the green light duration $\mathcal{G}_{ij}(k)$ of phase 2 is $\mathcal{G}_{ij}(k)=GR_{ij}(k)\cdot CT_{ij}=u_{ij}(k)\cdot CT_{ij}$ where $CT_{ij}=120$ (sec) is the cycle time of the perimeter intersections for $u_{ij}$. For instance, if the computed result of a control input is 0.6, the green light duration is set as 72 seconds for the 120-second cycle. All the normal intersections have four phases and the same cycle time 100 seconds. An identical static signal plan is adopted for the normal intersections. The total simulation time is 3 hours.

\begin{figure}[!htb]
\centering
\subfigure[Demand pattern]{
\begin{minipage}[t]{0.33\linewidth}\label{fig:sim_dm}
\centering
\includegraphics[width=2.1in]{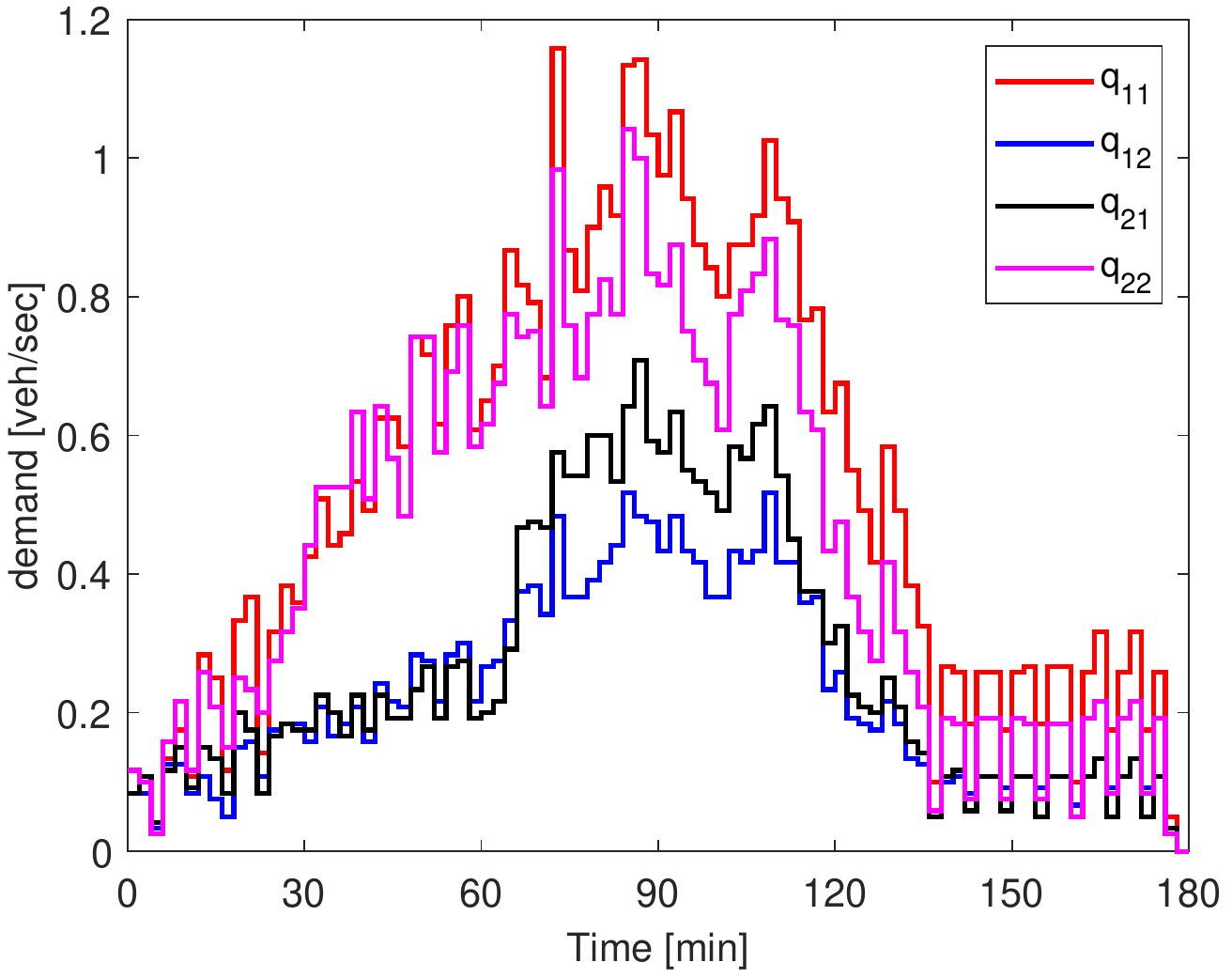}
\end{minipage}
}%
\subfigure[State evolution]{
\begin{minipage}[t]{0.33\linewidth}\label{fig:sim_state}
\centering
\includegraphics[width=2.1in]{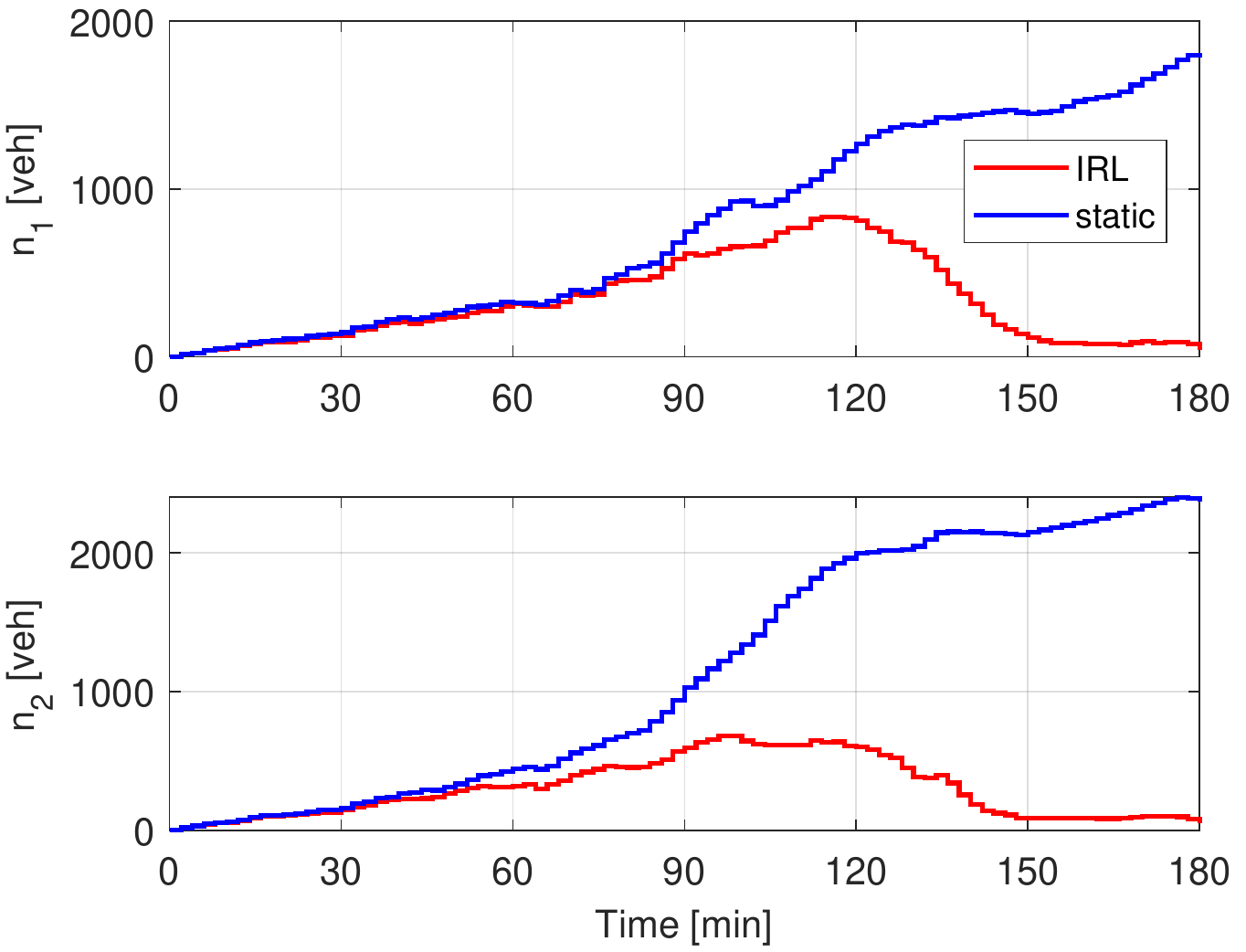}
\end{minipage}
}%
\subfigure[TTS]{
\begin{minipage}[t]{0.33\linewidth}\label{fig:sim_TTS}
\centering
\includegraphics[width=2.1in]{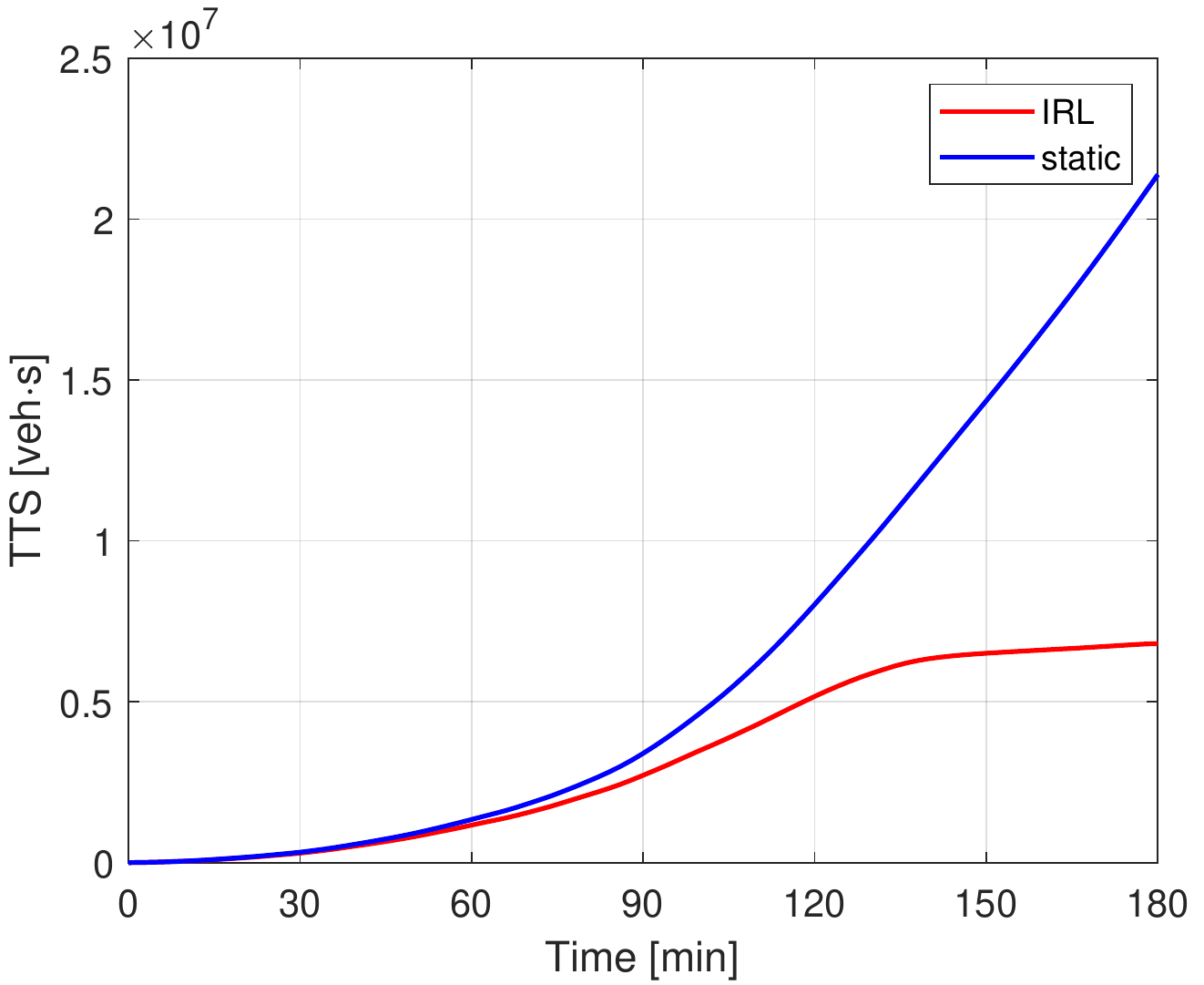}
\end{minipage}
}%
\\
\subfigure[flow-accumulation plots using IRL method]{
\begin{minipage}[t]{0.4\linewidth}\label{fig:sim_mfdirl}
\centering
\includegraphics[width=2.7in]{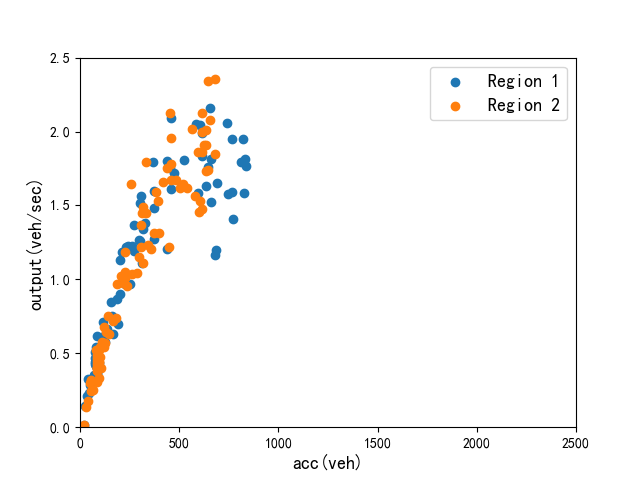}
\end{minipage}
}%
\subfigure[flow-accumulation plots using static scheme]{
\begin{minipage}[t]{0.4\linewidth}\label{fig:sim_mfdsp}
\centering
\includegraphics[width=2.7in]{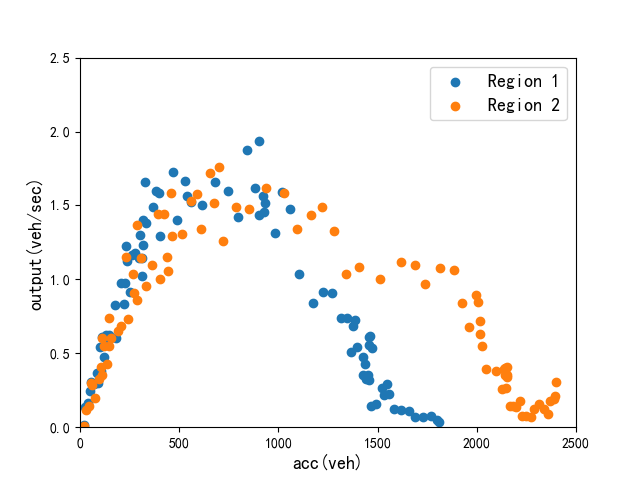}
\end{minipage}
}%
\centering
\caption{The microscopic simulation results}
\end{figure}

The perimeter control objective in this microscopic simulation is to minimize the TTS. Both regions are initially empty at the beginning. A time-varying travel demand pattern associated with a $10\%$ coefficient of variation to represent the stochasticity is adopted, which mimics the morning-peak traffic with congestion onset and dissolving processes (see \autoref{fig:sim_dm}). The accumulation evolutions are depicted in \autoref{fig:sim_state}. As observed from the results, by merely manipulating the green duration ratios of the four perimeter intersections, the IRL scheme can regulate the accumulation states below the critical accumulation and achieve a significant improvement over the static scheme in avoiding congestion. \autoref{fig:sim_TTS} shows that the IRL scheme can guarantee a low TTS at around 6.81$\times 10^6$ (veh$\cdot$sec), while the static scheme results in a very high TTS at around 2.14$\times 10^7$ (veh$\cdot$sec). The flow-accumulation plots by the proposed IRL approach and the static scheme are shown in \autoref{fig:sim_mfdirl} and \autoref{fig:sim_mfdsp}, respectively. One can see that the IRL scheme results in a higher maximal throughput than the static scheme, whereas observed from the simulation process, the static scheme induces severe congestion and even gridlocks in both regions. These microscopic simulation results validate the effectiveness of the proposed IRL method in optimal perimeter control for MFD based traffic networks.

\section{Conclusion} \label{Conclu}

This paper developed a data-driven IRL based framework for learning macroscopic urban traffic dynamics for adaptive constrained optimal perimeter control. An online adaptive optimal perimeter control scheme with continuous-time control and discrete gain updates was established to adapt to the discrete-time nature of traffic data. To further consider the heterogeneous traffic sensors with different resolutions of data measurements, the reinforcement interval of the proposed IRL based perimeter control could be selected online to ensure data richness for the data-driven RL algorithms and allow adaptive online learning to guarantee real-time performance. An actor-critic dual neural network structure was developed to approximate the optimal control and the objective function, respectively. The actor-critic dual neural networks could be used to circumvent the ``curse of dimensionality'' in solving the HJB equations involved. Integrating the experience replay technique, the proposed online learning approach could adapt to the real-time traffic conditions by using the historical and real-time data simultaneously in a ``smart'' manner. This proposed optimal perimeter control did not explicitly employ the knowledge of traffic network dynamics, i.e., ``model-free''.  The convergence of learning algorithms and the stability of the traffic dynamics under control were proven via the Lyapunov theory. The proposed online iterative learning approach was tested under various traffic conditions (e.g., constant demand, time-varying demand with and without uncertainties, unknown MFD model), where the convergence, adaptiveness and robustness of the network traffic state were achieved. Both numerical examples and microscopic simulation experiments were presented to validate the applicability of the proposed method to optimal perimeter control for MFD based networks. In addition, the comparison results indicated that the proposed IRL approach could achieve both good control performance and computational efficiency.

Considering the dynamic nature of travel demand and supply, future efforts can be dedicated to investigating a novel trajectory stability concept, instead of the stability of the desired equilibrium point, to fit such dynamic travel demand and supply.

\vspace{6pt}
\noindent \textbf{Acknowledgement}
\vspace{6pt}

The work in this paper was jointly supported by research grants from the National Natural Science Foundation of China (No. 72071214), the National Key R\&D Program of China (No. 2018YFB1600500) and the Research Grants Council of Hong Kong (Nos. R5029-18 \& 15211518E).

\appendix

\section{Macroscopic traffic dynamics in the affine form}\label{appenA}

This appendix presents the conservation equations and dynamics in the affine form of two cases investigated in the literature, i.e., the two-region and the three-region MFD systems, as shown in \autoref{twotopo} and \autoref{trtopo}, respectively. To begin with, let $M_{ii}(t)=\frac{n_{ii}(t)}{n_i(t)}G_i(n_i(t))$ and $M_{ij}(t)=\frac{n_{ij}(t)}{n_i(t)}G_i(n_i(t))$ denote the within-region flow and cross-boundary flow at time $t$, respectively.

\vspace{6pt}
\noindent\textbf{Case 1: The two-region MFD system}
\vspace{6pt}

Let $L=2$ \citep[i.e., the two-region MFD dynamics as shown by \autoref{twotopo} defined in][]{geroliminis2013optimal}, $n=[n_{11}, n_{12}, n_{21}, n_{22}]^T\in \mathbb{R}^4$ and $u=[u_{12}, u_{21}]^T\in \mathbb{R}^2$. The flow conservation equations are given as
\begin{equation*}
\begin{split}
  \frac{\text{d}n_{11}(t)}{\text{d}t} & =-M_{11}(t)+M_{21}(t)u_{21}(t)+q_{11}(t) \\
  \frac{\text{d}n_{12}(t)}{\text{d}t} & =-M_{12}(t)u_{12}(t)+q_{12}(t)    \\
  \frac{\text{d}n_{21}(t)}{\text{d}t} & =-M_{21}(t)u_{21}(t)+q_{21}(t)    \\
  \frac{\text{d}n_{22}(t)}{\text{d}t} & =-M_{22}(t)+M_{12}(t)u_{12}(t)+q_{22}(t)
\end{split}
\end{equation*}

For this case, the new state and control are $\tilde{n}=[\tilde{n}_1, \tilde{n}_2, \tilde{n}_3, \tilde{n}_4]^T\in \mathbb{R}^4$ and $\tilde{u}=[\tilde{u}_1, \tilde{u}_2]^T\in \mathbb{R}^2$, respectively. The drift dynamics $\mathbf{F}\in \mathbb{R}^4$ and input dynamics $\mathbf{S}\in \mathbb{R}^{4\times 2}$ of their affine-form traffic dynamics are

\begin{equation*}
  \mathbf{F}(\tilde{n})\triangleq \left[
  \begin{array}{c}
      -M_{11} +M_{21}u^*_{21}+q_{11} \\
      -M_{12}u^*_{12}+q_{12} \\
      -M_{21}u^*_{21}+q_{21} \\
      -M_{22} +M_{12}u^*_{12}+q_{22}
  \end{array} \right],\
  \mathbf{S}(\tilde{n})\triangleq \left[
  \begin{array}{cc}
   0 & M_{21} \\
   -M_{12} & 0 \\
   0 & -M_{21} \\
   M_{12} & 0
  \end{array} \right]
\end{equation*}

\vspace{6pt}
\noindent\textbf{Case 2: The three-region MFD system}
\vspace{6pt}

Let $L=3$ \citep[i.e., three-region MFD dynamics as shown by \autoref{trtopo}, see example in][]{ZHONG2018327}, $n=[n_{11}, n_{12}, n_{21}, n_{22}, n_{23}, n_{32}, n_{33}]^T\in \mathbb{R}^7$ and $u=[u_{12}, u_{21}, u_{23}, u_{32}]^T\in \mathbb{R}^4$. The flow conservation equations are given as
\begin{equation*}
   \begin{split}
     \frac{\text{d}n_{11}(t)}{\text{d}t} &=-M_{11}(t)+M_{21}(t)u_{21}(t)+q_{11}(t)   \\
     \frac{\text{d}n_{12}(t)}{\text{d}t} &=-M_{12}(t)u_{12}(t)+q_{12}(t)   \\
     \frac{\text{d}n_{21}(t)}{\text{d}t} &=-M_{21}(t)u_{21}(t)+q_{21}(t) \\
     \frac{\text{d}n_{22}(t)}{\text{d}t} &=-M_{22}(t)+M_{12}(t)u_{12}(t)+M_{32}(t)u_{32}(t)+q_{22}(t) \\
     \frac{\text{d}n_{23}(t)}{\text{d}t} &=-M_{23}(t)u_{23}(t)+q_{23}(t) \\
     \frac{\text{d}n_{32}(t)}{\text{d}t} &=-M_{32}(t)u_{32}(t)+q_{32}(t) \\
     \frac{\text{d}n_{33}(t)}{\text{d}t} &=-M_{33}(t)+M_{23}(t)u_{23}(t)+q_{33}(t)
   \end{split}
\end{equation*}

For this case, $\tilde{n}=[\tilde{n}_1,\ldots,\tilde{n}_7]^T\in \mathbb{R}^7$ and $\tilde{u}=[\tilde{u}_1, \ldots, \tilde{u}_4]^T\in \mathbb{R}^4$. $\mathbf{F}\in \mathbb{R}^7$ and $\mathbf{S}\in \mathbb{R}^{7\times 4}$ of their affine-form traffic dynamics are

\begin{equation*}
  \mathbf{F}(\tilde{n})\triangleq \left[
  \begin{array}{c}
      -M_{11} +M_{21}u^*_{21}+q_{11} \\
      -M_{12}u^*_{12}+q_{12} \\
      -M_{21}u^*_{21}+q_{21} \\
      -M_{22} +M_{12}u^*_{12}+M_{32}u^*_{32}+q_{22} \\
      -M_{23}u^*_{23}+q_{23} \\
      -M_{32}u^*_{32}+q_{32} \\
      -M_{33} +M_{23}u^*_{23}+q_{33}
  \end{array} \right],\
  \mathbf{S}(\tilde{n})\triangleq \left[
  \begin{array}{cccc}
   0 &M_{21} &0 &0 \\
   -M_{12} &0 &0 &0  \\
   0 &-M_{21} &0 &0 \\
   M_{12} &0 &0 &M_{32} \\
   0 &0 &-M_{23} &0 \\
   0 &0 &0 &-M_{32}  \\
   0 &0 &M_{23} &0
  \end{array} \right]
\end{equation*}

\section{Proof of \autoref{lem1}}\label{appenB}

This appendix presents the proof of \autoref{lem1}.

\bproof
1) Assume that $V^*$ is the optimal value function which satisfies \eqref{vbelleq}, then it yields the following HJB equation
\begin{align}\label{HJBmin}
  H\left(\tilde{n},\tilde{u},\frac{\partial V^*}{\partial \tilde{n}}\right) & = \min_{\tilde{u}} \left[ \mathcal{L}(\tilde{n},\tilde{u}) + \left(\frac{\partial V^*}{\partial \tilde{n}}\right)^T(\mathbf{F}(\tilde{n})+\mathbf{S}(\tilde{n})\tilde{u}) \right] \nonumber \\
  & = \min_{\tilde{u}} \left[ \tilde{n}^TQ\tilde{n} + 2\underline{v}^TR\int^{\tilde{u}}_{\overline{v}}\tanh^{-1}\left(\frac{1}{\underline{v}}\odot (v-\overline{v})\right)\mathrm{d}v + \left(\frac{\partial V^*}{\partial \tilde{n}}\right)^T(\mathbf{F}(\tilde{n})+\mathbf{S}(\tilde{n})\tilde{u}) \right] =0 \nonumber \\
  & = \min_{\tilde{u}} \left[ \tilde{n}^TQ\tilde{n} + \sum^{\alpha_{u}}_{k_u=1} 2\underline{v}_{k_u}\gamma_{k_u}\int^{\tilde{u}_{k_u}}_{\overline{v}_{k_u}}\tanh^{-1}\left(\frac{v_{k_u}-\overline{v}_{k_u}}{\underline{v}_{k_u}}\right) \mathrm{d}v_{k_u} + \left(\frac{\partial V^*}{\partial \tilde{n}}\right)^T\mathbf{F}(\tilde{n}) + \sum^{\alpha_u}_{k_u=1}\sum^{\alpha_n}_{k_n=1} \frac{\partial V^*}{\partial\tilde{n}_{k_n}}S_{k_n,k_u}\tilde{u}_{k_u}\right]
\end{align}
where $S_{k_n,k_u}$ denotes the $k_n$-th element of the $k_u$-th column of $\mathbf{S}$.

The optimal constrained perimeter control $\tilde{u}^*_{k_u}$ is calculated by applying the stationary (optimal) condition $\partial H/\partial \tilde{u}^*_{k_u}=0$, i.e.,
\begin{equation*}
  \frac{\partial H}{\partial \tilde{u}^*_{k_u}} = 2\underline{v}_{k_u}\gamma_{k_u}\tanh^{-1}\left(\frac{\tilde{u}^*_{k_u}-\overline{v}_{k_u}}{\underline{v}_{k_u}}\right) + \sum^{\alpha_n}_{k_n=1}\frac{\partial V^*}{\partial \tilde{n}_{k_n}}S_{k_n,k_u} = 0
\end{equation*}
Then it follows that
\begin{eqnarray*}
    &           & \tanh^{-1}\left(\frac{\tilde{u}^*_{k_u}-\overline{v}_{k_u}}{\underline{v}_{k_u}}\right) = -\frac{1}{2\underline{v}_{k_u}\gamma_{k_u}}\sum^{\alpha_n}_{k_n=1}\frac{\partial V^*}{\partial \tilde{n}_{k_n}}S_{k_n,k_u} \nonumber \\
    &\Rightarrow& \frac{\tilde{u}^*_{k_u}-\overline{v}_{k_u}}{\underline{v}_{k_u}} = \tanh\left(-\frac{1}{2\underline{v}_{k_u}\gamma_{k_u}}\sum^{\alpha_n}_{k_n=1}\frac{\partial V^*}{\partial \tilde{n}_{k_n}}S_{k_n,k_u}\right) = -\tanh\left(\frac{1}{2\underline{v}_{k_u}\gamma_{k_u}}\sum^{\alpha_n}_{k_n=1}\frac{\partial V^*}{\partial \tilde{n}_{k_n}}S_{k_n,k_u}\right) \nonumber \\
    &\Rightarrow& \tilde{u}^*_{k_u} = -\underline{v}_{k_u}\tanh\left(\frac{1}{2\underline{v}_{k_u}\gamma_{k_u}}\sum^{\alpha_n}_{k_n=1}\frac{\partial V^*}{\partial \tilde{n}_{k_n}}S_{k_n,k_u}\right)+\overline{v}_{k_u}
\end{eqnarray*}

Let $D^*_{k_u}=\frac{1}{2\underline{v}_{k_u}\gamma_{k_u}}\sum^{\alpha_n}_{k_n=1}\frac{\partial V^*}{\partial \tilde{n}_{k_n}}S_{k_n,k_u}$ be the $k_u$-th unconstrained optimal control input. The optimal control $\tilde{u}^*_{k_u}$ is obtained as
\begin{equation}\label{optipt}
  \tilde{u}^*_{k_u} = -\underline{v}_{k_u}\tanh(D^*_{k_u})+\overline{v}_{k_u}
\end{equation}

Therefore, the constrained optimal perimeter control is given by
\begin{equation}\label{optiptvec}
  \tilde{u}^*=-\underline{v}\odot\tanh(D^*)+\overline{v},\ \mathrm{with}\ D^*=\frac{1}{2\underline{v}}\odot\left(R^{-1}\mathbf{S}^T\frac{\partial V^*}{\partial \tilde{n}}\right)
\end{equation}

2) Let $\hat{v}=\frac{1}{\underline{v}}\odot (v-\overline{v})$. Substituting \eqref{optiptvec} into \eqref{conipt}, we have
\begin{align}\label{nconipt}
  U(\tilde{u}^*) & = 2\underline{v}^T\odot \underline{v}^TR\int^{\frac{1}{\underline{v}}\odot (\tilde{u}^*-\overline{v})}_{0}\tanh^{-1}(\hat{v})\mathrm{d}\hat{v} = 2\underline{v}^{2T}R\int^{-\tanh(D^*)}_{0}\tanh^{-1}(\hat{v})\mathrm{d}\hat{v} \nonumber \\
                 & = 2\underline{v}^{2T}R\cdot \left(\hat{v}\odot\tanh^{-1}(\hat{v})+\frac{1}{2}\ln(\mathbf{1}_{\alpha_u}-\hat{v}^2)\right)\Big|^{-\tanh(D^*)}_{0} = 2\underline{v}^{2T}R \left(\tanh(D^*)\odot D^*+\frac{1}{2}\ln(\mathbf{1}_{\alpha_u}-\tanh^2(D^*))\right) \nonumber \\
                 & = 2\underline{v}^{2T}R(D^*\odot\tanh(D^*)) + \underline{v}^{2T}R\ln(\mathbf{1}_{\alpha_u}-\tanh^2(D^*)) \nonumber \\
                 & = 2\underline{v}^{2T}R \left(\frac{1}{2\underline{v}}\odot\left(R^{-1}\mathbf{S}^T\frac{\partial V^*}{\partial \tilde{n}}\right)\odot\tanh(D^*)\right) + \underline{v}^{2T}R\ln(\mathbf{1}_{\alpha_u}-\tanh^2(D^*)) \nonumber \\
                 & = \left(\frac{\partial V^*}{\partial \tilde{n}}\right)^T\mathbf{S}(\tilde{n})(\underline{v}\odot\tanh(D^*)) + \underline{v}^{2T}R\ln(\mathbf{1}_{\alpha_u}-\tanh^2(D^*))
\end{align}

Then substituting \eqref{optiptvec}-\eqref{nconipt} into \eqref{HJBmin}, the HJB equation can further be expressed by
\begin{align}
0 & = \tilde{n}^TQ\tilde{n} + \left(\frac{\partial V^*}{\partial \tilde{n}}\right)^T(\mathbf{F}+\mathbf{S}\tilde{u}^*) + \left(\frac{\partial V^*}{\partial \tilde{n}}\right)^T\mathbf{S}(\underline{v}\odot\tanh(D^*)) + \underline{v}^{2T}R\ln(\mathbf{1}_{\alpha_u}-\tanh^2(D^*)) \nonumber \\
& = \tilde{n}^TQ\tilde{n} + \left(\frac{\partial V^*}{\partial \tilde{n}}\right)^T\mathbf{F} + \left(\frac{\partial V^*}{\partial \tilde{n}}\right)^T\mathbf{S}\tilde{u}^* + \left(\frac{\partial V^*}{\partial \tilde{n}}\right)^T\mathbf{S}\cdot(\underline{v}\odot\tanh(D^*)) + \underline{v}^{2T}R\ln(\mathbf{1}_{\alpha_u}-\tanh^2(D^*)) \nonumber \\
& = \tilde{n}^TQ\tilde{n} + \left(\frac{\partial V^*}{\partial \tilde{n}}\right)^T\mathbf{F} + \left(\frac{\partial V^*}{\partial \tilde{n}}\right)^T\mathbf{S}\cdot(-\underline{v}\odot\tanh(D^*)+\overline{v}) + \left(\frac{\partial V^*}{\partial \tilde{n}}\right)^T\mathbf{S}\cdot(\underline{v}\odot\tanh(D^*)) + \underline{v}^{2T}R\ln(\mathbf{1}_{\alpha_u}-\tanh^2(D^*)) \nonumber \\
& = \tilde{n}^TQ\tilde{n} + \left(\frac{\partial V^*}{\partial \tilde{n}}\right)^T\mathbf{F} + \left(\frac{\partial V^*}{\partial \tilde{n}}\right)^T\mathbf{S}\overline{v} + \underline{v}^{2T}R\ln(\mathbf{1}_{\alpha_u}-\tanh^2(D^*)) \label{sUpHJB}
\end{align}
That is, if $(V^*,D^*)$ is the solution to the COPCP, $(V^*,D^*)$ should satisfy \eqref{sUpHJB}. This completes the proof.
\eproof

\section{Proof of \autoref{thm1}}\label{appenC}

This appendix presents the proof of \autoref{thm1}. To prove this lemma, we need the following \autoref{lemma}.

\baplemma \label{lemma}
For a monotonically increasing odd function $\varrho(x)$, we have
\begin{enumerate}[1)]
  \item $\varrho(x)\cdot (y-x)-\int^{y}_{x}\varrho(s) \textrm{d}s\leq 0, \forall x,y$;
  \item $\varrho(x)\cdot (y-x)-\int^{y}_{0}\varrho(s) \textrm{d}s\leq 0, \forall x,y$.
\end{enumerate}
\eaplemma

\bproof
1) Without loss of generality, we assume that $y\geq x$.

Note that $\varrho(x)$ is monotonically increasing and odd. Thus,
\begin{align*}
  \varrho(x)\left\{
             \begin{array}{lr}
             <0, & x<0 \\
             =0, & x=0 \\
             >0, & x>0
             \end{array}
  \right.,\
  \varrho(-x)=-\varrho(x)
\end{align*}
Then we have $\int^{y}_{x}\varrho(s) \textrm{d}s\geq 0$. Moreover, $\int^{y}_{x}\varrho(s) \textrm{d}s=0$ if and only if $y=x$.

Let $\varphi(y)=\varrho(x)\cdot (y-x)-\int^{y}_{x}\varrho(s) \textrm{d}s$, then
\begin{align*}
  \frac{\textrm{d}\varphi}{\textrm{d}y}=\varrho(x)-\varrho(y)\left\{
             \begin{array}{lr}
             >0, & y<x \\
             =0, & y=x \\
             <0, & y>x
             \end{array}
  \right.
\end{align*}
The extreme value of $\varphi(y)$ is obtained at $y=x$, i.e., $\varphi(x)=0-\int^{x}_{x}\varrho(s) \textrm{d}s\leq 0$. $\varphi(y)=0$ if and only if $y=0$. Thus, we have
\begin{align*}
  \varrho(x)\cdot (y-x)-\int^{y}_{x}\varrho(s) \textrm{d}s\leq 0
\end{align*}
and $\varrho(x)\cdot (y-x)-\int^{y}_{x}\varrho(s) \textrm{d}s=0$ if and only if $y=x$.

2) The left side of the second part of \autoref{lemma} can be written as
\begin{equation}\label{relemp2}
  \varrho(x)\cdot (y-x)-\int^{y}_{0}\varrho(s) \textrm{d}s = \varrho(x)\cdot (y-0)-\int^{y}_{0}\varrho(s) \textrm{d}s - \varrho(x)x
\end{equation}
Based on the first part, we have $\varrho(x)\cdot (y-0)-\int^{y}_{0}\varrho(s) \textrm{d}s\leq 0$. Because $\varrho(x)$ and $x$ are monotonically increasing and odd, one has $\varrho(x)x\geq 0$, i.e., $-\varrho(x)x\leq 0$. Thus, we have
\begin{equation*}
  \varrho(x)\cdot (y-x)-\int^{y}_{0}\varrho(s) \textrm{d}s\leq 0
\end{equation*}
Moreover, $\varrho(x)\cdot (y-x)-\int^{y}_{0}\varrho(s) \textrm{d}s = 0$ if and only if $y=x=0$. This completes the proof.
\eproof

Now we present the proof of \autoref{thm1}.

\bproof
1) Taking the derivative of $V^{k+1}$ along the system $\mathbf{F}+\mathbf{S}\tilde{u}^{k+1}$ trajectory, we have
\begin{equation}\label{eq10}
  \dot{V}^{k+1}= \left(\frac{\partial V^{k+1}}{\partial \tilde{n}}\right)^{T} \mathbf{F}+ \left(\frac{\partial V^{k+1}}{\partial \tilde{n}}\right)^{T} \mathbf{S} \tilde{u}^{k+1}
\end{equation}
Based on \eqref{valfun} and \eqref{eq8}, we have
\begin{align}\label{eq11}
    & N(\tilde{n}) + 2\underline{v}^TR\int^{\tilde{u}^k}_{\overline{v}}\tanh^{-1}\left(\frac{1}{\underline{v}}\odot(v-\overline{v})\right)\mathrm{d}v + \left(\frac{\partial V^{k+1}}{\partial \tilde{n}}\right)^T\mathbf{F} + \left(\frac{\partial V^{k+1}}{\partial \tilde{n}}\right)^T\mathbf{S}\tilde{u}^{k} =0 \nonumber \\
  \Rightarrow & \left(\frac{\partial V^{k+1}}{\partial \tilde{n}}\right)^T\mathbf{F} = -\left(\frac{\partial V^{k+1}}{\partial \tilde{n}}\right)^T\mathbf{S}\tilde{u}^{k} - N(\tilde{n}) - 2\underline{v}^TR\int^{\tilde{u}^k}_{\overline{v}}\tanh^{-1}\left(\frac{1}{\underline{v}}\odot(v-\overline{v})\right)\mathrm{d}v
\end{align}
From \eqref{eq9}, one has
\begin{align}\label{eq11_2}
    & \tilde{u}^{k+1}-\overline{v} = -\underline{v}\odot\tanh\left(\frac{1}{2\underline{v}}\odot\left(R^{-1}\mathbf{S}^T\frac{\partial V^{k+1}}{\partial \tilde{n}}\right)\right) \nonumber \\
\Rightarrow & \frac{1}{\underline{v}}\odot(\tilde{u}^{k+1}-\overline{v}) = -\tanh\left(\frac{1}{2\underline{v}}\odot\left(R^{-1}\mathbf{S}^T\frac{\partial V^{k+1}}{\partial \tilde{n}}\right)\right) \nonumber \\
\Rightarrow & \tanh^{-1}\left(\frac{1}{\underline{v}}\odot(\tilde{u}^{k+1}-\overline{v})\right) = -\frac{1}{2\underline{v}}\odot\left(R^{-1}\mathbf{S}^T\frac{\partial V^{k+1}}{\partial \tilde{n}}\right) \nonumber \\
\Rightarrow & -2\underline{v}\odot\tanh^{-1}\left(\frac{1}{\underline{v}}\odot(\tilde{u}^{k+1}-\overline{v})\right) = R^{-1}\mathbf{S}^T\frac{\partial V^{k+1}}{\partial \tilde{n}} \nonumber \\
\Rightarrow & \left(\frac{\partial V^{k+1}}{\partial \tilde{n}}\right)^T\mathbf{S} = -2\underline{v}^T\odot\tanh^{-T}\left(\frac{1}{\underline{v}}\odot(\tilde{u}^{k+1}-\overline{v})\right)R
\end{align}
Substitute \eqref{eq11}-\eqref{eq11_2} into \eqref{eq10}, it follows that
\begin{align}\label{tsfdotVk}
   \dot{V}^{k+1} & = -N(\tilde{n}) - 2\underline{v}^TR\int^{\tilde{u}^{k}}_{\overline{v}}\tanh^{-1}\left(\frac{1}{\underline{v}}\odot(v-\overline{v})\right)\textrm{d}v - \left(\frac{\partial V^{k+1}}{\partial \tilde{n}}\right)^{T} \mathbf{S} \tilde{u}^{k} -2\underline{v}^T\odot\tanh^{-T}\left(\frac{1}{\underline{v}}\odot(\tilde{u}^{k+1}-\overline{v})\right)R \tilde{u}^{k+1} \nonumber\\
     & = -N(\tilde{n}) - 2\underline{v}^TR\int^{\tilde{u}^{k}}_{\overline{v}}\tanh^{-1}\left(\frac{1}{\underline{v}}\odot(v-\overline{v})\right)\textrm{d}v + 2\underline{v}^T\odot\tanh^{-T}\left(\frac{1}{\underline{v}}\odot(\tilde{u}^{k+1}-\overline{v})\right)R \tilde{u}^{k}
     -2\underline{v}^T\odot\tanh^{-T}\left(\frac{1}{\underline{v}}\odot(\tilde{u}^{k+1}-\overline{v})\right)R \tilde{u}^{k+1} \nonumber \\
     & = -N(\tilde{n}) - 2\left(\underline{v}^TR\int^{\tilde{u}^{k}}_{\overline{v}}\tanh^{-1}\left(\frac{1}{\underline{v}}\odot(v-\overline{v})\right)\textrm{d}v - \underline{v}^T\odot\tanh^{-T}\left(\frac{1}{\underline{v}}\odot(\tilde{u}^{k+1}-\overline{v})\right)R \tilde{u}^{k} \right.
     +\underline{v}^T\odot\tanh^{-T}\left(\frac{1}{\underline{v}}\odot(\tilde{u}^{k+1}-\overline{v})\right)R \tilde{u}^{k+1}\Bigg) \nonumber \\
     & = -N(\tilde{n}) + 2\left(\underline{v}^T\odot\tanh^{-T}\left(\frac{1}{\underline{v}}\odot(\tilde{u}^{k+1}-\overline{v})\right)R (\tilde{u}^{k}- \tilde{u}^{k+1}) - \underline{v}^TR\int^{\tilde{u}^{k}}_{\overline{v}}\tanh^{-1}\left(\frac{1}{\underline{v}}\odot(v-\overline{v})\right)\textrm{d}v\right) \nonumber \\
     & = -N(\tilde{n}) + 2\left(\sum^{\alpha_u}_{k_u=1}\underline{v}_{k_u}\gamma_{k_u}\tanh^{-1}\left(\frac{\tilde{u}^{k+1}_{k_u}-\overline{v}_{k_u}}{\underline{v}_{k_u}}\right) (\tilde{u}^{k}_{k_u}-\tilde{u}^{k+1}_{k_u}) - \sum^{\alpha_u}_{k_u=1}\underline{v}_{k_u}\gamma_{k_u} \int^{\tilde{u}^k_{k_u}}_{\overline{v}_{k_u}}\tanh^{-1}\left(\frac{v_{k_u}-\overline{v}_{k_u}}{\underline{v}_{k_u}}\right)\mathrm{d}v_{k_u} \right) \nonumber \\
     & = -N(\tilde{n}) + 2\sum^{\alpha_u}_{k_u=1}\underline{v}_{k_u}\gamma_{k_u}\left( \tanh^{-1}\left(\frac{\tilde{u}^{k+1}_{k_u}-\overline{v}_{k_u}}{\underline{v}_{k_u}}\right) (\tilde{u}^{k}_{k_u}-\tilde{u}^{k+1}_{k_u}) - \int^{\tilde{u}^k_{k_u}}_{\overline{v}_{k_u}}\tanh^{-1}\left(\frac{v_{k_u}-\overline{v}_{k_u}}{\underline{v}_{k_u}}\right)\mathrm{d}v_{k_u} \right)
\end{align}

Let $\varrho(x)\triangleq\tanh^{-1}(x/\underline{v}_{k_u})$ for $\forall x\in \mathbb{R}$, $s^k_{k_u}=\tilde{u}^k_{k_u}-\overline{v}_{k_u}$ and $\hat{v}_{k_u}=v_{k_u}-\overline{v}_{k_u}$. Then \eqref{tsfdotVk} is rewritten as follows
\begin{equation*}
  \dot{V}^{k+1} = -N(\tilde{n}) + 2\sum^{\alpha_{u}}_{k_u =1}\underline{v}_{k_u}\gamma_{k_u} \left(\varrho(s^{k+1}_{k_u}) (s^{k}_{k_u}-s^{k+1}_{k_u})-\int^{s^{k}_{k_u}}_{0}\varrho(\hat{v}_{k_u}) \textrm{d}\hat{v}_{k_u}\right)
\end{equation*}

Since $\tanh^{-1}(\cdot)$ is a monotonically increasing odd function, $\varrho(x)$ is monotonically increasing and odd. By \autoref{lemma}, the following inequality holds
\begin{equation*}
  \varrho(s^{k+1}_{k_u}) (s^{k}_{k_u}-s^{k+1}_{k_u})-\int^{s^{k}_{k_u}}_{0}\varrho(\hat{v}_{k_u}) \textrm{d}\hat{v}_{k_u} \leq 0
\end{equation*}
Recall that $\underline{v}>0$ and $\gamma>0$, we have $\dot{V}^{k+1}\leq 0$ and $V^{k+1}(\tilde{n})$ is a Lyapunov function for $\tilde{u}^{k+1}$ on $\Omega$.

Because the nonlinear function $\mathbf{S}$ is continuous and $V^{k+1}(0)=0$, $\tilde{u}^{k+1}(\tilde{n})\in C^1(\Omega)$ and $\tilde{u}^{k+1}(0)=0$, i.e., the obtained control policies $\tilde{u}^{k+1}(\tilde{n})$ in \eqref{eq9} are admissible as per \autoref{defAC} for system \eqref{eqmulti} on $\Omega$.

2) First, we prove that $V^{k+2}(\tilde{n}(t))\leq V^{k+1}(\tilde{n}(t))$.

Considering $V(\tilde{n})$ defined in \eqref{valfun} along the system $\mathbf{F}+\mathbf{S}\tilde{u}^{k+1}$ trajectory, we have
\begin{equation}\label{delVkp}
  V^{k+2}(\tilde{n})-V^{k+1}(\tilde{n}) = -\int^{\infty}_{t} \left(\frac{\partial (V^{k+2}-V^{k+1})^T}{\partial \tilde{n}}(\mathbf{F}+\mathbf{S}\tilde{u}^{k+1})\right) \textrm{d}\tau
\end{equation}
Since $(V^{k+1}(\tilde{n}), \tilde{u}^{k}(\tilde{n}))$ and $(V^{k+2}(\tilde{n}), \tilde{u}^{k+1}(\tilde{n}))$ both satisfy \eqref{vbelleq}, we can obtain
\begin{subequations}
\begin{align}
  \left(\frac{\partial V^{k+1}}{\partial \tilde{n}}\right)^T\mathbf{F} & = -N(\tilde{n}) - 2\underline{v}^TR\int^{\tilde{u}^{k}}_{\overline{v}}\tanh^{-1}\left(\frac{1}{\underline{v}}\odot (v-\overline{v})\right)\textrm{d}v - \left(\frac{\partial V^{k+1}}{\partial \tilde{n}}\right)^{T} \mathbf{S} \tilde{u}^{k} \label{VkpnF} \\
  \left(\frac{\partial V^{k+2}}{\partial \tilde{n}}\right)^T\mathbf{F} & = -N(\tilde{n}) - 2\underline{v}^TR\int^{\tilde{u}^{k+1}}_{\overline{v}}\tanh^{-1}\left(\frac{1}{\underline{v}}\odot (v-\overline{v})\right)\textrm{d}v - \left(\frac{\partial V^{k+2}}{\partial \tilde{n}}\right)^{T} \mathbf{S} \tilde{u}^{k+1} \label{VkppnF}
\end{align}
\end{subequations}
Substituting \eqref{VkpnF}-\eqref{VkppnF} into \eqref{delVkp}, we get
\begin{align}\label{NdelVkp}
  V^{k+2}(\tilde{n})-V^{k+1}(\tilde{n}) & = -\int^{\infty}_{t} \left(\left(\frac{\partial V^{k+2}}{\partial \tilde{n}}\right)^T\mathbf{F}-\left(\frac{\partial V^{k+1}}{\partial \tilde{n}}\right)^T\mathbf{F} + \left(\frac{\partial V^{k+2}}{\partial \tilde{n}}\right)^T\mathbf{S}\tilde{u}^{k+1}-\left(\frac{\partial V^{k+1}}{\partial \tilde{n}}\right)^T\mathbf{S}\tilde{u}^{k+1}\right)\textrm{d}\tau \nonumber \\
    & = -\int^{\infty}_{t} \left( -N(\tilde{n}) - 2\underline{v}^TR\int^{\tilde{u}^{k+1}}_{\overline{v}}\tanh^{-1}\left(\frac{1}{\underline{v}}\odot (v-\overline{v})\right)\textrm{d}v - \left(\frac{\partial V^{k+2}}{\partial \tilde{n}}\right)^{T} \mathbf{S} \tilde{u}^{k+1} \right. \nonumber \\
    &\qquad\qquad +N(\tilde{n}) + 2\underline{v}^TR\int^{\tilde{u}^{k}}_{\overline{v}}\tanh^{-1}\left(\frac{1}{\underline{v}}\odot (v-\overline{v})\right)\textrm{d}v + \left(\frac{\partial V^{k+1}}{\partial \tilde{n}}\right)^{T} \mathbf{S} \tilde{u}^{k} \nonumber \\
    &\qquad\qquad \left. +\left(\frac{\partial V^{k+2}}{\partial \tilde{n}}\right)^T\mathbf{S}\tilde{u}^{k+1}-\left(\frac{\partial V^{k+1}}{\partial \tilde{n}}\right)^T\mathbf{S}\tilde{u}^{k+1} \right)\textrm{d}\tau \nonumber \\
    & = -\int^{\infty}_{t} \left( 2\underline{v}^TR\int^{\tilde{u}^{k}}_{\tilde{u}^{k+1}}\tanh^{-1}\left(\frac{1}{\underline{v}}\odot (v-\overline{v})\right)\textrm{d}v + \left(\frac{\partial V^{k+1}}{\partial \tilde{n}}\right)^T\mathbf{S}(\tilde{u}^{k}-\tilde{u}^{k+1}) \right)\textrm{d}\tau
\end{align}

Substituting \eqref{eq11_2} into \eqref{NdelVkp}, one obtains
\begin{align}\label{uNdelVkp}
  V^{k+2}(\tilde{n})-V^{k+1}(\tilde{n}) &= -\int^{\infty}_{t} \left( 2\underline{v}^TR\int^{\tilde{u}^{k}}_{\tilde{u}^{k+1}}\tanh^{-1}\left(\frac{1}{\underline{v}}\odot (v-\overline{v})\right)\textrm{d}v \right.
  \left. - 2\underline{v}^T\odot\tanh^{-T}\left(\frac{1}{\underline{v}}\odot(\tilde{u}^{k+1}-\overline{v})\right)R(\tilde{u}^{k}-\tilde{u}^{k+1}) \right)\textrm{d}\tau \nonumber \\
  &= 2\int^{\infty}_{t} \left( \underline{v}^T\odot\tanh^{-T}\left(\frac{1}{\underline{v}}\odot(\tilde{u}^{k+1}-\overline{v})\right)R(\tilde{u}^{k}-\tilde{u}^{k+1}) - \underline{v}^TR\int^{\tilde{u}^{k}}_{\tilde{u}^{k+1}}\tanh^{-1}\left(\frac{1}{\underline{v}}\odot (v-\overline{v})\right)\textrm{d}v \right)\textrm{d}\tau \nonumber \\
  &= 2\int^{\infty}_{t} \left( \sum^{\alpha_u}_{k_u=1}\underline{v}_{k_u}\gamma_{k_u}\tanh^{-1}\left(\frac{\tilde{u}^{k+1}_{k_u}-\overline{v}_{k_u}}{\underline{v}_{k_u}}\right) (\tilde{u}^{k}_{k_u}-\tilde{u}^{k+1}_{k_u}) - \sum^{\alpha_u}_{k_u=1}\underline{v}_{k_u}\gamma_{k_u} \int^{\tilde{u}^k_{k_u}}_{\tilde{u}^{k+1}_{k_u}}\tanh^{-1}\left(\frac{v_{k_u}-\overline{v}_{k_u}}{\underline{v}_{k_u}}\right)\mathrm{d}v_{k_u} \right)\textrm{d}\tau \nonumber \\
  &= 2\sum^{\alpha_u}_{k_u=1}\underline{v}_{k_u}\gamma_{k_u} \int^{\infty}_{t} \left( \tanh^{-1}\left(\frac{\tilde{u}^{k+1}_{k_u}-\overline{v}_{k_u}}{\underline{v}_{k_u}}\right) (\tilde{u}^{k}_{k_u}-\tilde{u}^{k+1}_{k_u}) - \int^{\tilde{u}^k_{k_u}}_{\tilde{u}^{k+1}_{k_u}}\tanh^{-1}\left(\frac{v_{k_u}-\overline{v}_{k_u}}{\underline{v}_{k_u}}\right)\mathrm{d}v_{k_u} \right)\textrm{d}\tau \nonumber \\
  &= 2\sum^{\alpha_u}_{k_u=1}\underline{v}_{k_u}\gamma_{k_u} \int^{\infty}_{t} \left( \varrho(s^{k+1}_{k_u}) (s^{k}_{k_u}-s^{k+1}_{k_u})-\int^{s^{k}_{k_u}}_{s^{k+1}_{k_u}}\varrho(\hat{v}_{k_u}) \textrm{d}\hat{v}_{k_u} \right)\textrm{d}\tau
\end{align}
By \autoref{lemma}, the following inequality holds
\begin{equation*}
  \varrho(s^{k+1}_{k_u}) (s^{k}_{k_u}-s^{k+1}_{k_u})-\int^{s^{k}_{k_u}}_{s^{k+1}_{k_u}}\varrho(\hat{v}_{k_u}) \textrm{d}\hat{v}_{k_u} \leq 0
\end{equation*}
Thus, the right side of \eqref{uNdelVkp} is negative semi-definite. It follows that $V^{k+2}(\tilde{n})-V^{k+1}(\tilde{n})\leq 0$, i.e., $V^{k+2}(\tilde{n})\leq V^{k+1}(\tilde{n})$.

Next, we prove that $V^*(\tilde{n})\leq V^{k+2}(\tilde{n})$.

Since $(V^*,\tilde{u}^*)$, which satisfies \eqref{vbelleq}, is the optimal solution to the COPCP defined by \eqref{eqmulti}-\eqref{UPobj}, for $\forall \tilde{n}\in \Omega$, we have $\min_{\tilde{u}} V(\tilde{n}) = V^*(\tilde{n})$ and $\tilde{u}^* = \arg \min_{\tilde{u}} \int^{\infty}_{t} \mathcal{L}(\tilde{n}(\tau),\tilde{u}(\tau))\mathrm{d}\tau = \arg \min_{\tilde{u}} V(\tilde{n}(t))$.

Suppose $\exists\ k$ such that $V^{k+2}(\tilde{n})$, which also satisfies \eqref{vbelleq}, is smaller than $V^*(\tilde{n})$, i.e., $V^{k+2}(\tilde{n})<V^*(\tilde{n})$. This means that $V^*(\tilde{n})$ is not the optimal solution to the COPCP. By contradiction, we can deduce that $V^*(\tilde{n})\leq V^{k+2}(\tilde{n})$.

3) It follows from the second part of \autoref{thm1} that $\{V^{k}\}^{\infty}_{k=0}$ is a monotonically decreasing sequence with the lower bounded $V^{*}(\tilde{n})$, then $V^{k}$ converges pointwise to $V^{\infty}$. Because of the uniqueness of $V(\tilde{n})$ with $\tilde{n}\in \Omega$ \citep{lewis2012optimal,lyashevskiy1996constrained}, we can get that $V^{\infty}=V^{*}$, which means that $\lim_{k\rightarrow \infty}V^{k}(\tilde{n})=V^{*}(\tilde{n})$.

4) Since $\lim_{k\rightarrow \infty}V^{k}(\tilde{n})=V^{*}(\tilde{n})$, according to \eqref{eq9}, it can be deduced that $\lim_{k\rightarrow \infty}\tilde{u}^{k}(\tilde{n})=\tilde{u}^{*}(\tilde{n})$. The proof is completed.
\eproof

\bibliography{elsaexample}
\bibliographystyle{abbrvnat_custom}
\end{document}